\documentclass[useAMS,usenatbib]{mn2e}
\usepackage[caption=false]{subfig}
\usepackage{fancyhdr}
\usepackage{graphicx}
\usepackage{ifpdf}
\pdfoutput=1
\ifpdf
  \DeclareGraphicsExtensions{.pdf}
\else
  \DeclareGraphicsExtensions{}
\fi
\usepackage{longtable}
\usepackage{amsmath}
\usepackage{lscape}
\usepackage{subfig}

\newcommand{\Msun}{\mbox{\,$\rm M_{\odot}$}}
\newcommand{\Lsun}{\mbox{\,$\rm L_{\odot}$}}
 
\newcommand{\teff}{$T_{\rm eff}$}
\newcommand{\teffsed}{$T_{\rm eff,SED}$}

\newcommand{\logg}{$\log g$}

\title[Post-AGB/RGB stars and YSOs in the SMC]
{Optically visible post-AGB/RGB stars and young stellar objects in the Small Magellanic 
Cloud: candidate selection, 
spectral energy distributions and spectroscopic examination}
\author[D. Kamath et al.]{D. Kamath$^{1,2}$\thanks{E-mail:
Devika.Kamath@ster.kuleuven.be (DK)}, P. R. Wood$^{1}$ and H. Van Winckel$^{2}$\\
$^{1}$Research School of Astronomy and Astrophysics, Mount Stromlo 
Observatory, Weston Creek, ACT 2611, Australia\\
$^{2}$Instituut voor Sterrenkunde, K.U.Leuven, Celestijnenlaan 200D bus 
2401, B-3001 Leuven, Belgium}
\begin{document}

\date{Accepted 1/2014}

\pagerange{\pageref{firstpage}--\pageref{lastpage}} \pubyear{2014}

\maketitle

\label{firstpage}

\begin{abstract}
We have carried out a search for optically visible post-AGB candidates in the Small
Magellanic Cloud (SMC).  Firstly, we used mid-IR
observations from the Spitzer Space Telescope to select optically visible 
candidates with excess mid-IR flux and then we obtained low-resolution optical spectra
for 801 of the candidates.  After removing poor quality spectra and contaminants such as
M-stars, C-stars, planetary nebulae, quasi-stellar objects and background galaxies, we ended up with a final sample of 63 high
probability post-AGB/RGB candidates of A $-$ F spectral type. From the spectral observations, we
estimated the stellar parameters: effective temperature
($T_{\rm eff}$), surface gravity ($\log g$), and metallicity ([Fe/H]). We also
estimated the reddening and deduced the luminosity using the stellar parameters combined
with photometry. For the post-AGB/RGB candidates, we found that the metallicity distribution peaks at [Fe/H] $\approx$ -1.00 dex. 
Based on a luminosity criterion, 42 of these 63 sources were classified as 
post-RGB candidates and the remaining 21 as post-AGB candidates.
From the 
spectral energy distributions we were able to infer that  6 of the 63
post-AGB/RGB candidates have a surrounding circumstellar shell suggesting 
that they are single stars, while 27 of the post-AGB/RGB candidates have a surrounding disc, suggesting
that they lie in binary systems. For the remaining 30 post-AGB/RGB candidates 
the nature of the circumstellar environment was unclear. Variability is displayed by 
38 of the 63 post-AGB/RGB candidates with the most common 
variability types being the Population II Cepheids (including RV-Tauri stars) and semi-regular variables. 
This study has also revealed a new RV Tauri star in the SMC, J005107.19-734133.3, which 
shows signs of $s$-process enrichment. From the numbers of post-AGB/RGB stars in the SMC, 
we were able to estimate evolutionary rates. We find
that the number of post-AGB and post-RGB candidates that we have
identified are in good agreement with the stellar evolution models 
with some mass loss in the post-AGB phase and a small amount of re-accretion in the 
lower luminosity post-RGB 
phase. This study also resulted in a new sample of 40 young stellar objects (YSOs) of A $-$ F spectral type. The 40 YSO candidates for which we could estimate stellar parameters are luminous 
and of high mass ($\sim3-10 \Msun$). They lie on the cool side of the usually adopted birthline in the
HR-diagram. This line separates visually obscured protostars from optically
visible pre-main sequence stars, meaning that our YSO candidates have
become optically visible in the region of the HR-diagram usually
reserved for obscured protostars. Additionally, we also identified a group of 63 objects whose spectra are dominated by 
emission lines and in some cases, a significant UV continuum. These objects are very likely to be either hot post-AGB/RGB candidates or luminous YSOs.

\end{abstract}


\section{Introduction}
\label{intro}
The post-asymptotic giant branch (post-AGB) phase of evolution is a transient phase 
between the asymptotic giant branch (AGB) and planetary nebula (PNe) phases of stellar 
evolution. A super-wind with mass-loss rates up to $10^{-4}$M$_{\odot}$\,yr$^{-1}$ 
generally terminates the AGB phase, reducing the hydrogen-rich envelope 
to small values. Subsequently, the radius of the central 
star decreases and within about $10^{2}-10^{4}$yrs, the star evolves to 
higher temperatures (from 3$\times10^{3}$K on the AGB to $\sim$3$\times10^{4}$K where PN 
formation begins) with a constant luminosity \citep[e.g.][]{schoenberner83,vw93,vw94}. 
The ejected circumstellar 
matter expands and cools during the post-AGB phase, resulting in stars with a large mid-IR excess 
\citep[see][for reviews]{kwok93,vanwinckel03,habing03-bk}. Post-AGB stars emit radiation 
over a broad spectral range owing to a combination of high 
temperatures in  the photosphere and low temperatures in circumstellar dust. 
This enables the simultaneous study of the stellar
photosphere and the circumstellar environment, with the central star 
emitting in the ultra-violet(UV), optical and near-infrared (IR), and the cool circumstellar 
environment radiating in the near- and mid-IR. As the post-AGB star evolves to higher temperatures, the 
ejected circumstellar material gets ionised 
to form a PNe, after which the central star declines in luminosity and cools to become a white dwarf star. 
These phenomenon mark the end of stellar evolution for low-to intermediate-mass 
single stars. 

For stars in binary systems, a different mechanism can terminate the 
red giant evolution. The large expansions that occurs when a star 
becomes a red giant can cause the primary star to over-fill its 
Roche lobe. Theory suggests that Roche lobe overflow during the red giant 
phase will lead to run-away mass transfer on a dynamical or 
thermal timescale, until all that is left of the mass-losing star 
is the He or C/O core of the red giant, orbiting in a binary system: alternatively, 
complete merging 
may occur \citep[e.g.][]{paczynski72,iben96,han95a}. This process occurs 
for binaries with periods on the main sequence in the approximate range 20 $-$ 1000 days, from 
low on the red giant branch (RGB) to the tip of the AGB. The outcome of 
these systems are as of yet difficult to predict as many binary interaction 
process are poorly understood. A 
significant fraction of the ejected matter may end up in a circumbinary 
disc of dust and gas, and inside the disc is a binary system 
containing a post-AGB star or a post-RGB star. The distinction 
between these two cases is as follows: if the primary star fills its 
Roche lobe before reaching the RGB tip\footnote{According to observational studies 
in the Magellanic Clouds \citep[e.g.,][]{frogel83,wood99-pr,cioni99} and 
evolutionary tracks of \citet{bertelli08} 
corresponding to LMC and SMC 
metallicities, the bolometric luminosity of the RGB tip is found to 
be close to $M_{\rm bol} -3.6$ mag.}, it will be referred to as a 
post-RGB star while if the primary star successfully evolves past the 
RGB tip on its second ascent of the giant branch it will be 
referred to as a post-AGB star.

One of the challenges in the study of the post-AGB phase 
of evolution is the identification of post-AGB objects as they have a very short 
lifetime. Since they have dusty circumstellar envelopes, the 
detection of cold circumstellar dust using 
mid-IR photometry is an efficient method to select and study them. 
The first extensive 
search for these objects was initiated in the mid-80's using results from the 
Infrared Astronomical Satellite (IRAS). The large scale mid-IR IRAS 
mission enabled the identification of 
post-AGB stars in the Galaxy \citep{kwok93}. The Toru$\acute{\rm n}$ catalogue \citep{szczerba07} 
for Galactic post-AGB stars lists around 391 very likely post-AGB objects. The Galactic sample of 
optically visible post-AGB objects has 
revealed two highly distinct populations: one with cold, detached, expanding 
dust shells \citep[these probably arise from single stars and produce 'shell' or 'outflow sources',][]{vanwinckel03}, 
and another with hot dust and circumstellar discs \citep[these arise from 
binary stars and are called 'disc sources',][]{deruyter06,vanwinckel07,gielen09,vanwinckel09,dermine12}. 
This is as expected from the single and binary star evolution scenarios described above. 

So far, in the Galaxy, the luminosities (and 
hence initial masses) of the diverse group of post-AGB stars are badly 
affected by their unknown distances, making it difficult to use the observational 
characteristics of these interesting objects to throw light on the 
poorly-understood late stages of stellar evolution.

The Magellanic clouds are 
well suited for the identification and observation of post-AGB stars. The well-constrained 
distances to these extragalactic systems mean that distance-dependent parameters 
such as luminosities can be determined accurately. The Large Magellanic Cloud (LMC) and 
Small Magellanic Cloud (SMC) are both very suitable environments to 
locate individual post-AGB objects and study their evolution as a 
function of initial mass and metallicity.

In the SMC, only 5 possible post-AGB candidates have been identified previously. 
One IRAS source (IRAS 00350-7436) 
is believed to be a post-AGB star \citep{whitelock89}. \citet{kuzinskas00} identified a near-IR object  
in the field of the SMC cluster NGC\,330 using ISO (Infrared Space Observatory) 
observations. This object was classified as a likely post-AGB star in a 
proto-planetary nebula stage, or a Be-supergiant. 
Using low-resolution mid-IR spectra from the Spitzer Space Telescope (SST), 
\citet{kraemer06} identified a possible isolated post-AGB star (MSX SMC 029) in the SMC. 
\citet{volk11} identified two 21 micron sources in the SMC (J004441.04-732136.44 and J010546.40-714705.27). For the 
Galaxy the unidentified 21 micron feature only occurs around c-rich post-AGB stars. J004441.04-732136.44 was also identified in 
a preliminary version of this study by \citet{wood11}, who examined some of the 
brighter sources presented here and found a number of post-AGB candidates. \citet{desmedt12} carried out a detailed 
chemical abundance analysis on J004441.04-732136.44 and found it to be indeed C-rich and an extremely 
\emph{s}-process enriched shell-source.

In the LMC, post-AGB candidates have been identified by \citet{vanaarle11} and references therein. 
\citet{vanaarle11} constructed a catalogue of 1337 optically visible post-AGB candidates 
in the LMC based on mid-IR selection criteria and examination of 
spectral energy distributions (SEDs). They also carried out a spectroscopic examination of 105 candidates which 
resulted in 70 likely post-AGB candidates in the LMC.

To fully understand 
the post-AGB/RGB population in the Magellanic Clouds and to provide a better sample of post-AGB/RGB stars 
whose properties can be compared to models, a larger sample of these objects 
is definitely required. We have therefore 
carried out an extensive search for post-AGB/RGB stars in the SMC, to complement the survey of \citet{vanaarle11} in 
the LMC. In this paper, we present the 
results of our survey of post-AGB/RGB stars in the SMC. This includes a low-resolution 
optical spectral survey of the post-AGB candidates.\footnote{We have also carried out a 
low-resolution optical spectroscopic examination 
for the sample of the LMC post-AGB candidates identified by \citet{vanaarle11}. The results 
will be presented in a following publication (Kamath et al., in preperation).}

The structure of this paper is as follows: In Section~\ref{sampleselection} 
we present the selection criteria used to obtain an initial sample of post-AGB/RGB candidates in the 
SMC. We also present the photometric data, covering the UV, optical and 
IR wavelengths and use it to compute the SEDs for all the sources in our sample. 
Optical spectra along with 
the data reduction procedure and a preliminary spectral 
classification, are presented in Section~\ref{specobv}. In Section~\ref{STP}\,$-$\,~\ref{ms}, we describe  
the method used to estimate the stellar parameters (\teff\,, \logg\,, and [Fe/H]) from the 
spectra, and the reddening from SED fitting. 
Subsequently, in Section~\ref{classification} we present the catalogues of the post-AGB/RGB and YSO candidates. 
In Sections~\ref{sedanalysis}\,$-$~\ref{lcanalysis}, we analyse different characteristics 
of the two populations by examining the SEDs, the optical spectra and the light curves of the 
individual 
candidates. In Section~\ref{completeness} we present the completeness of the survey and in 
Section~\ref{evolrate} we empirically estimate the evolutionary rates of the transient post-AGB/RGB phase. 
Finally, we conclude with a summary of our survey.

\section{Sample Selection, Photometric Data, Extinction and Integrated Luminosities}
\label{sampleselection}

The successful completion of the 
large-scale mid-IR SST surveys: 
SAGE for the LMC \citep{meixner06,blum06} and S$^{3}$MC \citep{bolatto07} and 
SAGE-SMC \citep{gordon11} for the SMC, provides the opportunity to make a new search for post-AGB/RGB star candidates. 
Post-AGB/RGB stars are expected to show excess mid-IR flux, and can therefore be identified 
in the SMC using S$^{3}$MC and 
SAGE-SMC. These surveys cover the 
IRAC (3.6, 4.5, 5.8, and 8 $\mu$m) and the MIPS (24.0, 70.0, and 160.0$\mu$m) bands. 
The Spitzer sources have been merged 
with the Two Micron All Sky Survey (2MASS) $J$, $H$, and $K$ bands 
\citep[1.24, 1.66 and 2.16$\mu$m,][]{skrutskie06} in the S$^{3}$MC and SAGE-SMC 
catalogues.

In our survey, we used the SAGE-SMC catalogue\footnote{By 2010 S$^{3}$MC 
had been superseded by SAGE-SMC, so we did 
not use the former} (including the 2MASS near-IR photometry) along with additional optical and 
mid-IR photometry. Photometry in the $U, V, B,$ and $I$ bands was added from 
the Magellanic Clouds Photometric Survey \citep[MCPS;][]{zaritsky02}. We required the MCPS 
coordinate to lie within 2 arcsec of the SAGE-SMC coordinate for a match. 
For those candidates with no $I$ magnitude from the MCPS survey, we searched for an $I$ magnitude 
in the Denis (Deep Near-Infrared Survey of the Southern Sky) Catalog \citep{epchtein98}. We 
also added WISE photometry in the 3.4, 4.6, 12, and 22 $\mu$m bands 
$W1, W2, W3,$ and $W4$, respectively \citep{wright10}.

In the study by \citet{vanaarle11}, they define criteria to 
select optically visible post-AGB stars in the LMC based on an initial cross-correlation 
of the Spitzer SAGE catalogue with optical catalogues. Their selection criteria require that 
the sources have a MIPS 24$\mu$m detection and also an optical detection in the $U, B, V, R$ or $I$ filters. 
They further narrow their sample by applying a colour selection [8]$-$[24] $>$ 1.384, 
inspired by what is known from Galactic sources. To reject the obvious intruders such as supergiants 
and young stella objects (YSOs), which also have a large mid-IR excess, they impose a luminosity criterion which rejects 
sources with luminosities less than 1000\,\Lsun\, (thereby rejecting low mass YSOs) and greater than 
35000\,\Lsun (thereby rejecting super giants).

To select candidate post-AGB/RGB stars in the SMC we used a 
selection procedure similar to the one used by \citet{vanaarle11}. Our first criteria is that 
all objects should have $V$ $<$ 20, therefore restricting our search to optically visible 
post-AGB/RGB candidates. We also require that all objects should have a 
valid 8$\mu$m detection in the SAGE-SMC catalogue as opposed to the MIPS 24$\mu$m detection required 
for the LMC objects \citep{vanaarle11}, since the objects in the SMC are fainter than in the LMC.

We further narrowed this sample of objects by selecting those candidates that satisfy any one 
of the following five selection criteria, which includes a colour criterion and a luminosity criterion. 
These selection criteria remove extreme AGB stars, red supergiants and low mass YSOs.

\begin{enumerate}
\item{For those candidates that have a 24$\mu$m magnitude, we used the 
[8]$-$[24] $>$ 1.384 and 1000\,$<$\,$L$/L$_{\odot}$\,$< 35000$ criteria 
which were used in the LMC by \citet{vanaarle11}. 
We selected 150 objects in this way and gave them a priority 1 when assigning 
objects for spectroscopic observations.}

\item{To include lower luminosity objects, we also select those candidates that obey the 
same colour-colour criteria as for the priority 1 objects (item (i)) but with a slightly relaxed 
luminosity criterion of 500\,$<$\,$L$/L$_{\odot}$\,$< 35000$. 
We selected 54 objects in this way and gave them a priority 2 when assigning 
objects for spectroscopic observations.}

\item{To further relax the previous selection criteria, we also include 
stars that have [8]$-$ [24]$<$ 1.384 provided $[3.6]-[8] > 1$ with a luminosity criterion of 
500\,$<$\,$L$/L$_{\odot}$\,$< 35000$. This selection 
criteria resulted in 71 objects and we gave these objects also a priority 2 when assigning objects for 
spectroscopic observations.}

\item{Since for majority of the stars in the 
SMC there is no MIPS 24$\mu$m magnitude we also select those sources with 
$I-$[3.6] $>$ 1.5 and $I-$[8] $>$ 0.50 $+$ 1.06$\times$($I-[3.6]$) (a colour-colour cut 
which is defined to select objects such as post-AGB/RGB stars which have hot central stars) 
with a luminosity criterion of $500 <$ $L$/L$_{\odot}$\,$< 35000$. 
This selection criteria resulted in 178 objects and we gave them a 
priority 2 as well when assigning objects for spectroscopic observations.}

\item{To increase the sample of objects available to fill the fibres of the 
multifibre spectrograph (see Section~\ref{specobv}), we used the same colour-colour criterion as for the 
priority 2 objects (items (ii), (iii) and (iv) above) but relaxed the luminosity criterion to 
$100 <$ $L$/L$_{\odot}$\,$< 35000$, thereby including 
lower luminosity objects. This yielded 741 additional objects which were given a 
priority 3 when assigning objects for spectroscopic observations.}
\end{enumerate}

The luminosities for the selected 
sources were obtained by integrating under the SED defined by the photometry after correcting for the effects of reddening. 
All objects in the SMC are subjected to some level of reddening. This 
reddening can be attributed to three possible sources. The first is produced by 
the interstellar dust in our Galaxy, along the line of sight of the SMC. 
This extinction is small with an average of E(B-V) = 0.037 mag 
\citep{schlegel98}. The second source of reddening is from the SMC itself. 
\citet{keller06} derived a mean reddening E(B-V) = 0.12 mag for the combined SMC and 
Galactic components. 
The final source of reddening for the central stars in each object 
is the circumstellar dust envelope. The circumstellar reddening for 
each individual star differs, but it can be estimated from spectroscopic observations 
(see Section~\ref{reddening}).

For all the candidates in the sample the photometry was corrected 
for foreground (SMC plus Galactic) extinction corresponding to 
E(B-V) = 0.12 and Rv = 3.1 using the extinction law of \citet{cardelli89}.
A luminosity for each candidate was obtained by integrating the SED 
that is defined by the photometry corrected for foreground SMC extinction. The integration 
was not extended beyond the longest wavelength point in the flux distribution. Therefore, in objects where the 
flux is still rising at the longest wavelength of observation, the estimated luminosity will be a 
lower limit since large amounts of flux will be missing. The circumstellar 
extinction was not corrected for because it was assumed that all radiation absorbed by the 
circumstellar matter was re-radiated at a longer mid-IR wavelength still within the wavelength 
range of the observed SED. The observed luminosity ($L_{\rm obs}$/L$_\odot$) of each candidate 
was then obtained by applying a distance modulus for the SMC of 18.93 
\citep{keller06}.

\begin{figure}
\includegraphics[width=9cm,angle=0]{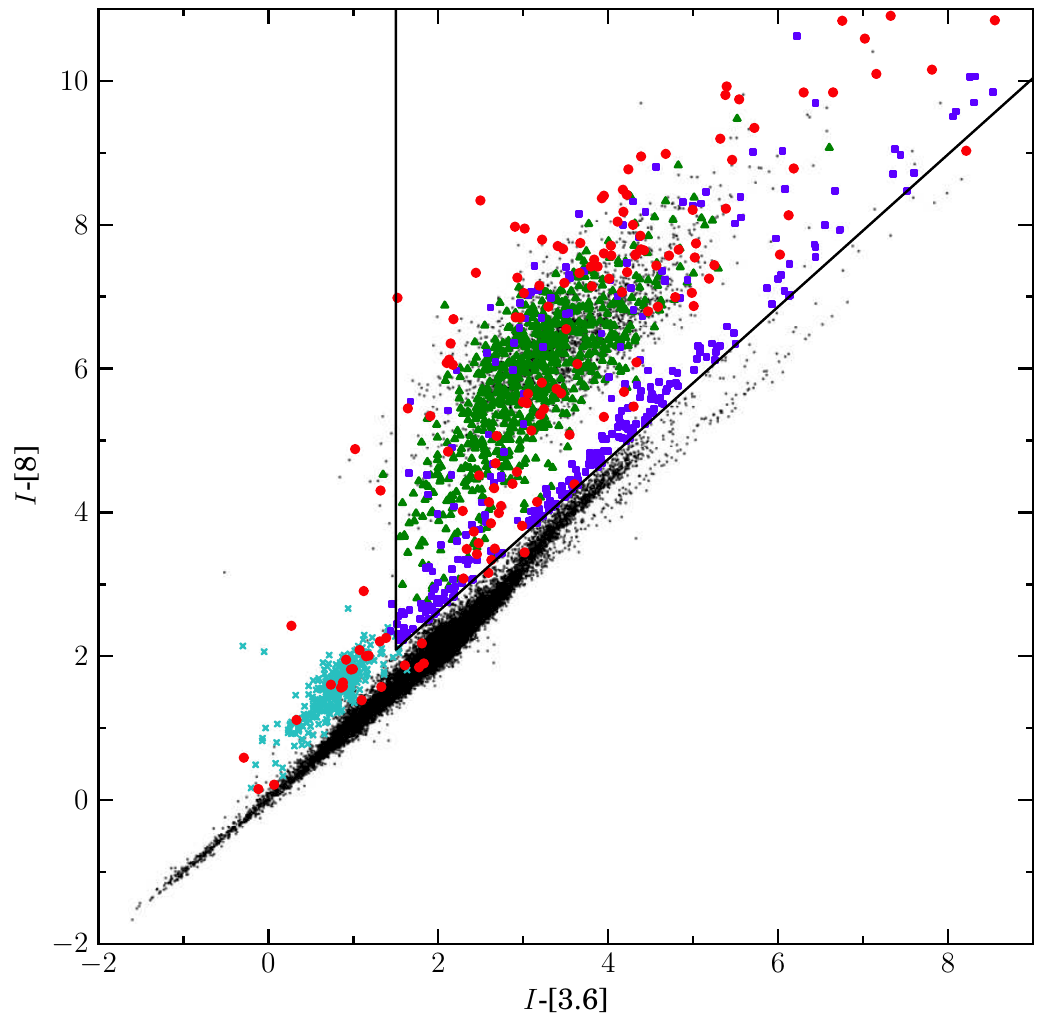}
\caption{The colour-colour plot of I-[8] vs I-[3.6]. The small black dots in the 
background represent the entire field 
SMC population with a valid [8] magnitude from the SAGE-SMC catalogue. The red filled circles represent the 150 priority 1 
candidates. The blue filled squares represent the 303 priority 2 objects and the green filled squares represent the 
741 priority 3 objects. The region within the 
black solid lines defines our selection criterion in this plane. The black dots in this region fail 
the luminosity or $V$ magnitude selection criteria. The cyan crosses near 
($I - [3.6], I - [8]$) = ($1, 1.5$) represent the probable Be star population. 
We note that some of our selected objects fall outside the 
boundaries of this region as these are sources with a MIPS [24] magnitude and were selected
using the colour criteria used for selecting sources with MIPS [24] magnitudes. 
See text for further details.}
\label{fig:sampleselection}
\end{figure}

At this stage we were left with a total of 1194 objects within which we expect to 
find post-AGB/RGB candidates. The positions of the 1194 
objects are marked on the $I$-[8] vs $I$-[3.6] colour-colour plot in 
Figure~\ref{fig:sampleselection}. We represent the priority 1 objects with red filled circles, the 
priority 2 objects with blue filled squares and the priority 3 objects with green filled triangles. 
We note in passing that we also selected a group of 352 probable 
Be star candidates for future study 
(represented with cyan crosses in Figure~\ref{fig:sampleselection}). 
These Be star candidates will not be discussed in this 
paper but will be examined in a following publication.

\begin{landscape}
\begin{table}
\small
\caption{Photometric data. Optical photometry ($U, V, B,$ and $I$ bands) is 
from MCPS, with the $I$ band from the Denis Catalog if the 
MCPS has no I magnitude. $J, H,$ and $K$ photometry is from 2MASS. Mid-IR photometry is 
from the SST survey SAGE-SMC in the 
IRAC (3.6, 4.5, 5.8 and, 8 $\mu$m) and the MIPS (24.0, 72.0, and 160$\mu$m) 
bands, and from WISE in the 3.4, 4.6, 12, and 22 $\mu$m bands ($W1$, $W2$, $W3$, and $W4$). 
The full table is available as online supporting information.}
\vspace{-3mm}
\tabcolsep=2pt
\begin{tabular}{rrrrrrrrrrrrrrrrrrrrr}
\hline
Object Name & RA($^\circ$) & DEC($^\circ$) & $U$ & $B$ & $V$ & $I$ & $J$ & $H$ & $K$ & $W1$ & $[3.6]$ & $[4.5]$ & $W2$ & $[5.8]$ & $[8.0]$ & $W3$ & $W4$ & $[24]$ & $[70]$ & $[160]$ \\
$L_{\rm obs}$/L$_\odot$ & $T_{\rm eff,SED}$(K) &&&&&&&&&&&&&&&&&&\\
\hline
J002417.12-734932.2 & 6.071333 & -73.825611 & 99.999 & 20.444 & 19.432 & 18.357 & 17.525 & 16.951 & 16.168 & 99.999 & 15.836 & 15.697 & 99.999&
14.2 & 12.703 & 99.999 & 99.999 & 99.999 & 99.999 & 99.999\\
101 & 4500 & &&&&\\
J002420.34-733611.2 & 6.08475 & -73.603111 & 99.999 & 21.888 & 17.621 & 14.667 & 13.177 & 12.001 & 11.078 & 99.999 & 9.608 & 9.129 & 99.999&
8.671 & 8.356 & 99.999 & 99.999 & 8.228 & 99.999 & 99.999\\
8141 & 3500 & &&&&\\
J002439.95-734016.8 & 6.166458 & -73.671333 & 19.693 & 20.213 & 19.387 & 18.407 & 16.464 & 15.898 & 15.024 & 99.999 & 14.788 & 14.231 & 99.999&
12.922 & 11.613 & 99.999 & 99.999 & 8.342 & 99.999 & 99.999 \\ 
272 & 4750 & &&&&\\
J002521.94-735519.8 & 6.341417 & -73.922167 & 20.401 & 20.712 & 19.964 & 18.359 & 17.417 & 16.644 & 16.167 & 99.999 & 15.652 & 15.128 & 99.999&
13.798 & 12.468 & 99.999 & 99.999 & 8.926 & 99.999 & 99.999\\
139 & 4750 & &&&&\\
J002546.36-740328.2 & 6.443167 & -74.057833 & 99.999 & 20.783 & 19.696 & 18.625 & 17.004 & 16.31 & 15.869 & 99.999 & 15.305 & 14.737 & 99.999&
14.421 & 11.593 & 99.999 & 99.999 & 8.906 & 99.999 & 99.999\\
177 & 4000 & &&&&\\
\hline
\end{tabular}
\vspace{-3mm}
\begin{flushleft}
Note: The object name is the SAGE name. The RA and DEC coordinates are given for the 
J2000 epoch. Null magnitudes are listed as 99.999. $L_{\rm obs}$/L$_\odot$ is the observed luminosity 
and $T_{\rm eff,SED}$(K) is the photometric temperature (see text for details).
\end{flushleft}
\label{photdatatable}
\normalsize
\end{table}
\end{landscape}

In Table~\ref{photdatatable}, we provide photometric 
magnitudes for the 5 candidates from the selected sample. 
Column (22) of Table~\ref{photdatatable} gives the 
observed 
luminosity of the candidates. We also estimate an effective temperature, 
(\teffsed) presented in Column (23) of Table~\ref{photdatatable}, by fitting 
ATLAS9 atmosphere models \citep{castelli03} 
to the $B, V, I,$ and $J$ bands corrected for foreground extinction. 
However, since the photometry has not been corrected for circumstellar reddening, this estimate of 
photometric \teff\, can only serve as a rough estimate. We determine the true \teff, as well as other 
stellar parameters, from the spectra of the candidates (see Section~\ref{STP}). The 
full table which contains the photometry of the 1194 candidates 
is available as online supporting information.

\section{Spectroscopic Observations}
\label{specobv}

\begin{figure}
\begin{center}
\includegraphics[bb= 0 0 270 248,width=8cm,angle=0]{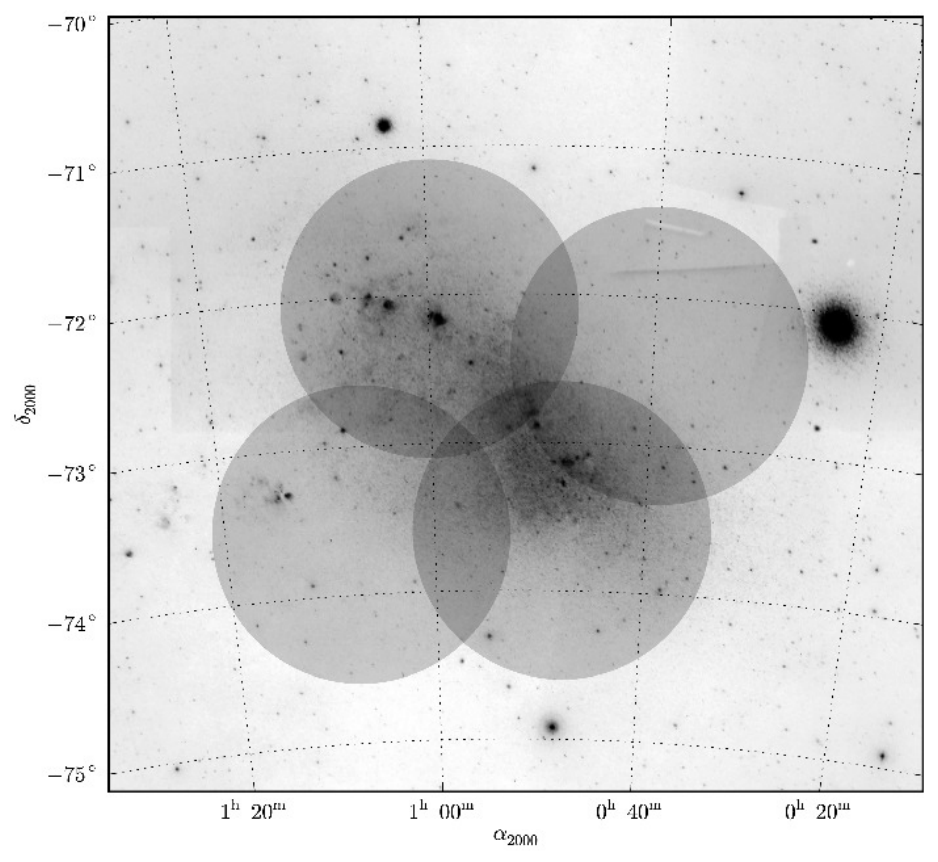} 
\caption{A Digitalised Sky Survey (DSS) image of a 5 degree field of view centered on the SMC. 
The four circles represent the observed 2 degree fields of the SMC that were covered in our survey.}
\label{fields}
\end{center}
\end{figure}

We have conducted a low-resolution optical spectroscopic survey of our SMC post-AGB/RGB candidates. The spectra were 
taken using the AAOmega double-beam multi-fibre spectrograph \citep{sharp06} mounted 
on the 3.9m Anglo Australian Telescope (AAT) at Siding Spring Observatory (SSO). 
AAOmega allows for the 
simultaneous observation of 392 targets 
(including science objects, sky-positions, and fiducial guide stars) over a 
2 degree field using the 2dF fibre positioner \citep{lewis02}. The fibres are of 2 arcsec projected 
diameter. Within each configuration there is a minimum target 
separation of $\sim$30 arcsec imposed by the 
physical size of the fibres \citep{miszalski06}. The positional information for our targets was taken from
the 2MASS Point Source Catalogue \citep{skrutskie06} which has an accuracy of $\sim$\,0.1 arcsecond.
The fibre allocations were done 
automatically using the AAOmega specific CONFIGURE program \citep{miszalski06} which ensures the 
optimal allocation of fibres in terms of maximising the number of objects observed. Thirty fibres 
were assigned to random positions in the sky to sample 
the background sky spectrum. About 5 to 8 fibres were allocated to guide stars. Excluding the defective fibres, 
the remaining fibres were assigned science targets. Altogether, 4 overlapping 2 degree 
diameter fields were chosen to cover the survey region of the SMC. 
The field centers of the SMC observations in our survey are given in Table~\ref{table:fields} and 
shown in Figure~\ref{fields}. A Ne-Ar arc and a quartz lamp flat field exposure were
 taken per field for calibration. We used the 580V grating 
with a central wavelength of 4800\AA\,\, and the 385R grating with a central wavelength of 
7250\AA. This combination provides a maximum spectral coverage of 3700-8800\AA\,\, at a 
resolution of 1300. 

\begin{table}
  \caption{The field centers of the SMC observations in our survey and the corresponding exposure times.}
   \centering
  \begin{tabular}{|cccc|}
  \hline
   Field & RA (2000) & Dec (2000) & Exposure \\
  \hline
   SMC1 & 01 00 00.00 & -72 06 00.0 & 4\,$\times$\,900s \\
   SMC2 & 00 48 00.00 & -73 36 00.0 & 4\,$\times$\,900s \\
   SMC3 & 01 07 12.00 & -73 36 00.0 & 3\,$\times$\,900s \\
   SMC4 & 00 39 36.00 & -72 24 00.0 & 4\,$\times$\,900s \\
  \hline
 \end{tabular}
\label{table:fields}
\end{table}

The AAOmega raw data were reduced using a combination of the AAOmega-2dFDR reduction 
pipeline\footnote{http://www.aao.gov.au/AAO/2df/aaomega/aaomega$\_$software.html} supplied and maintained
by the Anglo-Australian Observatory and 
IRAF\footnote{IRAF is distributed by the National Optical Astronomy Observatory, 
which is operated by the Association of Universities for Research in 
Astronomy (AURA) under cooperative agreement with the National Science Foundation} routines. Prior to reduction
with 2dFDR, bad pixel columns were identified and repaired.
We then used 2dFDR to perform a bias subtraction, cosmic ray rejection, 
flat-field, throughput calibration and sky subtraction on 
the data. For each CCD, the final co-addition of science frames was carried
out after excluding any single exposures with very low flux
relative to the other frames. The 2dFDR splicing routine was used to join 
the red and blue arms. As a consequence of the splicing, for most of the objects, a slight artificial 
slope difference was 
introduced in the blue spectra with respect to the red spectra. The spectra were therefore 
continuum normalised before spectral analysis (see Section~\ref{STP}). 
At the end of the data reduction procedure, a multi-fits file containing the spectra for all the objects observed 
in each field was produced. 

To remove the telluric absorption 
lines imprinted on the spectra, we performed a telluric line removal. For each 
field, we constructed an individual telluric spectrum by 
averaging around five to ten spectra of hot stars. 
All the stellar spectral features were manually removed from the telluric spectrum. 
Telluric line removal was then performed with this artificial telluric spectrum using the IRAF task 
\emph{TELLURIC}. 

We obtained a total of 1060 spectra. Not all the 1194 candidates could be observed since 
some were out of the fields observed and some objects were too close spatially for fibre assignments 
to be done simultaneously. Furthermore, owing to the overlap in the observed fields, some of the target sources 
were observed multiple times. For such sources we either averaged the multiple observations or rejected the 
observations with low signal to noise ratio, retaining only the spectrum with a higher signal to noise. In the end 
we were left with spectra of 801 unique candidates from the initial selected sample of 1194. 
For these 801 candidates, we found that a small fraction of their spectra were affected by sinusoidal continuum 
variations attributed to the fringing patterns exhibited by specific fibres. 
This fringing is caused by a gap between the front end of the fibres and their refractors. The fringing 
is not stable, resulting in unstable spectra \citep{sharp12}. A total of 
18 spectra with severe fringing were discarded, leaving 783 candidates. We 
also found that for 162 objects their spectra had low signal with less than 200 counts. 
We discarded these objects. At this stage we were left with a good spectra of 621 
individual objects.
  
\subsection{Preliminary Spectral Classification}
\label{preclass}

Since our colour and luminosity criteria are not very restrictive, it is very likely that 
the selected sample of post-AGB/RGB candidates is contaminated by objects with similar IR colours. 
Of the 621 candidates for which we have spectra, the intermingled contaminants 
include M stars, C stars, PNe, red shifted galaxies, quasi-stellar objects (QSOs), and 
YSOs which have $L_{\rm obs}$/L$_\odot$\, $>$ 100.

We performed a preliminary spectral analysis, by eye, to categorise the remaining spectra into bins based on the nature of the spectra. We found 20 M-stars identified based on 
the presence of 
strong molecular absorption features of titanium oxide (TiO) and vanadium oxide (VO) 
\citep[e.g.][]{kirkpatrick99}. We identified 
140 C-stars characterised by the presence of key molecules such as C$_{\rm 2}$, 
CN, and CH \citep[see][for a review on C stars]{wallerstein98}. A group of 
204 background galaxies and 36 QSOs were identified by their large redshifts and the width 
of the emission lines of hydrogen and ionic lines \citep[e.g.][]{field73, vandenberk01}. 
Based on the presence of an emission-line spectrum characterised by recombination lines 
of hydrogen and helium as well as various collisionally-excited forbidden lines of heavier elements 
such as O, N, C, Ne, and Ar, we were able to identify 46 PNe 
\citep[see][for further details on identifying PNe]{frew10}. Nine stars in our sample were identified 
to be stars with TiO bands in emission. These objects are discussed in  \citet{wood13} 
and are not considered further here. 

The sample of M-stars, 
C-stars and the PNe is presented in the 
Appendix~\ref{PN_M_C}. The sample of background galaxies and QSOs will 
be discussed in a following publication.

Of the remaining sample of 166 objects, we expect to find not just post-AGB/RGB candidates 
but also luminous massive YSOs. In general, the spectra of luminous massive YSOs are 
similar to those of post-AGB/RGB stars. However, owing to that fact that the post-AGB/RGB candidates 
have different masses (and hence surface gravities) from the 
luminous YSO candidates, one way to separate them is by carrying out a detailed spectral analysis to 
estimate their surface gravities. 

We found that 63 of the remaining 166 objects had prominent emission lines mostly along with a strong UV continuum. These two features are 
characteristic of both hot post-AGB/RGB objects as well as YSOs.  Hot-post AGB/RGB stars are likely to have an emission-line spectrum characterised by weak recombination lines of hydrogen and helium and various collisionally-excited forbidden lines of heavier elements \citep[e.g.,][]{vanwinckel03}. The spectra of YSO candidates are likely to show a broad H$\alpha$ line profile owing to the disc accretion  in YSOs \citep{natta02,jayawardhana02}. Furthermore the YSO objects show a flared SED peaking at longer wavelengths (mostly $>$100$\mu$m). Therefore, we classify these 63 objects as probable hot post-AGB/RGB or YSO candidates.  We do not carry out any further photospheric analysis on these objects, owing to their spectra being completely dominated by emission lines.  The properties of these objects are presented in 
Appendix~\ref{YSO}. Table A1 contains a list of all the objects along with their $L_{\rm obs}$, the
emission lines seen, previous identifications of the objects as
well as the most probable nature of the object (hot post-AGB/RGB or YSO). This classification was 
performed based on a visual inspection of the spectra and SEDs of the candidates resulting in 40 probable hot post-AGB/RGB candidates, 23 probable YSOs. Out of these 23 objects that we classified as YSOs, 6 have been identified previously as YSOs by \citet{2012MNRAS.tmp..229O}. 
The spectra and SEDs of these 63 objects are also shown in  Appendix~\ref{YSO}.

The remaining  103 objects are carried forward for detailed spectral analysis, to search for 
post-AGB/RGB candidates (mostly of A, F, G, K spectral types) and remove the luminous YSO candidates.

\begin{figure*}
\begin{center}
\includegraphics[width=18cm,angle=0]{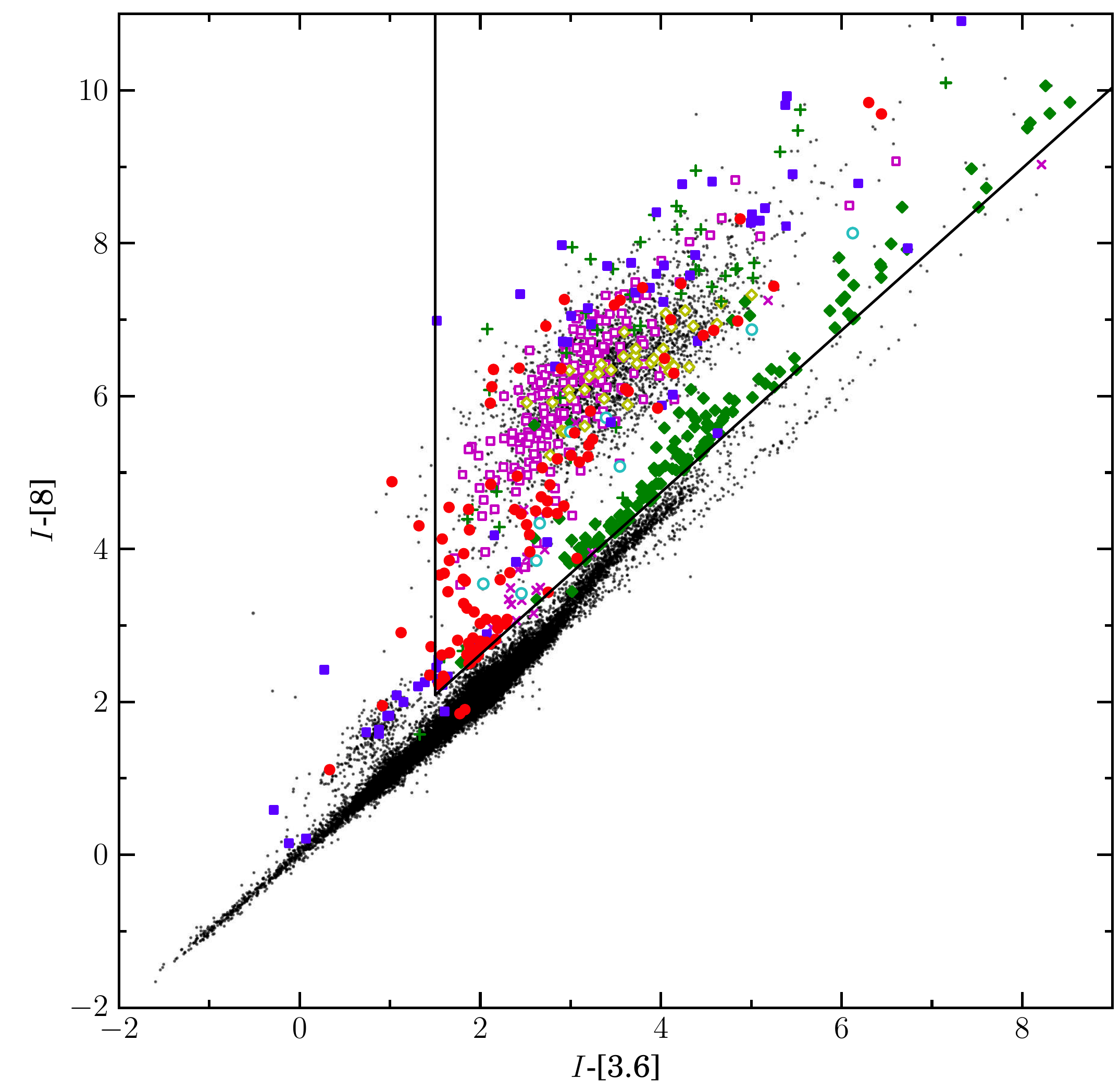} 
\caption{The colour-colour plot of I-[8] vs I-[3.6]. The black dots in the background represent the entire field 
SMC population with a valid 8$\mu$m magnitude from the SAGE-SMC IRAC catalogue. The red 
filled circles represent the 
sample of 103 probable post-AGB/RGB and luminous YSO candidates. The blue filled squares represent the sample of 
63 objects that have strong emission lines and a significant UV continuum. The cyan open circles represent the 9 objects with TiO bands in emission. 
The green plus symbols represent the PN population. The magenta open squares represent the sample that has been classified as background galaxies. The 
yellow open diamonds represent the sources identified as QSOs. The green filled diamonds represent the 
C-stars, and the magenta crosses represent the sources that were classified as M-stars.}
\label{sampleselect2}
\end{center}
\end{figure*}

Figure~\ref{sampleselect2} shows all of the 621 sources with good spectra plotted on 
the colour-colour plot of I-[8] vs I-[3.6] used for our sample selection (see Section~\ref{sampleselection}), 
with the symbol type showing the type of source, as described above.

\subsection{Establishing SMC Membership of the Probable Post-AGB/RGB and YSO Candidates}
\label{rv}

\begin{figure}
\begin{center}
\includegraphics[bb= 0 0 288 216,width=8cm,angle=0]{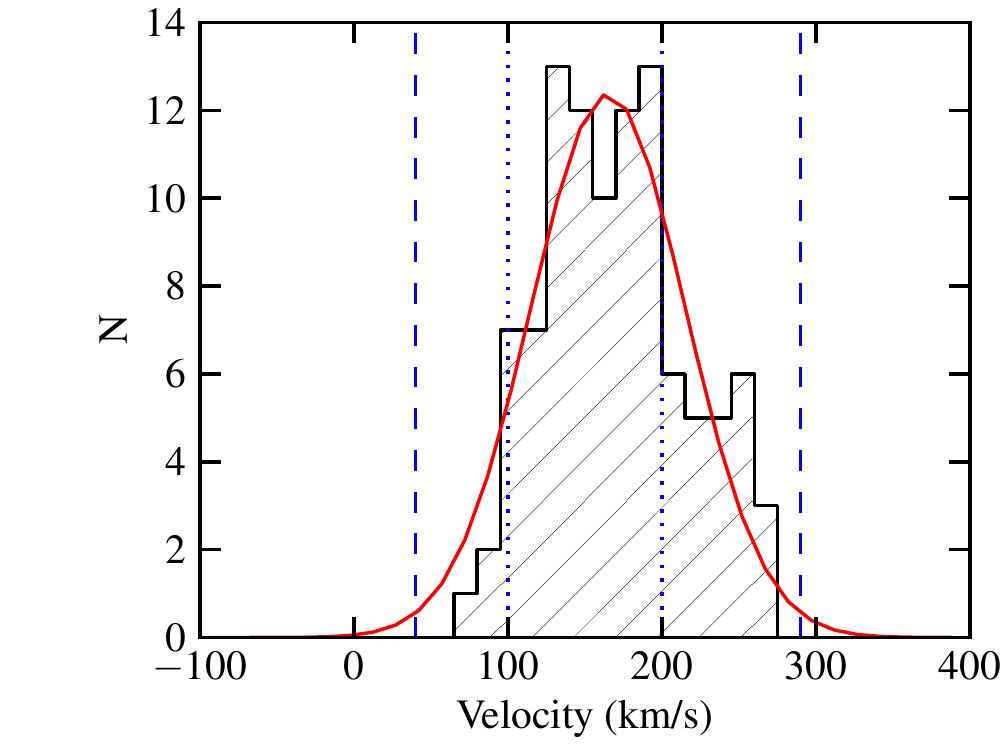}
\caption{Radial velocity histogram for the group of 103 post-AGB/RGB and YSO candidates. 
The red curve denotes a Gaussian with a mean of 165\,km/s and a standard deviation of 50\,km/s. 
The formal standard error in the mean velocity, computed from the individual velocity errors, 
is 6\,km/s. The blue dotted lines mark the low velocity component (at $\sim$106\,km/s) 
and the high velocity component (at 200\,km/s) 
observed by \citet{depropris10} in the SMC. The blue dashed lines denote the radial velocity 
interval (defined by a 
2.5 sigma deviation from the mean) used in our study. See text for further details.}
\label{fig:rvplot}
\end{center}
\end{figure}

To establish SMC membership of the group of 103 probable post-AGB/RGB and YSO candidates, we 
derived heliocentric radial velocities using the Fourier cross-correlation technique. 
For cross-correlation we used only certain regions of the spectra. The regions used were the calcium 
IR triplet (CaT) region from 8400\AA\,\, to 8700\AA\,, 
the Balmer line region from from 3700\AA\,\, to 4000\AA\,, and the H$\alpha$ region from 6250\AA\,\, to 6450\AA\,.
The cross-correlation was done using the routine \emph{FXCOR} in IRAF, using as templates three 
separate individual synthetic spectra that covered the CaT, the Balmer and the H$\alpha$ regions, 
respectively. For each object we estimated three radial velocities resulting from the 
cross-correlation of the three regions of the spectra with the three templates. 
Radial velocity errors for each cross-correlation were 
computed by \emph{FXCOR} based on the fitted peak height and the antisymmetric noise as described by 
\citet{tonry79}. The adopted heliocentric velocity was chosen to be the one with the minimum 
radial velocity error. A typical error in the adopted radial velocity is 6\,km/s. 
We note that in some objects, instead of the CaT absorption lines, there are CaT lines in 
emission. For such objects only two regions (Balmer line region and H$\alpha$ region) 
were used for cross-correlation. This also applies to stars of an earlier spectral type that have 
Paschen lines that lie in the CaT region.

Figure~\ref{fig:rvplot} shows the histogram of the heliocentric velocities for the 103
probable post-AGB/RGB and YSO candidates. A Gaussian fit to the histogram results in an average velocity of 
165\,km/s and standard deviation 
of 50\,km/s. The estimated average velocity agrees well the velocity expected for stars in the SMC. In a radial 
velocity study of red giants in the 
SMC, \citet{depropris10} observe a double peak in the radial velocities of stars, with one component 
peaking at the classical SMC recession velocity of $\sim$160\,km/s and the second component peaking at a higher 
velocity of about $\sim$200\,km/s. They also find a low-velocity component at $\sim$106\,km/s in the most 
distant field of their study (3.9\,kpc from the SMC center). In Figure~\ref{fig:rvplot} the blue dotted lines denote the 
high and low velocity peaks from \citet{depropris10}. Based on these results and also the shape of the velocity distribution 
of our sample of stars, we found that all objects lie well within the range of 40 to 290\,km/s (a 2.5 sigma deviation from the mean radial velocity), 
resulting in a sample of 103 possible post-AGB/RGB and YSO candidates with confirmed SMC membership. The blue dashed lines in Figure~\ref{fig:rvplot} denotes the radial 
velocity interval that we use in our study. 

\section{ Spectroscopic Analysis}
\label{STP}

We developed a fully automated spectral typing pipeline (STP) to simultaneously 
determine the stellar parameters (\teff, \logg\,, and [Fe/H]) of the candidates. The STP matches each
individual observed spectrum to a library of synthetic templates, and finds the minimum RMS 
deviation over a restricted \teff, \logg\,, and [Fe/H] grid. 

The synthetic templates were obtained from the Munari synthetic library 
\citep{munari05} which consists of a grid of nearly 60000 spectra, 
based on the local thermodynamical equilibrium (LTE) Kurucz-Castelli atmosphere models \citep{castelli03}. 
For the purpose of our study, we choose the grid characterised by the following ranges 
of stellar parameters: 3500 $\leq$ \teff\, $\leq$ 47500K, 0.0 $\leq$ \logg\, 
$\leq$ 5.0, -2.5 $\leq$ [Fe/H] $\leq$ 0.5, the micro-turbulence velocity fixed at 2 km/s, 
V$_{rot}$ fixed at 0 km/s, and a scaled solar composition. The STP is designed for normalised 
spectra (to avoid the effects of reddening) of low-resolution (R = 1300) and with a 
wavelength range of 3700 $-$ 8800\AA\,. The synthetic templates used within the STP 
are convolved to match the resolution of the observed spectra.

The spectra of our selected objects are often affected by significant amounts of reddening. This, 
coupled with inherent systematic difficulties in the flux calibration of AAOmega spectra, 
makes the overall continuum slope unreliable for estimating the \teff\, and we restrict 
the use of the slope to only a small interval $\lambda$\,$<$\,4000\AA.

The Balmer lines and slope of the continuum in the region $<$ 4000\AA\,\, provide an 
excellent \teff\, diagnostic for cool stars due to their virtually nil gravity dependence 
\citep{gray92,heiter02}. So for those stars with photometric temperatures 
(see Section~\ref{sampleselection}) cooler than 8000\,K, we make use of the slope of the 
continuum region $<$ 4000\AA. For these spectra the normalisation was carried out 
by splitting each spectrum into two parts: the region blue-ward of 4000\AA\,\, and the 
region beyond 4000\AA. For the region $\leq$ 4000\AA\,, we simply divided the 
spectral flux by the value at 4000\AA. For region of spectrum beyond 4000\AA, we normalised as 
usual using a low order polynomial fit. The two regions were then joined back together. For stars 
with effective temperatures $>$ 8000K we performed a standard 
continuum normalisation using the IRAF task \emph{CONTINUUM}. Finally, the spectra of all the 
candidates were corrected to zero radial velocity using the IRAF task \emph{DOPCOR}. 
The normalised and shifted spectra were then used to derive the stellar parameters. 
The templates were also normalised based on their temperatures, using the same normalisation 
procedure as used for the observed spectra.

For each individual spectrum, the STP performs a grid search over a restricted \teff\, range 
initially centered at the photometric temperature of the candidate\footnote{We note that the 
photometric \teff\, does not skew the final estimated \teff\, in any way. In fact we find that for stars with 
large reddening, the estimated \teff\, by the STP is larger than the photometric \teff\, and for stars with small 
reddening the photometric \teff\, and the estimated \teff\, are similar, as expected.}. The \logg\, values for 
models are constrained to lie between the theoretical values of \logg\, expected for a star in the 
post-AGB/RGB phase and a star in the pre-main sequence phase. To derive the \logg\, the star would 
have in the post-AGB phase, the mass of the post-AGB star is derived using the observed 
luminosity and the luminosity-core mass relation \citep{wood81} for AGB stars 
(candidates with $L$/\Lsun\, $>$ 2500 are assumed to be post-AGB stars). For the post-RGB stars 
($L$/\Lsun $\leq$ 2500), a similar procedure is used but using a luminosity-core mass 
relation derived from a fit to the evolutionary tracks of \citet{bertelli08} with $Z$ = 0.004. 
We note that for post-AGB/RGB stars, the stellar mass is essentially the core mass of the 
progenitor AGB/RGB star. Using the mass, the observed luminosity and the photometric temperature, 
the theoretical \logg,\ is calculated. Similarly, using the PISA pre-main sequence evolutionary tracks 
\citep{tognelli11} for $Z$ = 0.004 (and an extrapolation to higher masses since the 
maximum mass of the PISA tracks is 7\Msun), we can estimate the theoretical value of \logg\, a star would 
have on a pre-main sequence evolutionary track at the observed luminosity and photometric \teff\,. 
We find that for a star of a given luminosity and \teff\,, the value of \logg\, differs by a 
factor of $\sim$\,1.3 between the post-AGB/RGB stars and the pre-main sequence stars of K to G spectral types. 

The restricted grid search reduces computational time and also reduces the occurrence of degeneracies 
between stellar parameters. No restriction is placed on [Fe/H]. Before the spectral matching 
is carried out, each synthetic template is shifted in velocity to that of the observed 
spectrum by cross-correlating the template with the observed spectrum. The template is then resampled 
at the observed wavelength points. For each grid point in the \teff, \logg\,, and [Fe/H] 
space, the RMS deviation between the template and the observed spectrum is calculated as described below. 
The minimum of the RMS deviation gives the best fitting \teff, \logg\,, and [Fe/H]. 

Specific spectral regions are sensitive to specific stellar parameters.  
Our RMS calculations preferentially weight
specific spectral regions to help break the degeneracy that often plagues 
automated spectral classification algorithms. We use weights
$w_1$, $w_2$ and $w_3$ such that,
\begin{equation}
   \label{eq:rms_weight}
   \text{RMS} = \text{RMS}_\text{wide} + w_1 \cdot 
\text{RMS}_\text{\teff} + w_2 \cdot \text{RMS}_{\log g} + w_3 \cdot 
\text{RMS}_\text{[Fe/H]}\,,
\end{equation}
where the weights were determined by trial and error such that the spectral typing
results are consistent with that of previously classified targets, and 
visual classification of selected targets ($w_1 = 70, w_2 = 60, w_3 = 40$).

For those objects whose spectra show the presence of CaT absorption lines or Paschen lines in absorption, 
$\text{RMS}_\text{wide}$ is calculated from the full spectrum, omitting the H$\alpha$ region at 
6500-6650\AA\,\, since the H$\alpha$ line at 6563\AA\,\, is predominantly in emission and it is 
dominated by non-LTE effects. For those objects whose spectra do not show CaT or Paschen lines or if these 
lines are in emission, then we consider only the 
region from 3700\AA\,\, to 6000\AA\,\, when calculating $\text{RMS}_\text{wide}$.

$\text{RMS}_\text{\teff}$, $\text{RMS}_{\log g}$, 
$\text{RMS}_\text{[Fe/H]}$ are calculated using the $T_\text{eff}$, $\log g$, and [Fe/H] 
sensitive spectral regions (see below), respectively. 
The weighted spectral regions we choose, differ for 
targets with different $T_\text{eff}$ estimates. They are listed in Table~\ref{tab:weighted_regions}. 
For example, the Mg b lines near 5170\AA\,\, are indicators of $\log g$ for
stars with $T_\text{eff} \sim 5000\,\text{K}$, while the CaT region from 8400\AA\,\, $-$ 8700\AA\,\, 
serves as a better $\log g$ indicator for stars with $T_\text{eff} \sim 6500\,\text{K}$.

Since our spectral resolution is low, the only preferentially weighted spectral region that can be used 
when calculating $\text{RMS}_\text{[Fe/H]}$ is the CaT region with the CaT lines in absorption, as they 
serve as good metallicity indicators. Therefore, for 
those stars whose spectra do not contain CaT absorption lines (this includes all stars 
with \teff\, $>$ 8000K) [Fe/H] is fixed to -1.0 as we have no [Fe/H] sensitive regions defined and $w_3 = 0$ in these cases.

To estimate the best \teff, \logg\,, and [Fe/H] value, the STP performs a local 
quadratic interpolation of the RMS deviation in a $3\,\times\,3\,\times\,3\,$ grid in \teff, 
\logg\, and [Fe/H] space. At the borders of the parameter space, or in the case of 
missing grid points, we reduce the dimensions of the interpolation accordingly. 
The entire process is performed iteratively by feeding the estimated best-fitting \teff\, value of 
the previous iteration to the new iteration. This is done for a maximum of 20 iterations or 
until the best fitting model does not change on successive iterations.

\begin{table*}
   \caption{The preferentially weighted spectral regions. 
$\lambda_\text{Teff}$ and $\lambda_{\log g}$  indicate the regions 
sensitive to $T_\text{eff}$ and $\log g$, respectively, as a function 
of the \teff estimate. For stars with CaT lines in absorption and $\lambda_\text{[Fe/H]}$ is 
the CaT region from 8450-8700\AA. For stars without CaT absorption lines, their spectra 
have no [Fe/H] sensitive regions defined (w3=0) and [Fe/H] = $-$1 is assumed. See text for more details.}
   \label{tab:weighted_regions}
   \begin{tabular}{|c|c|c|c|}
     \hline
     $T_\text{eff}$ & All Spectra & Spectra with CaT absorption lines & 
Spectra without CaT absorption lines\\
     \hline
& $\lambda_\text{Teff}$ (\AA\,) & $\lambda_{\log g}$ (\AA\,) & $\lambda_{\log g}$ (\AA\,)\\
     \hline
     $<4300$K & 3750-4000&4720-4970&4270-4970\\
     4300-5000K & 3750-4000&5100-5200& 5100-5200\\
     5000-6500K 
&3750-4000,4240-4340&4240-4340,5100-5200&4240-4340,5100-5200\\
     6500-8000K &3750-4000&8350-8800&3750-3950\\
     8000-15000K &3750-6000&&3750-3950\\
     15000-30000K 
&4000-4050,4300-4500&&4000-4050,4300-4500\\
     $>30000$K 
&4000-4050,4300-4500,5610-5830&&4000-4050,4300-4500\\
     \hline
   \end{tabular}
\end{table*}

\begin{table*}
{\renewcommand{\arraystretch}{0.8}
\caption{The observational and stellar parameters for the Q1 post-AGB/RGB 
candidates.}
\medskip
\tabcolsep=2pt
\begin{tabular}{llllllllllllc}
\hline
$\#$& Name & $T_{\rm eff}$(K) & $\log g$ & [Fe/H] & E(B-V) & ($L_{\rm obs}$/L$_\odot$) & ($L_{\rm phot}$/L$_\odot$) & Type & RV\,(km/s) & $M$/M$_\odot$ & SED & Catalogue \\ 
\hline
\multicolumn{13}{c}{Candidates with [Fe/H] estimates from spectra}\\ 
\hline
1 & J003441.01-732908.0$^{\rm c}$ & 4485 & 0.50 & -1.66 & 0.00 & 337 & 226 & p-RGB & $162.1 \pm 6.2$ & 0.32 & Disc & - \\ 
2 & J003643.94-723722.1 & 7458 & 0.50 & -0.63 & 0.48 & 3699 & 8495 & p-AGB & $195.8 \pm 7.4$ & 0.64 & Shell & - \\ 
3 & J003829.99-730334.1 & 5080 & 0.48 & -1.03 & 0.18 & 3919 & 4361 & p-AGB & $152.5 \pm 2.0$ & 0.57 & Uncertain & RSG$^{1}$ \\ 
4 & J003908.89-724314.9  &  5099 & 1.00 &  -0.79 &  0.23  & 562 & 684 & p-RGB & $ 164.7 \pm 6.7$ & 0.38 &  Uncertain & - \\
5 & J003941.74-714708.5$^{\rm b}$ & 4147 & 0.00 & -1.24 & 0.12 & 1778 & 1633 & p-RGB & $155.6 \pm 1.7$ & 0.43 & Uncertain & - \\ 
6 & J004050.18-733226.6 & 5706 & 1.05 & -1.31 & 0.12 & 885 & 894 & p-RGB & $149.9 \pm 2.9$ & 0.39 & Uncertain & - \\ 
7 & J004114.10-741130.1 & 5722 & 0.50 & -1.04 & 0.96 & 3924 & 12000 & p-AGB & $172.5 \pm 5.0$ & 0.70 & Shell & FIR$^{1}$ \\ 
8 & J004441.03-732136.0 & 6168 & 1.00 & -1.07 & 0.82 & 4504 & 12729 & p-AGB & $167.1 \pm 5.5$ & 0.71 & Shell & RGB$^{1}$, p-AGB$^{2}$, [21] $^{3}$\\ 
9 & J004906.52-733136.1 & 5068 & 1.00 & -1.25 & 0.15 &  411 & 452 & p-RGB & $107.5 \pm 6.0$ & 0.35 & Disc & - \\
10 & J004909.72-724745.4 & 6271 & 0.50 & -0.86 & 0.83 & 2882 & 7641 & p-AGB & $203.0 \pm 4.7$ & 0.62 & Uncertain & aO-AGB$^{1}$ \\ 
11 & J004944.15-725209.0 & 5787 & 0.50 & -1.09 & 0.70 & 912 & 2834 & p-AGB & $129.4 \pm 3.0$ & 0.54 & Uncertain & - \\ 
12 & J005107.19-734133.3$^{\rm c}$ & 5767 & 0.72 & -1.56 & 0.09 & 3465 & 1773 & p-RGB & $187.5 \pm 4.3$ & 0.43 & Disc & x-AGB$^{1}$ \\ 
13 & J005159.04-734214.4 & 4240 & 0.66 & -1.01 & 0.20 & 1915 & 2163 & p-RGB & $142.4 \pm 2.5$ & 0.44 & Uncertain & RGB$^{1}$\\ 
14 & J005252.87-722842.9 & 7651 & 1.39 & -1.97 & 0.02 & 8093 & 7338 & p-AGB & $252.7 \pm 10.6$ & 0.62 & Uncertain & - \\ 
15 & J005307.35-734404.5$^{\rm b}$ & 4134 & 0.00 & -1.04 & 0.11 & 1545 & 1472 & p-RGB & $177.8 \pm 2.6$ & 0.42 & Uncertain & RGB$^{1}$ \\ 
16 & J005310.08-722921.0 & 4941 & 0.23 & -1.05 & 0.20 & 7740 & 8456 & p-AGB & $112.5 \pm 1.5$ & 0.64 & Uncertain & RSG$^{1}$ \\ 
17 & J005803.08-732245.1 & 6348 & 0.50 & -1.03 & 1.12 & 4633 & 11868 & p-AGB & $182.0 \pm 5.6$ & 0.70 & Shell & FIR$^{1}$ \\ 
18 & J005914.20-723327.1 & 5002 & 0.50 & -1.48 & 0.06 & 939 & 886 & p-RGB & $125.3 \pm 3.2$ & 0.39 & Uncertain & - \\ 
19 & J010056.93-715551.3$^{\rm b}$ & 4295 & 0.00 & -0.85 & 0.12 & 10767 & 9908 & p-AGB & $130.7 \pm 1.6$ & 0.66 & Uncertain & RSG$^{1}$ \\ 
20 & J010247.72-740151.6$^{\rm c}$ & 4762 & 0.00 & -1.39 & 0.09 & 1211 & 1247 & p-RGB & $164.3 \pm 2.0$ & 0.41 & Uncertain & - \\ 
21 & J010333.93-724405.1$^{\rm c}$ & 4621 & 0.00 & -0.89 & 0.10 & 14969 & 10327 & p-AGB & $190.6 \pm 1.7$ & 0.67 & Disc & O-AGB$^{1}$ \\ 
22 & J011219.69-735125.9 & 6716 & 0.99 & -1.11 & 0.03 & 12334 & 3819 & p-AGB & $274.2 \pm 7.7$ & 0.56 & Disc &  x-AGB$^{1}$, EmO$^{4,5}$\\ 
23 & J011222.88-715820.4$^{\rm b}$ & 4691 & 0.00 & -1.04 & 0.04 & 1445 & 1085 & p-RGB & $131.0 \pm 5.3$ & 0.40 & Disc & - \\ 
\hline
\multicolumn{13}{c}{Candidates with assumed [Fe/H] = -1.00}\\ 
\hline
24 & J004534.36-734811.8  & 5493 &  0.50 & -1.00 &  0.25 &  700 &  787 & p-RGB & $137.5 \pm 4.7$ &  0.38 &  Disc & - \\
25 & J004456.21-732256.6 & 13931 & 2.50 & -1.00 & 0.22 & 5173 & 22243 & p-AGB & $97.9 \pm 20.1$ & 0.87 & Uncertain & FIR$^{1}$, Em*$^{4,5}$\\ 
26 & J004614.67-723519 & 8313 & 1.00 & -1.00 & 0.08 & 1105 & 555 & p-RGB & $202.5 \pm 13.9$ & 0.36 & Shell & - \\ 
27 & J004629.29-731552.3  & 7640 & 1.00 &  -1.00  & 1.07 &  855 & 1679 & p-RGB &$108.0 \pm 15.0$ & 0.43  & Disc & - \\
28 & J004644.05-735944.7 & 16910 & 3.00 & -1.00 & 0.64 & 967 & 4742 & p-AGB & $254.5 \pm 14.9$ & 0.58 & Shell & - \\ 
29 & J004854.24-735651.9 & 5353 & 1.00 & -1.00 & 0.08 & 331 & 223 & p-RGB & $127.0 \pm 7.1$ & 0.32 & Disc & - \\ 
30 & J005104.61-722058.5 & 5652 &  1.50 &  -1.00 &  0.40 &  346 &  542 & p-RGB & $164.1 \pm 4.2$ & 0.36  & Disc & - \\
31 & J005113.04-722227.0  & 5795 & 1.00  & -1.00  & 0.17 &  311 &  327 & p-RGB & $48.5 \pm 20.5$ & 0.34  & Disc &  - \\
32 & J005136.79-722818.0 & 10615 & 2.00 & -1.00  & 0.08 & 3026 &  5180 & p-AGB & $224.1 \pm 8.8 $ & 0.58 & Uncertain & - \\
33 & J005310.47-732800.4 & 5240 & 0.50  & -1.00  & 0.39 & 744 &  1160 & p-RGB & $112.5 \pm  1.5$ &  0.40 & Uncertain & - \\
34 & J005327.75-733339.6 & 7653 & 1.50  & -1.00  & 1.17 & 149 &   854  & p-RGB & $239.0 \pm 15.8$ &0.39  & Disc & - \\
35 & J005506.24-731347.6*  & 11000 & 2.50  & -1.00  &  0.10 & 1814 & 3212 & p-AGB & $252.6 \pm 11.6$ & 0.55 & Uncertain & RGB$^{1}$, Em$^{4}$\\
36 & J005553.75-720859.7 & 8268 & 1.00 & -1.00 & 0.06 & 17601 & 16025 & p-AGB & $152.3 \pm 5.5$ & 0.77 & Disc & RSG$^{1}$, Em*$^{4,5}$\\ 
37 & J005908.99-710648.6  & 5410 & 0.00 & -1.00 &  0.56 & 3178 &  8165  & p-AGB & $204.8 \pm 8.8$ &0.63  & Disc & - \\
38 & J010342.34-721342.7 & 8265 & 1.00 & -1.00 & 0.13 & 14521 & 10944 & p-AGB & $176.0 \pm 4.9$ & 0.68 & Disc & x-AGB$^{1}$, C*$^{4}$\\ 
\hline
\label{tab:pagb1_param}
\end{tabular}}
\begin{flushleft}
Notes: ($L_{\rm obs}$/L$_\odot$) is the observed luminosity corrected for foreground 
extinction, ($L_{\rm phot}$/L$_\odot$) 
is the photospheric luminosity of the central star, Type is the estimated evolutionary 
status (p-RGB for a post-RGB star, p-AGB for a post-AGB star), $M$/M$_\odot$ is the derived 
mass of the post-AGB/RGB candidate (see Section~\ref{STP}).\\
*For J005506.24-731347.6 the spectra shows slight traces of the HeI lines at 
4471\AA and 5876\AA, which is not clearly detected by 
the spectral tryping pipeline. Therefore the 
estimated \teff\, for this object is probably a lower limit for the real 
\teff\, of the star.\\
The subscripts 'b', 'c', and 'd' represent the methods used to
estimate the E(B-V) value. See text for further details.\\
A positional matching was found with the following catalogues: 
$^{1}$\citet{boyer11},
$^{2}$\citet{desmedt12},
$^{3}$\citet{volk11},
$^{4}$\citet{1993A&amp;AS..102..451M},
$^{5}$\citet{2000MNRAS.311..741M}.
The results of the matching are listed in the last column. 
Catalogue identifications: C* - Carbon star; Em*,EmO - object with emission features; p-AGB - post-AGB star, [21] - 21 micron source. 
The following objects are defined in \citet{boyer11}: RGB - red giant branch star, RSG - 
red supergiant, FIR - far-IR object, 
x-AGB - dusty AGB star with superwind mass loss, aO-AGB - anomalous O-rich AGB star, O-AGB - O rich AGB, 
C-AGB - C rich AGB. 
Throughout this contribution, positional cross-matching was performed
with the following catalogues: 
\citet{desmedt12},
\citet{volk11},
\citet{2012MNRAS.tmp..229O}, 
\citet{2009AJ....138.1003B}, 
\citet{2000A&amp;A...363..901G},
\citet{1997A&amp;AS..125..419L},
\citet{1999A&amp;A...346..843T},
\citet{2001A&amp;A...369..932K},
\citet{2007MNRAS.376.1270L},
\citet{2010AJ....139.1553V},
\citet{1978ApJ...221..586S},
\citet{1980ApJS...42....1J},
\citet{2002AJ....123..269J},
\citet{1985MNRAS.213..491M},
\citet{1995A&amp;AS..112..445M},
\citet{1981PASP...93..431S},
\citet{1993A&amp;AS..102..451M},
\citet{2000MNRAS.311..741M},
\citet{2003A&amp;A...401..873W} (12$\mu$m),
\citet{2003A&amp;A...401..873W} (25$\mu$m),
\citet{2003A&amp;A...401..873W} (60$\mu$m),
\citet{2003A&amp;A...401..873W} (100$\mu$m),
\citet{2003A&amp;A...401..873W} (170$\mu$m), 
\citet{boyer11},
\citet{simon07}.\\
\end{flushleft}
\end{table*}

\begin{table*}
{\renewcommand{\arraystretch}{0.8}
\caption{The observational and stellar parameters for the Q2 post-AGB/RGB 
candidates.}
\medskip
\tabcolsep=2pt
\begin{tabular}{llllllllllllc}
\hline
$\#$ & Name & $T_{\rm eff}$(K) & $\log g$ & [Fe/H] & E(B-V) & ($L_{\rm obs}$/L$_\odot$) & ($L_{\rm phot}$/L$_\odot$) & Type & RV\,(km/s) & $M$/M$_\odot$ & SED & Catalogue \\ 
\hline
\multicolumn{13}{c}{Candidates with [Fe/H] estimates from spectra}\\ 
\hline
39 & J003611.06-730447 & 5392 & 0.50 & -1.50 & 0.26 & 497 & 617 & p-RGB & $126.6 \pm 9.4$ & 0.37 & Disc & - \\ 
40 & J003818.36-731120.7 & 5150 & 1.48 & -1.76 & 0.48 & 277 & 261 & p-RGB & $120.3 \pm 12.4$ & 0.32 & Disc & - \\ 
41 & J003946.58-730433.5$^{\rm b}$ & 4500 & 0.49 & -1.10 & 0.04 & 670 & 588 & p-RGB & $141.0 \pm 2.5$ & 0.37 & Uncertain & - \\ 
42 & J004215.31-740219.1$^{\rm c}$ & 4827 & 0.50 & -0.81 & 0.06 & 414 & 348 & p-RGB & $99.7 \pm 4.5$ & 0.34 & Disc & - \\ 
43 & J004431.23-730549.3$^{\rm c}$ & 4509 & 0.00 & -1.49 & 0.10 & 1268 & 1250 & p-RGB & $151.5 \pm 1.4$ & 0.41 & Uncertain & - \\ 
44 & J005222.19-733537.6 & 4176 & 0.57 & -1.03 & 0.09 & 967 & 949 & p-RGB & $187.0 \pm 2.4$ & 0.39 & Uncertain & - \\ 
45 & J005311.41-740621.2 & 4209 & 0.00 & -0.58 & 0.89 & 3974 & 3861 & p-AGB & $215.3 \pm 3.5$ & 0.56 & Disc & C-AGB$^{1}$ \\ 
46 & J005447.59-740121.4$^{\rm c}$ & 4342 & 0.00 & -0.90 & 0.13 & 1732 & 1510 & p-RGB & $133.3 \pm 3.7$ & 0.42 & Uncertain & FIR$^{1}$ \\ 
47 & J005515.71-712516.9 & 4695 & 1.33 & -1.13 & 0.22 & 811 & 992 & p-RGB & $129.3 \pm 4.0$ & 0.40 & Uncertain & - \\ 
48 & J005526.37-723248.7 & 5136 & 0.84 & -1.23 & 0.26 & 851 & 1064 & p-RGB & $136.8 \pm 3.0$ & 0.40 & Uncertain & - \\ 
49 & J005658.04-735059.7 & 5169 & 0.50 & -1.36 & 0.30 & 384 & 536 & p-RGB & $135.6 \pm 4.7$ & 0.36 & Uncertain & RGB$^{1}$ \\ 
50 & J005925.13-741309.6 & 3951 & 0.00 & -1.14 & 0.10 & 3167 & 3006 & p-AGB & $135.5 \pm 2.6$ & 0.55 & Uncertain & aO-AGB$^{1}$ \\ 
51 & J005941.66-742842.9 & 4983 & 1.50 & -2.21 & 0.13 & 282 & 259 & p-RGB & $198.8 \pm 14.2$ & 0.33 & Disc & - \\ 
52 & J010021.78-730901.3 & 4834 & 0.00 & -1.32 & 0.32 & 1067 & 954 & p-RGB & $174.7 \pm 3.4$ & 0.39 & Disc & - \\ 
53 & J010049.88-723459.7$^{\rm c}$ & 4564 & 0.00 & -1.30 & 0.16 & 1174 & 1164 & p-RGB & $151.0 \pm 2.4$ & 0.41 & Uncertain & - \\ 
54 & J010254.90-722120.9$^{\rm b}$ & 4111 & 0.00 & -1.08 & 0.04 & 1632 & 1114 & p-RGB & $116.6 \pm 2.7$ & 0.40 & Disc & - \\ 
55 & J010304.72-721245.3$^{\rm c}$ & 4578 & 0.62 & -1.10 & 0.12 & 766 & 796 & p-RGB & $151.1 \pm 3.3$ & 0.38 & Uncertain & RGB$^{1}$ \\ 
56 & J010310.25-730602.7 & 5812 & 1.51 & -0.65 & 0.18 & 493 & 540 & p-RGB & $221.6 \pm 4.0$ & 0.36 & Disc & - \\ 
57 & J010404.07-723521.5 & 5519 & 0.50 & -1.00 & 0.50 & 751 & 1530 & p-RGB & $156.0 \pm 1.9$ & 0.42 & Uncertain & RGB$^{1}$\\ 
58 & J010549.25-725028.9 & 5490 & 0.00 & -1.31 & 0.44 & 1359 & 1620 & p-RGB & $191.1 \pm 2.1$ & 0.42 & Disc & - \\ 
59 & J010623.71-724413.5 & 4178 & 0.00 & -0.79 & 0.11 & 1962 & 1291 & p-RGB & $193.9 \pm 2.0$ & 0.41 & Disc & - \\ 
60 & J010814.67-721306.2 & 6340 & 0.50 & -0.63 & 1.14 & 1200 & 4707 & p-AGB & $157.0 \pm 7.4$ & 0.57 & Disc & - \\ 
61 & J011133.41-733300.6$^{\rm b}$ & 4444 & 0.00 & -1.17 & 0.13 & 910 & 914 & p-RGB & $149.9 \pm 2.3$ & 0.39 & Uncertain & - \\ 
62 & J011149.89-720822.4 & 4366 & 0.00 & -1.17 & 0.03 & 788 & 625 & p-RGB & $157.1 \pm 4.4$ & 0.37 & Uncertain & - \\ 
\hline
\multicolumn{13}{c}{Candidates with assumed [Fe/H] = -1.00}\\ 
\hline
63 & J003549.26-740737.9 & 8250 & 1.00 & -1.00 & 0.17 & 799 & 564 & p-RGB & $223.8 \pm 10.1$ & 0.36 & Disc & - \\ 
\hline
\label{tab:pagb2_param}
\end{tabular}}
\begin{flushleft}
Notes: As for Table~\ref{tab:pagb1_param}. A positional cross-matching
was found with the following catalogue: $^{1}$\citet{boyer11}. Catalogue identifications: The following objects are defined in \citet{boyer11}: RGB - red giant branch star, 
FIR - far-IR object, aO-AGB - anomalous O-rich AGB star, C-AGB - C rich AGB.\\
\end{flushleft}
\end{table*}

\begin{table*}
{\renewcommand{\arraystretch}{0.8}
\caption{The observational and stellar parameters for the Q1 YSO 
candidates.}
\medskip
\tabcolsep=2pt
\begin{tabular}{lllllllllc}
\hline
$\#$ &Name & $T_{\rm eff}$(K) & $\log g$ & [Fe/H] & E(B-V) & ($L_{\rm obs}$/L$_\odot$) & ($L_{\rm phot}$/L$_\odot$) & RV\,(km/s) & Catalogue \\ 
\hline
\multicolumn{10}{c}{Candidates with [Fe/H] estimates from spectra}\\ 
\hline
1 & J004927.26-724738.4 & $ $7596 & 2.46 & -1.22 & 0.11 & 2175 & 2126 & $248.3 \pm 9.7$ & RGB$^{1}$\\ 
2 & J004905.36-721959.9 & $ $5042 & 2.50 & -0.50 & 0.41 &  867 &  1398 & $116.6 \pm  2.8$ & RGB$^{1}$\\
3 & J004949.43-731327.3 & $ $5303 & 2.13 & -0.61 & 0.37 & 761 & 1360  & $182.5 \pm 2.4$ & RGB$^{1}$ \\ 
4 & J010134.91-720605.4 & $ $5162 & 2.00 & -0.81 & 0.24 & 1988 & 2572 & $193.1 \pm 1.6$ & - \\ 
5 & J010222.29-724502.6 & $ $5460 & 2.50 & -0.97 & 0.27 & 842 & 1242  & $189.2 \pm 3.7$ & RGB$^{1}$ \\ 
6 & J010648.26-734305.4 & $ $7020 & 2.50 & -0.16 & 0.10 & 2950 & 2860 & $248.1 \pm 5.9$ & - \\ 
7 & J011316.84-733130.9 & $ $4500 & 2.00 & -0.56 & 0.18 &  762 & 854  & $161.8 \pm 2.5$ & RGB$^{1}$ \\
\hline
\multicolumn{10}{c}{Candidates with assumed [Fe/H] = -1.00}\\ 
\hline
8 & J003640.64-740747.2 & $ $5901 & 3.00 & -1.00 & 0.50 &    401 &   794 & $208.1 \pm  9.6$ & - \\
9 & J004208.74-733108.4 & $ $16000 & 3.36 & -1.00 & 0.57 & 2327 &  6205 & $191.0 \pm 16.3$ & IRAS60$^{2}$ \\ 
10 & J004221.85-732417.5 & $ $7627 & 3.00 & -1.00 & 0.38 &  1336 &  2813 & $198.4 \pm  8.0$ & - \\ 
11 & J004301.63-732050.9 & $ $7338 & 3.50 & -1.00 & 0.12 &  400 &   238 & $189.9 \pm 12.7$ & - \\ 
12 & J004451.87-725733.6 & $ $7625 & 2.50 & -1.00 & 1.14 & 6027 &1193&$273.9 \pm 30.3$ & FIR$^{1}$, YSO$^{3}$, IRAS60$^{4,5}$\\
13 & J004501.19-723321.0 & $ $7631 & 3.00 & -1.00 & 0.15 &  846 &   966 & $274.0 \pm 18.5$ & - \\
14 & J004503.51-731627.4 & $ $7625 & 2.50 & -1.00 & 0.31 & 3417 & 479 & $ 83.7 \pm 15.0$ & FIR$^{1}$, IRAS25$^{6}$\\
15 & J004657.45-731143.4 & $ $6674 & 3.00 & -1.00 & 0.39 &  613 &   335 & $149.1 \pm  6.0$ & - \\
16 & J004831.82-720535.7 & $ $7540 & 4.00 & -1.00 & 0.12 &  129 &    75 & $144.8 \pm 11.1$ & - \\ 
17 & J004840.55-730101.3 & $ $6310 & 2.50 & -1.00 & 0.01 & 2129 &  1651 & $153.7 \pm 13.8$ & Ce*$^{7}$ \\ 
18 & J004950.02-734011.5 & $ $5407 & 2.50 & -1.00 & 0.04 &  757 &   692 & $ 65.1 \pm  3.5$ & -\\
19 & J005101.48-733100.4 & $ $7639 & 3.00 & -1.00 & 1.11 & 1568 & 10101 & $102.9 \pm  2.0$ & - \\
20 & J005112.29-722552.7 & $ $5481 & 2.00 & -1.00 & 0.61 & 1080 &  2495 & $176.5 \pm  3.0$ & o-AGB$^{1}$\\
21 & J005159.81-723511.1 & $ $7633 & 3.00 & -1.00 & 0.00 &  993 &   783 & $182.9 \pm 10.1$ & - \\
22 & J005318.28-733528.7 & $ $7643 & 2.98 & -1.00 & 0.12 &  356 &   325 & $116.6 \pm 17.7$ & - \\
23 & J005606.53-724722.7 & $ $6733 & 2.50 & -1.00 & 0.20 & 2616 &   527 & $ 89.8 \pm  9.3$ & YSO$^{3}$,Em*$^{7}$\\
24 & J005934.21-733025.2 & $ $7628 & 3.50 & -1.00 & 0.64 &  197 &   501 & $184.2 \pm 15.5$ & RGB$^{1}$\\  
25 & J010242.25-720306.0 & $ $5797 & 3.00 & -1.00 & 0.32 &  303 &   440 & $210.6 \pm  7.4$ & - \\
26 & J010309.59-715354.2 & $ $7661 & 3.00 & -1.00 & 0.48 & 2428 &  2199 & $232.3 \pm 20.0$ & FIR$^{1}$ \\
27 & J010427.62-721037.0 & $ $7638 & 3.50 & -1.00 & 0.20 &  464 &   334 & $204.1 \pm  6.6$ & - \\
\hline

\label{tab:yso1_param}
\end{tabular}}
\begin{flushleft}

Notes: As for Table~\ref{tab:pagb1_param}. A positional cross-matching
was found with the following catalogues: 
$^{1}$\citet{boyer11},
$^{2}$\citet{2003A&amp;A...401..873W} (60$\mu$m),
$^{3}$\citet{2012MNRAS.tmp..229O},
$^{4}$\citet{1997A&amp;AS..125..419L},
$^{5}$\citet{2010AJ....139.1553V},
$^{6}$\citet{2003A&amp;A...401..873W} (25$\mu$m),
$^{7}$\citet{1993A&amp;AS..102..451M}.
Catalogue identifications: Ce* - Cepheid variable; Em* - object with 
emission features; IRAS25 - IRAS source at 25$\mu$m; IRAS60 - IRAS source at 60$\mu$m; Y*O - Young stellar object; RGB - red giant branch star; 
o-AGB - O-rich AGB; FIR -Far-IR object \citep[defined in][]{boyer11}.\\
\end{flushleft}
\end{table*}

\begin{table*}
{\renewcommand{\arraystretch}{0.8}
\caption{The observational and stellar parameters for the Q2 YSO 
candidates.}
\medskip
\tabcolsep=2pt
\begin{tabular}{lllllllllc}
\hline
$\#$ & Name & $T_{\rm eff}$(K) & $\log g$ & [Fe/H] & E(B-V) & ($L_{\rm obs}$/L$_\odot$) & ($L_{\rm phot}$/L$_\odot$) & RV\,(km/s) & Catalogue \\ 
\hline
\multicolumn{10}{c}{Candidates with [Fe/H] estimates from spectra}\\ 
\hline
28 & J004423.32-733343.5 & 5928 & 0.25 & -0.12 & 0.57 & 574 & 1422 & $182.7 \pm 4.3$ & - \\ 
29 & J004830.67-735428.0$^{\rm b}$ & 4643 & 3.00 & -1.55 & 0.08 & 310 & 275 & $123.6 \pm 10.2$ & \\ 
30 & J004853.37-714952.5 & 4191 & 1.50 & -0.68 & 0.06 & 736 & 655 & $150.3 \pm 2.2$ & RGB$^{1}$\\ 
31 & J005143.07-721700.7 & 5288 & 2.50 & -0.29 & 0.47 & 695 & 1140 & $189.6 \pm 3.7$ & - \\ 
32 & J005800.62-721439.8 & 5160 & 2.50 & -0.54 & 0.32 & 598 & 854 & $134.1 \pm 3.0$ & - \\ 
33 & J010051.48-710844.9 & 5062 & 2.50 & -0.60 & 0.10 & 929 & 932 & $117.5 \pm 2.9$ & RGB$^{1}$ \\ 
34 & J010441.50-734021.5 & 5919 & 2.44 & -0.64 & 0.60 & 229 & 484 & $174.9 \pm 4.1$ & - \\ 
\hline
\multicolumn{10}{c}{Candidates with assumed [Fe/H] = -1.00}\\ 
\hline
35 & J004547.50-735331.7 & 6847 & 3.50 & -1.00 & 0.12 & 177 & 81 & $224.1.0 \pm 13.0$ & - \\ 
36 & J004707.49-730259.2 & 6250 & 3.00 & -1.00 & 0.51 & 744 & 463 & $179.1 \pm 14.6$ &- \\ 
37 & J010634.50-721505.0 & 7650 & 3.00 & -1.00 & 0.17 & 907 & 986 & $234.3 \pm 16.5$ & Em*$^{2}$ \\ 
38 & J011109.79-714226.9 & 7634 & 2.85 & -1.00 & 1.19 & 196 & 1008 & $231.3 \pm 27.7$ & - \\ 
39 & J011229.23-724511.6 & 7633 & 3.00 & -1.00 & 0.12 & 1311 & 226 & $242.7 \pm 22.9$ & - \\ 
40 & J011302.68-724852.5 & 7532 & 3.00 & -1.00 & 0.19 & 823 & 521 & $105.3 \pm 17.1$ & - \\ 
\hline
\label{tab:yso2_param}
\end{tabular}}
\begin{flushleft}
Notes: As for Table~\ref{tab:pagb1_param}. A positional cross-matching
was found with the following catalogues: $^{1}$\citet{boyer11},
$^2$\citet{1993A&amp;AS..102..451M}. 
Catalogue identifications: Em* - object with emission features; RGB - red giant branch star.\\
\end{flushleft}
\end{table*}

\begin{figure}
\begin{center}
\includegraphics[bb=0 0 454 1008,width=7cm,height=15cm,angle=0]{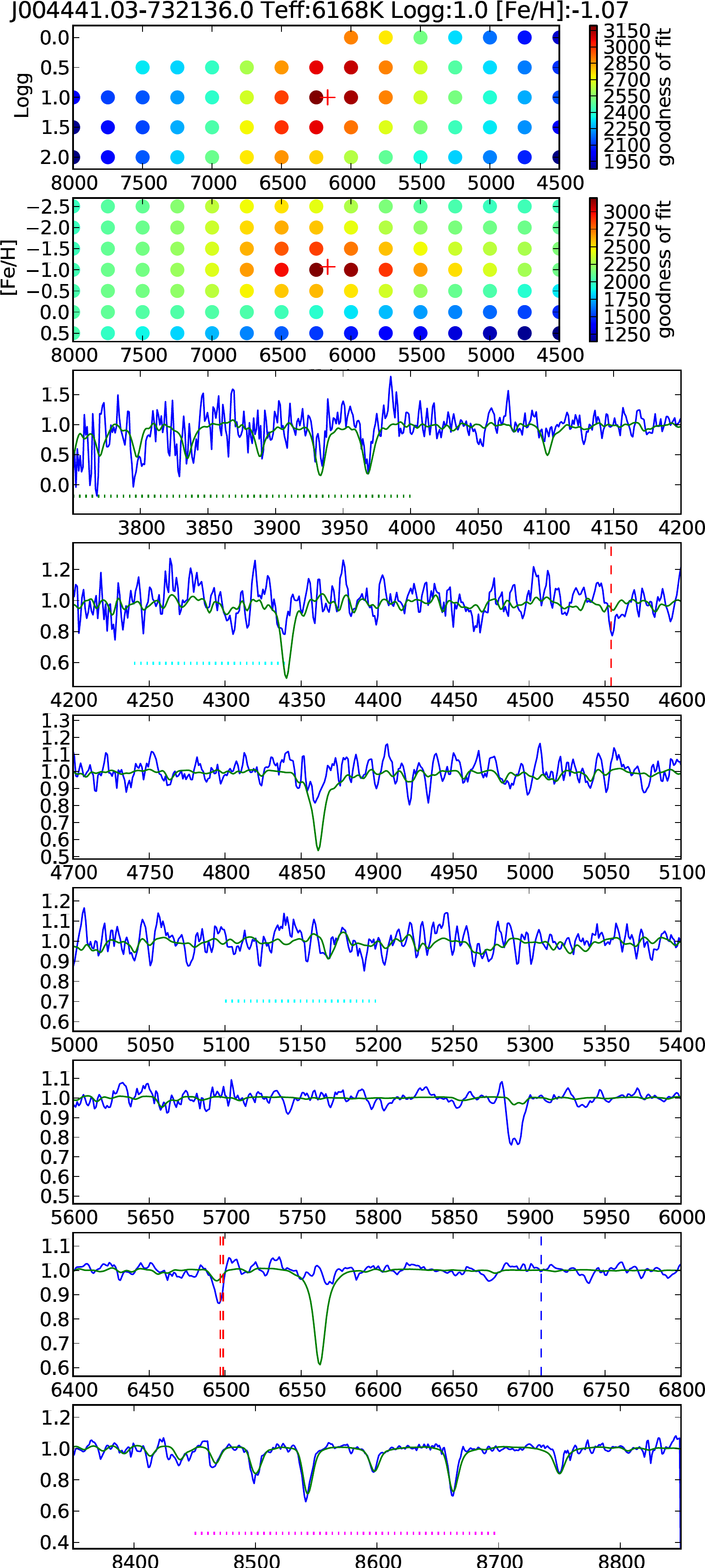}
\caption{Spectral typing result for J004441.04-732136.4. The first 
two subplots show the inverse RMS distribution (goodness of fit) in the 
\teff$-$\logg\, space and \teff$-$[Fe/H] space with a red plus representing the 
final interpolated value. In the remaining subplots, the observed spectrum is shown in blue, 
whilst the green line 
represents the best-fitting synthetic spectrum. The  green, cyan and pink
 dotted horizontal dotted lines represents the 
preferentially weighted \teff, \logg\, and [Fe/H] regions (see text for details). 
The blue dashed vertical line marks the position of the LiI lines. The single red dashed 
marks the position of the BaII line at 4554.03\AA\,\, and the double red dashed lines mark 
the positions of the BaII line at 6496.89\AA\,, and the BaI line at 6498.76\AA.}
\label{stpplot}
\end{center}
\end{figure}

As an example, the fitting result for a candidate (J004441.04-732136.4) is shown in Figure~\ref{stpplot}. 
The plot presents the inverse RMS  distribution (goodness of fit) in the
\teff$-$\,\logg\, space and the \teff$-$\,[Fe/H] space, with the interpolated final values marked.
The preferentially weighted spectral 
regions used during the spectral typing process are plotted and important spectral 
features are also indicated. Depending on the strength of the spectral features, lines such 
as the BaII line at 4554.03\AA\,, the BaII line at 6496.89\AA\,, and the BaI line at 6498.76\AA\,\, 
can be detected. These are indicators of 
$s$-process enrichment and are marked on each plot for the identification of the sources in which we can 
identify $s$-process 
enrichment. Similarly, LiI lines at 6707.77\AA\,\, and 6707.92\AA\,\, are also marked for identification of 
Li in the candidates (although the two lines are unresolved at the resolution of 
our spectra and their position is indicated by a single blue dashed vertical line in Figure~\ref{stpplot}.)

The 103 objects that were fed into the STP had their \teff, \logg\,, and [Fe/H] derived. 
We obtained 63 post-AGB/RGB candidates and 
40 YSO candidates (see Section~\ref{classification}). The estimated \teff, \logg\,, and 
[Fe/H] for our final sample of high probability post-AGB/RGB 
and YSO candidates are listed in Tables~\ref{tab:pagb1_param}\,$-$~\ref{tab:yso2_param}. 
Furthermore, the post-AGB/RGB and YSO candidates have been grouped into 
Quality one (Q1) and Quality two (Q2) (see Section~\ref{classification}), 
giving the 4 tables. The spectra of the 103 final objects are shown in  
Appendix~\ref{specstuff} (Figures~\ref{pagb1spec} $-$~\ref{yso2spec}) 
and Tables~\ref{tab:pagb1lines} $-$~\ref{tab:yso2lines} list their spectral features. 
Plots similar to Figure~\ref{stpplot} 
for the final sample of high-probability post-AGB/RGB and YSO candidates are 
available as online supporting 
information.

\subsection{Estimating the Error of the Spectral Typing Routine}
\label{error}

As a first check of uncertainty in our spectral typing method, we identified 
the stars in our sample that had been previously examined using high-resolution spectra. 
In the SMC, so far, a detailed 
chemical analysis has been carried out on only one of our stars J004441.04-732136.4 
\citep{desmedt12}. A comparison of our derived 
stellar parameters to those of \citet{desmedt12}, who obtained high-resolution 
optical UVES spectra, resulted in 
$\Delta$\teff = 82K, $\Delta$\logg\, = 0.5, $\Delta$[Fe/H] = 0.27 dex. 
High-resolution chemical analysis studies have been carried out for 2 LMC stars \citep{vanaarle13} 
in our LMC sample 
(Kamath et al., in preperation). Comparing the stellar parameters 
that we derived for the 2 stars (J053250.69-713925.8 and J053253.51-695915.1) to the values estimated 
from the high resolution chemical abundance study, we find that for 
J053250.69-713925.8, $\Delta$\teff = 394K, $\Delta$\logg\, = 0,and $\Delta$[Fe/H] = 0.23 and for 
J053253.51-695915.1, $\Delta$\teff = 57K, $\Delta$\logg\, = 1.0, and $\Delta$[Fe/H] = 0.07 dex. 
Thus the mean RMS difference between our measurements and the literature values 
are $\Delta$\teff = 135K, $\Delta$ \logg\, = 0.37, and $\Delta$[Fe/H] = 0.12 dex.

\begin{table}
  \caption{Error estimates in \teff, \logg\,, and [Fe/H] as a function of the number of counts in the spectra}
   \centering
  \begin{tabular}{|cccc|}
  \hline
   Counts & $\Delta$\teff & $\Delta$\logg\, & $\Delta$[Fe/H] \\
  \hline
   100 & 157 & 0.47 & 0.16 \\
   500 & 104 & 0.20 & 0.09 \\
  1500 & 52 & 0.15 & 0.05 \\
  3000 & 59 & 0.15 & 0.06 \\
  \hline
 \end{tabular}
\label{counts}
\end{table}

The low-resolution optical AAOmega spectra have a range of signal from 100 to 3000 counts. 
To test the reliability of our spectral 
typing pipeline as a function of signal, we took a set of synthetic spectra varying in temperatures from 
3500K to 9500K (the \teff\, region over which we expect most of the post-AGB/RGB and YSO candidates to lie), in 
\logg\,\, from 0.5 to 1.5, with fixed [Fe/H] = -1.0. We added to the synthetic spectra 
AAOmega detector
read noise and varying levels of photon noise, resulting in artificial spectra with 
quality equivalent to measured spectra of 100, 500, 1500, and 3000 counts. The artificial spectra were then passed through the 
spectral typing pipeline. Table~\ref{counts} shows the 
mean errors in the measurements of \teff, \logg\, and [Fe/H] as a function of counts. 
From this error estimation exercise and comparison with 
high resolution studies, we conclude that the mean errors in our derived 
parameters are smaller than the grid spacings of the synthetic template spectra: 
250K in \teff, 0.5 in \logg, and 0.5 in [Fe/H]. 

\section{Reddening Estimates}
\label{reddening}

The total reddening, which includes both the interstellar 
and circumstellar reddening, can be determined by estimating the difference between the 
intrinsic colour of the candidate (derived from the \teff\, estimated from the spectrum) and 
the measured colour (derived from the raw photometry). We calculated the E(B-V) for each 
individual candidate by estimating the value of E(B-V) that minimised the sum of the 
squared differences between the de-reddened observed and the intrinsic 
$B$, $V$, $I$ and $J$ magnitudes (at longer wavelengths, emission from 
dust can contribute to the observed magnitudes). We used the \citet{cardelli89} extinction law, 
assuming Rv = 3.1. It is possible that the circumstellar extinction law is different from the 
interstellar extinction law but we have not explored this possibility.
The derived E(B-V) values were used to correct the observed magnitudes for extinction. Then the 
$BVIJ$ fluxes of the best-fit model atmosphere (derived from the STP) were normalised 
to the corrected $BVIJ$ fluxes.

Typically, the uncertainty 
in the E(B-V) estimate is dominated by the errors in the derived \teff\, values. 
The uncertainty in the E(B-V) estimate 
owing to other errors, such as errors in the photometry, are small.
Given that the uncertainty in the estimated \teff\, could be up to $\pm$\,250K, we estimated the maximum error in E(B-V) to be the difference between 
E(B-V) at the estimated \teff, and at \teff\, values of $\pm$\,250. The 
error $\Delta$E(B-V) $\approx$ 0.2 mag at \teff $\sim$\, 4000K and declines with increasing \teff\, to 
$\Delta$E(B-V) $\approx$ 0.1 mag at \teff\, $\sim$\, 5000K and $\Delta$E(B-V) $\approx$ 0.05 mag at \teff\, $\sim$\, 6500K

The total E(B-V) estimated for 
the final sample of post-AGB/RGB and YSO candidates is listed in 
Tables~\ref{tab:pagb1_param} $-$~\ref{tab:yso2_param}. For some stars 
the estimated E(B-V) was negative which indicates that the \teff\, estimated from the spectra 
is likely to be cooler than the actual \teff\, of the star. For such stars, we estimated the E(B-V) 
using \teff\, values increased by 250K and 500K. These stars are denoted using a suffix 'b' and 'c', respectively in 
Tables~\ref{tab:pagb1_param} $-$~\ref{tab:yso2_param}. The SEDs, both as observed and de-extincted, are 
plotted in Appendix~\ref{specstuff}, Figures~\ref{fig:pagb1_sed} $-$~\ref{fig:yso2_sed}.

\section{Luminosity of the Central Star}
\label{ms}

For post-AGB/RGB stars and YSOs the central 
star is surrounded by circumstellar dust that is not necessarily spherically symmetric. For such cases, the 
observed luminosity $L_{\rm obs}$ could either be over-estimated or under-estimated. For this reason 
it is essential to estimate the photospheric luminosity of the central star $L_{\rm phot}$. This 
photospheric luminosity can be derived from the bolometric correction for the model atmosphere 
corresponding to each individual candidate normalised to the de-extincted $V$ magnitude, coupled with the distance modulus to the SMC. 
In Tables~\ref{tab:pagb1_param}\,$-$~\ref{tab:yso2_param} we list the photospheric luminosities ($L$$_{\rm phot}$) for the central star of the 
post-AGB/RGB and YSO candidates.

For some of our objects we encounter an energy problem as the available luminosity $L$$_{\rm phot}$ from 
the central star is too small to account for the 
luminosity $L_{\rm obs}$ derived by integrating the total SED thus resulting in  $L_{\rm obs}$ being 
$\sim$\,1.5\,$-$\,5 times $L_{\rm phot}$. It is possible that if scattered light dominates in the optical part of the 
spectrum, the reddening as well as the luminosity estimate $L_{\rm phot}$ are not correct and the 
luminosity $L_{\rm obs}$ could be a better tracer of the total luminosity of the object. Models 
by \citet{menshchikov02} show that $L_{\rm obs}$ can be several times $L_{\rm phot}$ which is consistent with the 
luminosity ratios we obtain for some objects. Another possibility is that there is another independent object coincident on the sky with 
the optically observed star.

\section{Separating the Post-AGB/RGB and YSO Candidates}
\label{classification}

Disentangling the post-AGB/RGB candidates from the YSO candidates is a concern since these 
unrelated objects lie in the same region of the HR diagram. One of the 
ways to distinguish between these objects is to use the derived \logg\, values from the spectral fit for the 
individual objects. As mentioned in Section~\ref{STP}, the \logg\, values that a star would have in the 
post-AGB/RGB phase and in the pre-main sequence phase differ by $\sim$\,1.3, hence we separated the post-AGB/RGB candidates from the YSO candidates by comparing the 
individual derived \logg\, values from the spectral fitting to the theoretical value of \logg\, a star would have in the 
post-AGB/RGB phase and in the pre-main sequence phase. Based on this separation scheme, we formed the final sample which 
consisted of 63 post-AGB/RGB candidates and 40 YSO candidates.

We note that in determining the 
stellar parameters such as \teff, \logg, and [Fe/H], the \logg\, estimates have the highest 
uncertainty since the \logg\, value least affects the spectra when compared to the \teff\, and [Fe/H]. 
Therefore, despite the criteria used to separate the post-AGB/RGB candidates from the YSOs, 
there remains a degree of uncertainty in our classification method. 
Detailed studies based on high resolution spectra are needed to confirm the 
nature of the individual objects.

The 63 post-AGB/RGB and 40 YSO candidates were then further classified into two groups: 
Q1 (quality 1) and Q2 (quality 2) 
based on the signal of the observed spectrum and also a visual inspection of the results of the spectral 
typing routine. We classified those 
candidates with a relatively high signal and a good spectral matching fit as Q1 and the remaining 
candidates were classified as Q2 candidates. Therefore our final sample of 63 post-AGB/RGB candidates were split into 2 groups of 
38 Q1 and 25 Q2 objects. Similarly the final sample of 40 YSO candidates were split into 2 groups 
consisting of 27 Q1 and 13 Q2 objects.

In Tables~\ref{tab:pagb1_param} and~\ref{tab:pagb2_param} we list the final sample of 
post-AGB/RGB candidates, along with their stellar parameters and their 
SED classification (see Section~\ref{sedanalysis}). We also make the distinction between post-AGB and post-RGB candidates 
using a luminosity criterion based on the 
expected luminosity of the RGB-tip for stars in the LMC and SMC (see Section~\ref{intro}). 
We consider post-AGB stars to be those objects 
with $L_{\rm phot}$/L$_{\odot}$\, $>$ 2500 and post-RGB stars to be those objects with $L_{\rm phot}$/L$_{\odot}$\, 
$\leq$ 2500. We find that our sample 
consists of 42 post-RGB candidates and 21 post-AGB candidates.

Similarly, in Tables~\ref{tab:yso1_param} and~\ref{tab:yso2_param} we present the final sample of YSO candidates. 

\begin{figure}
\begin{center}
\includegraphics[bb= 0 0 263 206,width=8cm,angle=0]{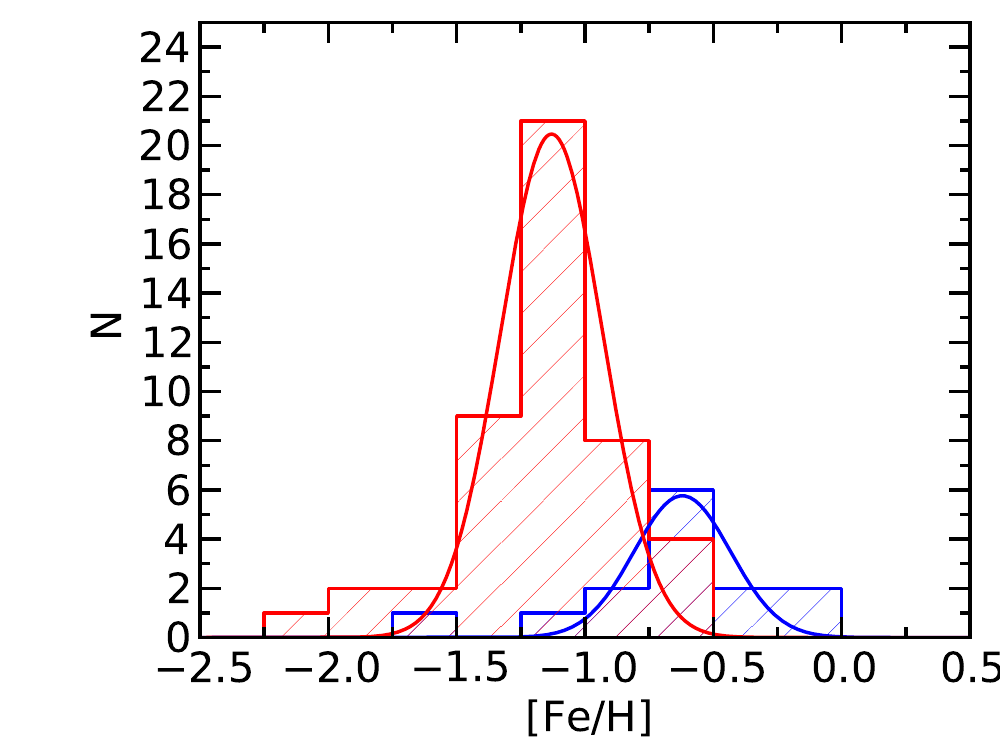}
\caption{Histograms showing the [Fe/H] distribution for the post-AGB/RGB and YSO candidates in the sample. The red 
histogram corresponds to the post-AGB/RGB objects, and blue to the YSOs. The red and blue curves denote individual 
Gaussian fits to the histograms for the post-AGB/RGB and YSO candidates, respectively. See text for further details.}
\label{fehplot}
\end{center}
\end{figure}

Since post-AGB/RGB stars are an old to intermediate age population, we expect 
them to be more metal poor than the YSOs which belong to the young SMC 
population, which has a mean metallicity of 
[Fe/H] $\simeq$ -0.7 \citep{luck98}. Figure~\ref{fehplot} shows the [Fe/H] distribution for the post-AGB/RGB candidates (represented 
by the red histogram)
and YSO candidates (represented by the blue histogram)\footnote{Note: The post-AGB/RGB and YSO candidates 
for which we imposed a [Fe/H] value of -1.00 
(see Tables~\ref{tab:pagb1_param} $-$~\ref{tab:yso2_param}) have not 
been considered while plotting the [Fe/H] distribution.}. From Gaussian fits to the histograms (Figure~\ref{fehplot}) we 
find that the post-AGB/RGB sample peaks at a [Fe/H] = -1.14 with a standard deviation 
of 0.20, whilst the YSO sample, peaks at a higher metallicity of [Fe/H] = -0.62 
with a standard deviation of 0.18. Using the 2-sided Kolmogorov-Smirnov (KS) test, we find that 
the post-AGB/RGB candidates are more metal poor than the YSOs with high confidence (probability of the two samples to be drawn from the 
same distribution $P$ $\sim$\, 10$^{-4}$). The existence of this bimodal metallicity distribution supports our separation of 
post-AGB/RGB from the YSO candidates.

\section{Classification of Spectral Energy Distributions}
\label{sedanalysis}

\begin{figure*}
\begin{center}
\begin{minipage}{150pt}
\resizebox{\hsize}{!}{\includegraphics[bb = 0 0 288 216,width=5cm]{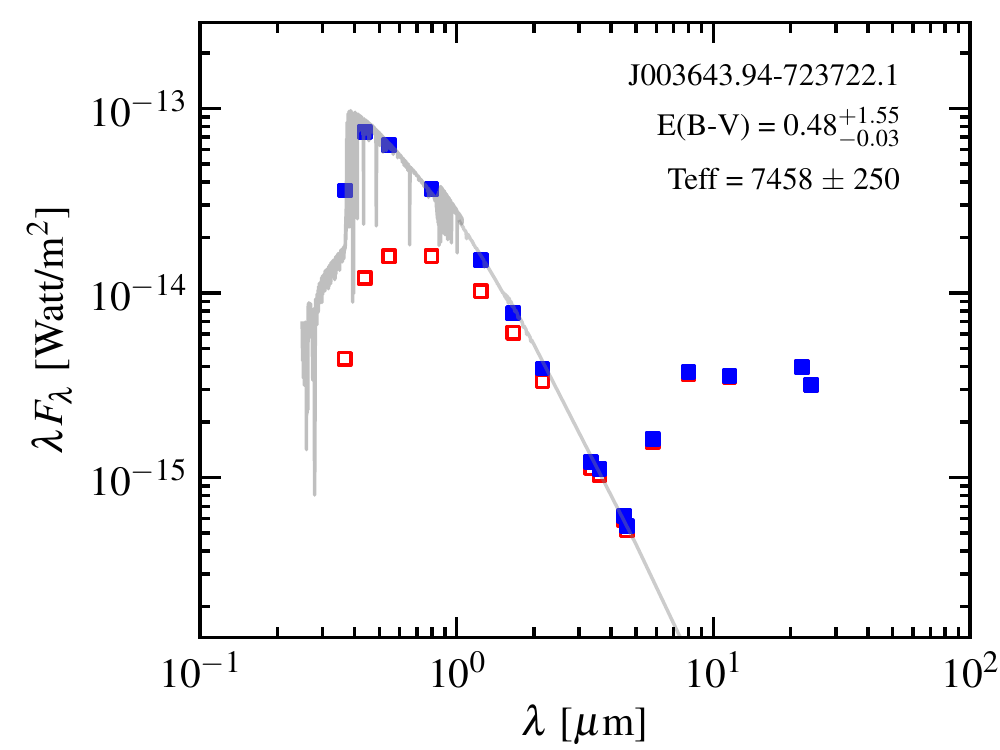}}
\end{minipage}
\hspace{0.1cm}
\begin{minipage}{150pt}
\resizebox{\hsize}{!}{\includegraphics[bb = 0 0 288 216,width=5cm]{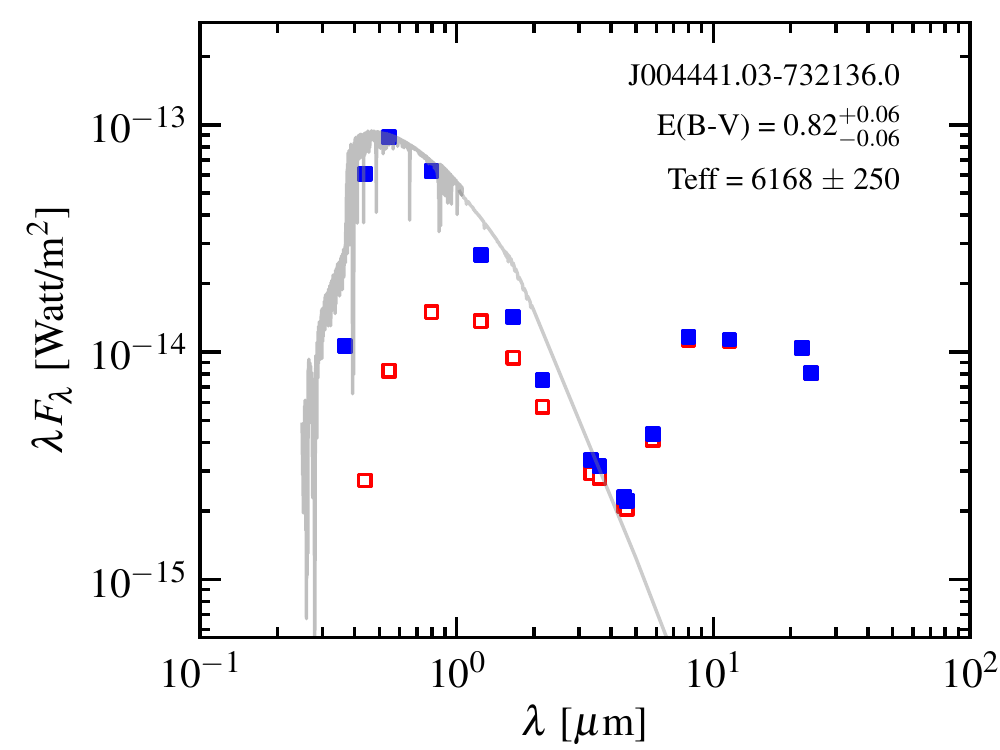}}
\end{minipage}
\hspace{0.1cm}
\begin{minipage}{150pt}
\resizebox{\hsize}{!}{\includegraphics[bb = 0 0 288 216,width=5cm]{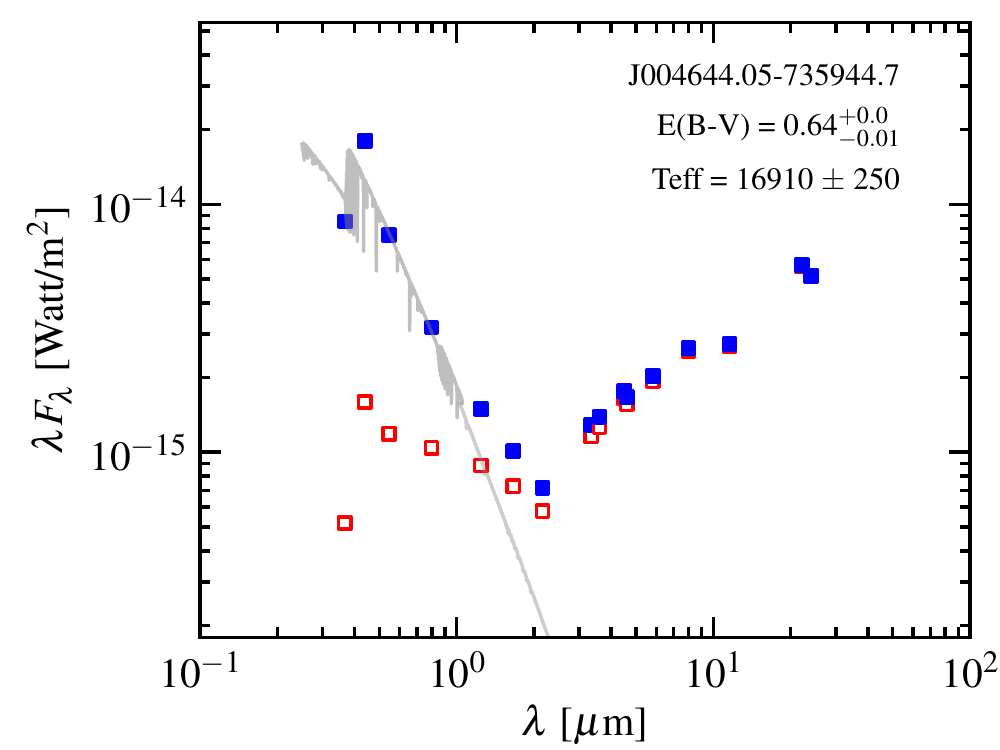}}
\end{minipage}
\hspace{0.1cm}
\end{center}
\caption{Example SEDs of the post-AGB/RGB candidates classified as shell sources.}
\label{pagbsed_eg1}
\end{figure*}

\begin{figure*}
\begin{center}
\begin{minipage}{150pt}
\resizebox{\hsize}{!}{\includegraphics[bb = 0 0 288 216,width=5cm]{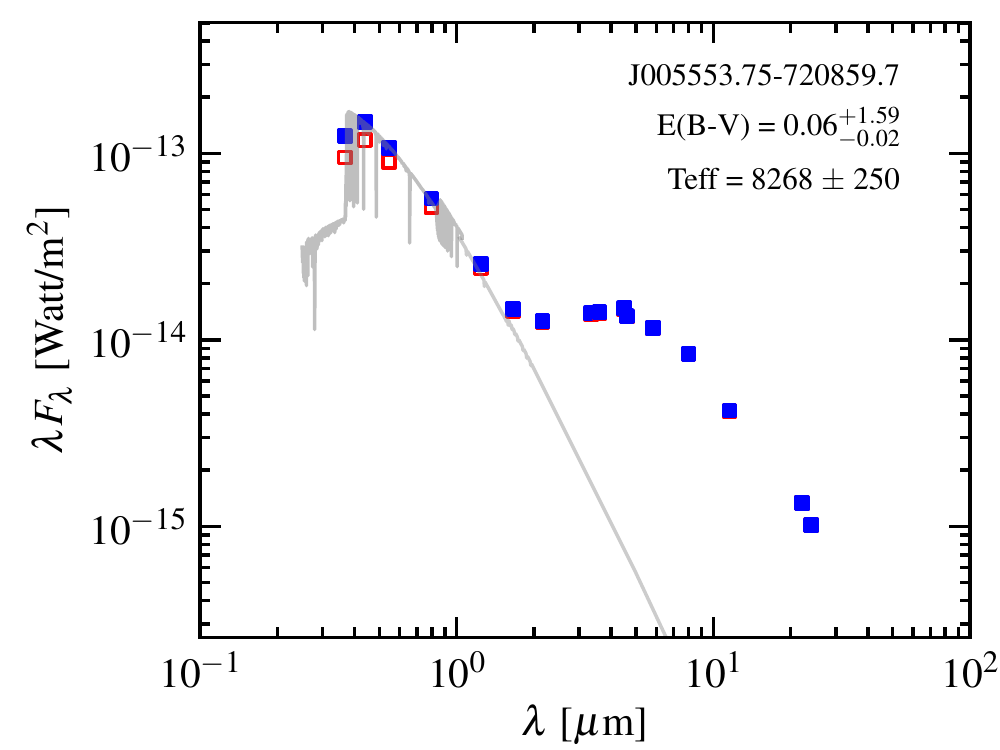}}
\end{minipage}
\hspace{0.1cm}
\begin{minipage}{150pt}
\resizebox{\hsize}{!}{\includegraphics[bb = 0 0 288 216,width=5cm]{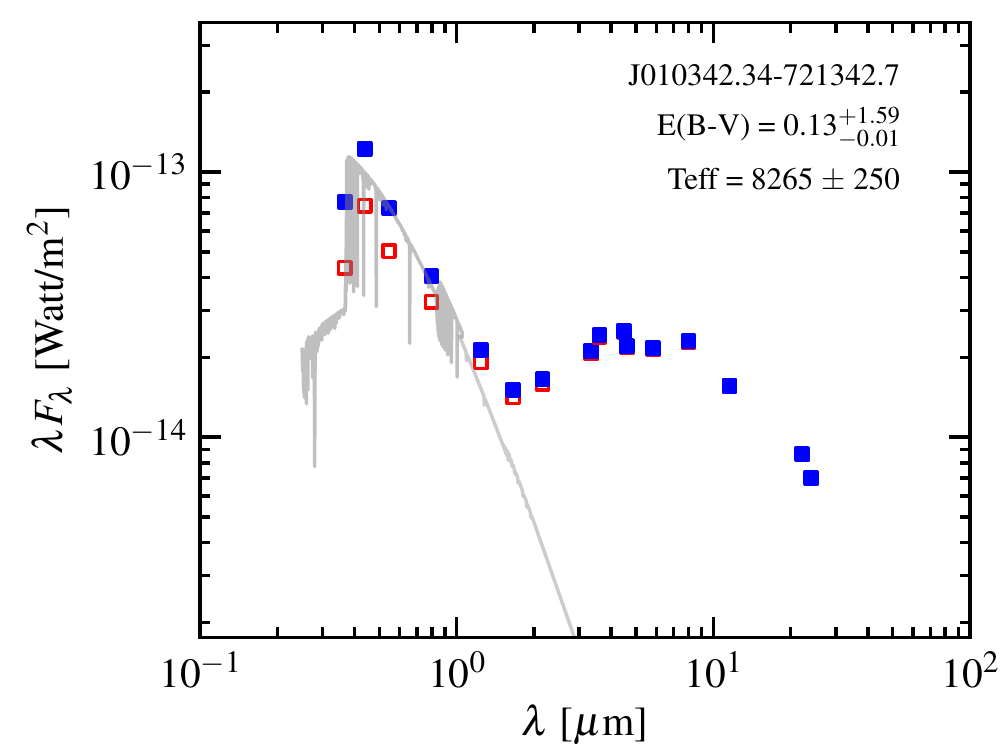}}
\end{minipage}
\hspace{0.1cm}
\begin{minipage}{150pt}
\resizebox{\hsize}{!}{\includegraphics[bb = 0 0 288 216,width=5cm]{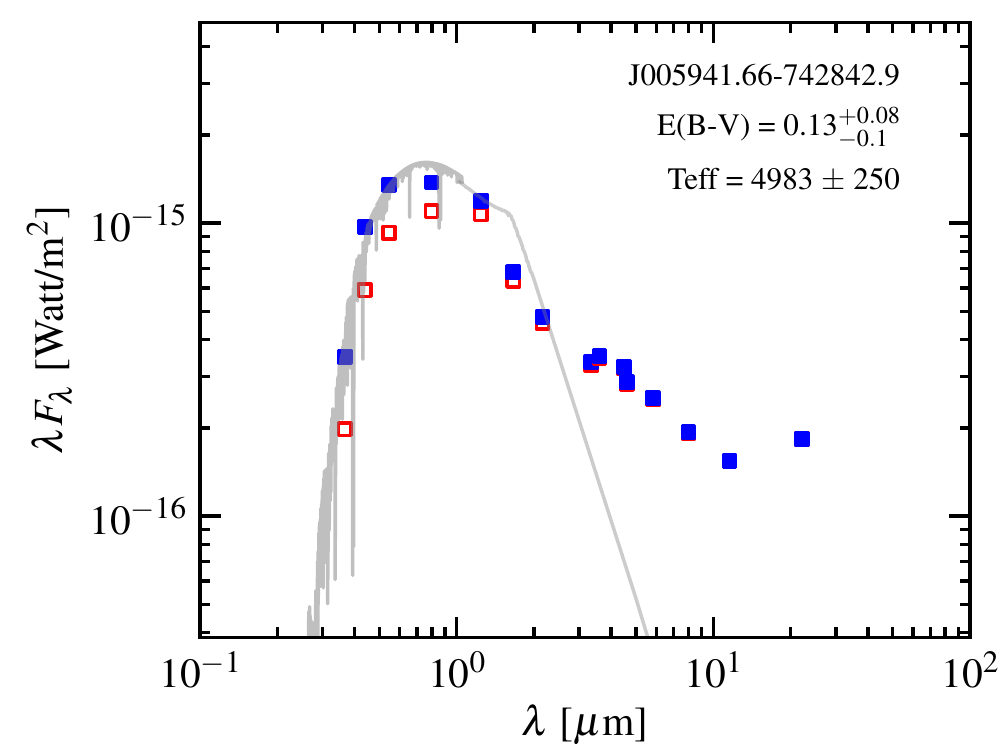}}
\end{minipage}
\hspace{0.1cm}
\end{center}
\caption{Example SEDs of the post-AGB/RGB candidates classified as disc sources.}
\label{pagbsed_eg2}
\end{figure*}

\begin{figure*}
\begin{center}
\begin{minipage}{150pt}
\resizebox{\hsize}{!}{\includegraphics[bb = 0 0 288 216,width=5cm]{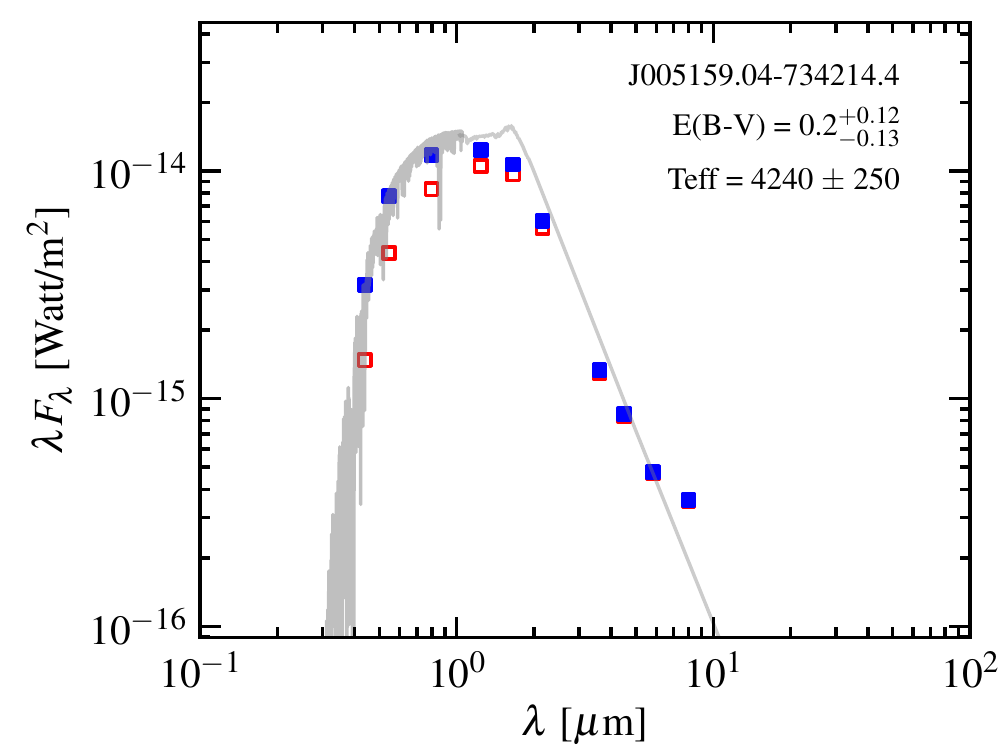}}
\end{minipage}
\hspace{0.1cm}
\begin{minipage}{150pt}
\resizebox{\hsize}{!}{\includegraphics[bb = 0 0 288 216,width=5cm]{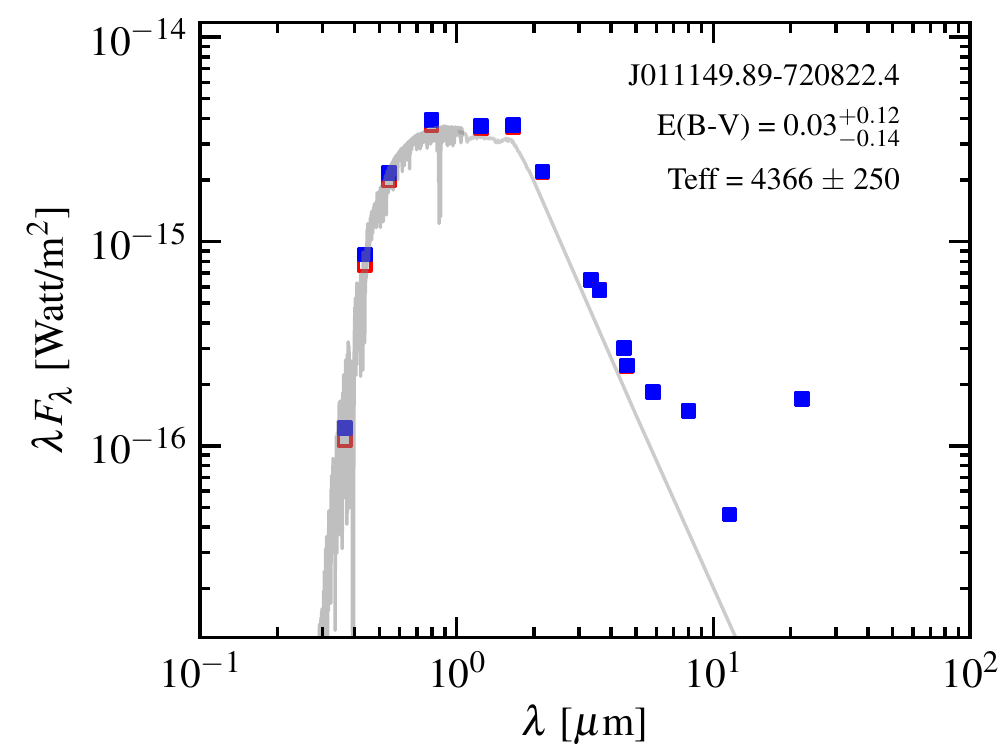}}
\end{minipage}
\hspace{0.1cm}
\begin{minipage}{150pt}
\resizebox{\hsize}{!}{\includegraphics[bb = 0 0 288 216,width=5cm]{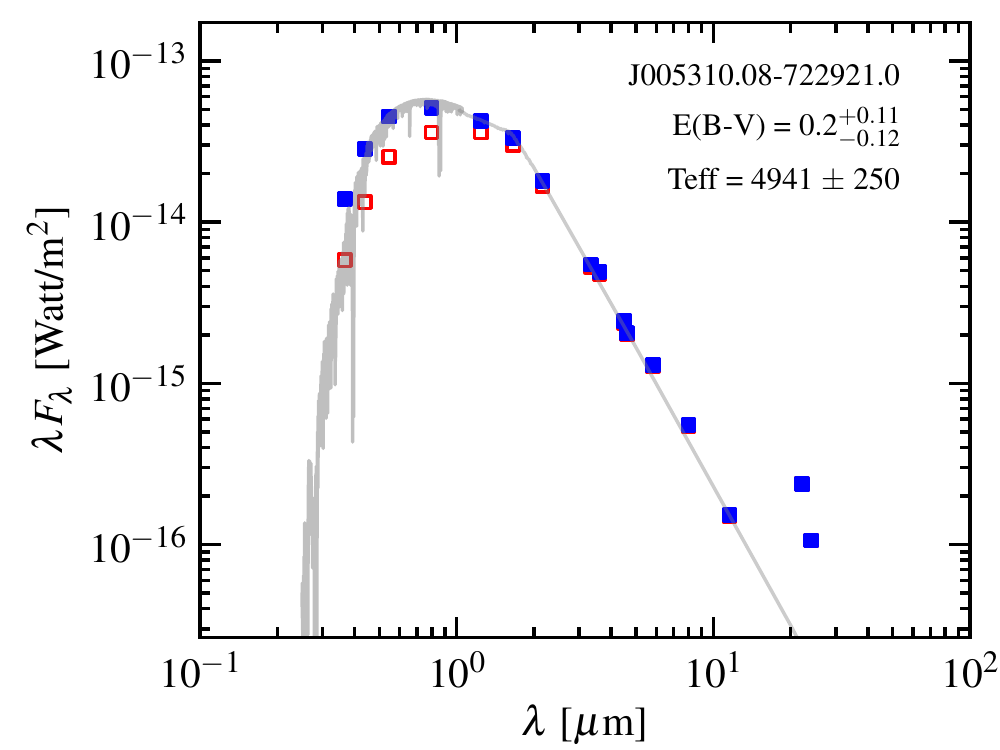}}
\end{minipage}
\hspace{0.1cm}
\end{center}
\caption{Example SEDs of the post-AGB/RGB candidates that are classified as 'uncertain'.}
\label{pagbsed_eg3}
\end{figure*}

Post-AGB/RGB and YSO candidates consists of two components: 
the central star and the circumstellar environment associated with the central star. 

In single optically visible post-AGB stars, there exists a circumstellar shell which corresponds to an 
optically thin expanding circumstellar envelope that is the remnant AGB mass-loss. The shell 
continues to move outwards, gradually exposing the central star. This results in a double-peaked SED with one 
peak due to the stellar emission and the other due to circumstellar dust \citep{vanwinckel03} which 
peaks greater than 10 $\mu$m \citep[as shown from radiative transfer models of a 
well known expanding shell source HD161796, where the peak of the dust SED is at around 30$\mu$m, see][for details]{min13}. These 
objects are considered to be shell sources.

Some post-AGB/RGB stars have SEDs with a strong near-IR emission, indicating the 
presence of hot dust in the system. Some of the these stars have been found to reside 
in binary systems which have a stable circumbinary disc \citep{waters92,deruyter06,gielen08,vanwinckel07,vanwinckel09}. 
It is assumed that all objects with such a SED, have a circumbinary disc. 
A general characteristic of these sources is that the dust energy distribution peaks at relatively 
high temperatures and the peak of the dust SED lies around 10\,$\mu$m and in some cases even bluer 
\citep{deruyter06,gielen11}.

Based on a visual inspection of the position of the peak of the dust excess  in the SEDs, for majority of the 
sources, we were able to identify whether the SEDs were representative of a shell or disc 
source.

\begin{figure*}
\begin{center}
\includegraphics[width=16cm]{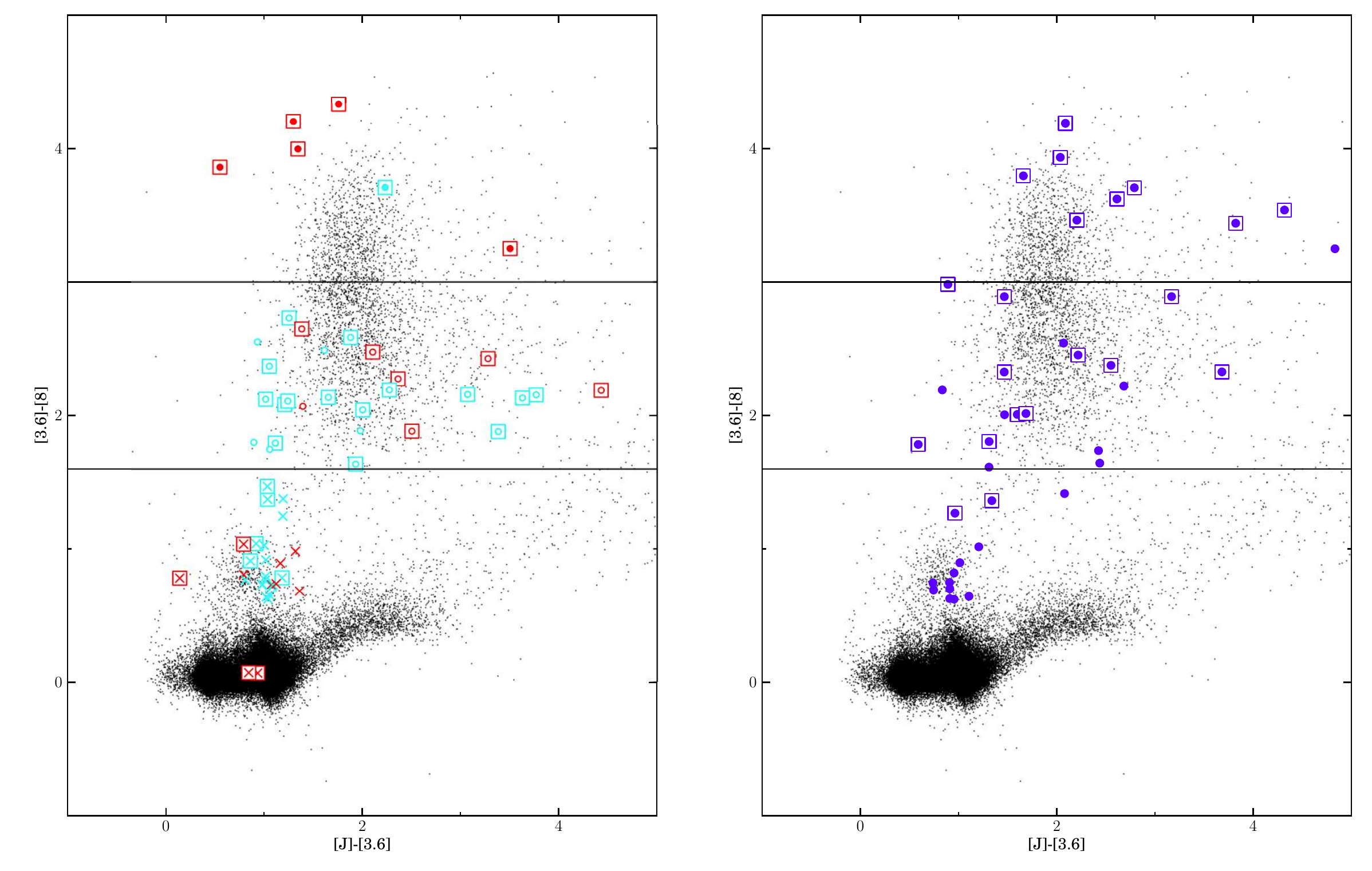}
\end{center}
\caption{The left panel 
shows the $J-$[3.6] vs [3.6]$-$[8] colour-colour plot for the post-AGB/RGB candidates. 
The red symbols denote the post-AGB candidates and the cyan symbols denoted the 
post-RGB candidates. Disc sources are represented as open circles, shells as filled circles and 
uncertain sources as crosses. Those candidates that show the presence 
of a [24] micron excess are enclosed in an open square. 
The region 1.6\,$<$\,[3.6]$-$[8]\,$<$\,3.0 is where the majority 
of the post-AGB/RGB discs sources lie. 
The right panel shows the $J-$[3.6] vs [3.6]$-$[8] colour-colour plot for the YSO candidates, represented as blue 
filled circles. Those candidates that show the presence 
of a [24] micron excess are enclosed in an open square. The discs source region is transferred from the
left panel to the right panel for comparison purposes only.}
\label{sedclassification}
\end{figure*}

To confirm whether these objects are indeed likely shell or disc sources, we used the $J$$-$[3.6] 
colour to check the presence of a near-IR excess due to hot dust surrounding the central star. 
This is indicative of a stable dust structure as these objects have photospheres too hot to be in a dust producing phase. 
We plot the post-AGB/RGB candidates on a $J-$[3.6] vs [3.6]$-$[8] 
colour-colour plot in the left panel of Figure~\ref{sedclassification}. The cyan/grey symbols represent the 
post-RGB candidates and the red symbols represent the post-AGB candidates. We find that majority of the post-AGB/RGB candidates (27 sources in total) that we 
identified as disc sources based on the visual inspection of their SEDs, lie in the 
region 1.6\,$<$\,[3.6]$-$[8]\,$<$\,3.0 of the $J-$[3.6] vs [3.6]$-$[8] 
colour-colour plot. These sources appear to have redder $J - $[3.6] colours, 
mostly with $J$\,$-$[3.6]\,$\geq$\,1.0, which is indicative of hot dust in the system. Therefore, 
we classify these 27 objects to be likely disc sources. In Figure~\ref{sedclassification}, we represent these disc 
sources with open circles. 
The two black solid lines mark the region where 
1.6\,$\leq$\,[3.6]$-$[8]\,$\leq$\,3.0. 

We find that a small group of 6 post-AGB/RGB candidates that we identified to be shell sources 
based on the visual inspection of their SEDs, have a [3.6]$-$[8] excess with 
[3.6]$-$[8]\,$>$\,3.0 which indicates cool dust in the system. The SEDs of these 
objects show a very strong double peaked feature indicating a 
detached cool dust shell surrounding the central star. Therefore, we classify these 6 objects to be likely shell sources. 
In Figure~\ref{sedclassification}, we represent these shell sources with filled circles. 
We find that all 
these sources show detections at 24$\mu$m. We note that sources with a 24$\mu$m detection are enclosed within an open square 
symbol in Figure~\ref{sedclassification}.

These classifications, "disc"
or "shell", are given in the SED column of Tables~\ref{tab:pagb1_param} and ~\ref{tab:pagb2_param}.

The remaining 30 post-AGB/RGB candidates are found to lie in the region [3.6]$-$[8]\,$<$\,1.6. 
Majority of these objects show a mild $J$\,$-$[3.6] excess with 0.8\,$<$\,$J-$[3.6]\,$<$\,1.3, 
which indicates hot dust surrounding the central star, characteristic of disc sources. Based on the visual inspection of their SEDs, 
we were able to identify a few of these sources as likely shells or discs, however, for majority of these sources such an identification was not possible 
since we do not have enough information beyond 10\,$\mu$m for most of these sources. 
Studies at longer wavelengths to extend these SEDs will be needed to study the temperature 
distributions of the dust. Therefore, we classify these 30 sources as uncertain. 
In Figure~\ref{sedclassification}, we represent these 30 uncertain sources with cross symbols.  We find that the majority of these objects are post-RGB stars 
which are believed to be formed as a result of the termination of RGB evolution by binary interaction and hence we 
expect these objects to have dusty discs.  These objects are labelled as "uncertain" 
in the SED column of Tables~\ref{tab:pagb1_param} and ~\ref{tab:pagb2_param}. Figure~\ref{pagbsed_eg1}\,$-$\,~\ref{pagbsed_eg3} 
show examples of post-RGB/AGB candidates that 
we classified as discs, shells and uncertain. 

We conclude that out of the 63 post-AGB/RGB candidates, 27 are disc sources, 6 are shell sources and 
30 are uncertain. As expected, majority of the sources classified as discs are post-RGB candidates and 
majority of the sources classified as shells are post-AGB candidates. In Table~\ref{tab:pagb1_param} and 
Table~\ref{tab:pagb2_param} we list the nature of the SEDs (shell, disc or uncertain) 
for the Q1 and Q2 post-AGB/RGB candidates, respectively.

In the right panel of Figure~\ref{sedclassification} we also plot the YSO candidates 
(as blue filled circles) though the classification scheme for YSOs are different 
compared to that of post-AGB/RGB candidates. Sources with a 24$\mu$m detection 
are enclosed within an open square symbol. For our YSO candidates their SEDs are limited 
to 8$\mu$m in most cases and a SED based YSO classification is beyond the scope of this study.

Typically, YSOs in their early stages of evolution are bright at 70$\mu$m while for 
evolved YSOs, the peak of the SED is bluer than 70$\mu$m except when the disc is strongly flared. 
We inspected the individual 70$\mu$m MIPS images \citep{rieke04} of all the 
Q1 and Q2 YSO candidates. We looked for point source
detections as well as evidence for resolved ISM dust-emission coming
from star-forming regions. 
We find that 10 of the 40 YSO candidates (Q1 candidates: J004208.74-733108.4, J004451.87-725733.6, J004503.51-731627.4, J004657.45-731143.4, J004301.63-732050.9, J005606.53-724722.7; Q2 candidates:  J004547.50-735331.7, J004707.49-730259.2,
J011229.23-724511.6 and J011302.68-724852.5) show a detection in the 
70$\mu$m MIPS images. The SEDs of these objects are representative of a strongly flared disc 
(see Figures~\ref{fig:yso1_sed},~\ref{fig:yso2_sed} in Appendix~\ref{specstuff}). 
The majority of the 70$\mu$m images for the remaining 
YSO candidates show emission from dusty resolved nebulosity which might be 
evidence for a star forming region.

We note that the presence of a detection at 70$\mu$m or diffuse emission from 
the local environment does not necessarily confirm the YSO status of the candidates since we find that a few post-AGB/RGB 
candidates in our sample also appear bright at 70$\mu$m and for some the local environment shows diffuse emission. We find that three 
(J004614.67-723519.0, J004644.05-735944.7 and J010814.67-721306.2) out of 63 post-AGB/RGB candidates show a detection in the 
70$\mu$m MIPS images. The SEDs of J004614.67-723519.0 and J004644.05-735944.7 (see Figure~\ref{fig:pagb1_sed} in Appendix~\ref{specstuff}) 
represent that of a disc source with a strongly flared disc and therefore they are bright at 70$\mu$m. The SED of J010814.67-721306.2 
(see Figure~\ref{fig:pagb2_sed} in Appendix~\ref{specstuff}) 
represents a shell source with cold circumstellar dust surrounding the central star with emissions at 70$\mu$m. Furthermore, for some sources, 
including J004441.03-732136.0, the confirmed $s$-process rich post-AGB star 
\citep[from this study and also from][]{desmedt12}, the local environment shows diffuse emission.

Based on the inspection of the 70$\mu$m MIPS images we conclude that it is likely that early age YSOs show a detection at 70$\mu$m and also 
most YSOs show a local environment full of diffuse emission characteristic of a star formation region although 
this criterion is not exclusive to YSOs.

\section{Features in the Stellar Spectrum}
\label{specanalysis}

A wealth of information can be obtained from the spectrum of each candidate. 
In Figures~\ref{pagb1spec}\,$-$~\ref{yso2spec} (Appendix~\ref{specstuff}), we show the optical 
spectra of the sample of 
Q1 and Q2 post-AGB/RGB and YSO candidates. The individual spectral fits files are available as supporting information online. Also, in Appendix~\ref{specstuff}, we have summarised 
some of the most prominent features observable in the spectra of the final sample of post-AGB/RGB and YSO candidates 
(Tables~\ref{tab:pagb1lines}\,$-$\,~\ref{tab:yso2lines}). 

On analysing the spectra of the Q1 and Q2 post-AGB/RGB and YSO candidates we find that, 
in some cases, the hydrogen lines, 
the forbidden lines of oxygen ([OIII]), sulphur ([SII]), 
nitrogen ([NII]), the HeI lines, the CaT line and the Paschen lines are in emission, indicating 
either that the star is of an early spectra type capable of exciting circumstellar gas, or that there is 
unassociated nebulosity in line-of-sight to the candidate. We also find that in many cases, the lines show 
the presence of emission cores, indicating the 
presence of strong stellar winds or inflows or the presence of an accretion disc. Furthermore, among the spectra of the 40 YSO candidates, there 
are  27 that show a strong H$\alpha$ emission line.
H$\alpha$ emission and forbidden line emission indicate disc 
accretion in YSOs \citep{natta02,jayawardhana02}.

The presence of enhanced Ba lines in the stellar spectrum indicates a 
\emph{s}-process enriched post-AGB object.  
For low to intermediate mass stars, a significant amount of \emph{s}-process 
nucleosynthesis takes place prior to the post-AGB phases of 
stellar evolution. Therefore we expect to detect the presence of 
$\emph{s}$-process elements. However, chemical analysis studies of a sample of supposed Galactic 
post-AGB stars show that the abundance pattern in these stars is more diverse than expected 
\citep[][and references therein]{vanwinckel00,reyniers07}, with only some objects showing an 
enhancement of \emph{s}-process elements, whilst others are either mildly or 
not enhanced at all. We suggest that the un-enhanced Galactic objects could be post-RGB stars as well since the luminosities of our 
objects clearly shows that post-RGB stars are as common as post-AGB stars. 

On visually inspecting the spectra of the post-AGB/RGB candidates, we were unable to identify 
Barium in the majority of the candidates. This could be due to the low-resolution of our spectra. 
However, for strongly $s$-process enriched stars, we were able to detect the presence of the strong 
BaII line at 4554.03\AA. We found that 6 out of 63 stars (J003643.94-723722.1, J004114.10-741130.1, 
J004441.03-732136.0, J005107.19-734133.3, J005941.66-742842.9, and J010247.72-740151.6) 
showed the presence of the BaII line at 4554.03\AA. J004441.03-732136.0 has been previously identified, 
from abundance studies with high resolution spectra, as a \emph{s}-process enriched post-AGB star 
by \citet{desmedt12}.

Another element of interest is lithium, which can be detected by the presence of the 
LiI line at 6708 \AA. Lithium is abundant in the parent molecular cloud but it is 
destroyed in the stellar interior at relatively 
low temperatures ($\sim$ 2\,$\times$\,10$^{6}$K). If these interior temperatures are reached when 
the star is convective, Li will be depleted at the stellar surface during the 
pre-main sequence phase. During the evolution beyond the main sequence, lithium is 
further decreased owing to the first and second-dredge up processes that occur during the 
red-giant phase of evolution and the early-AGB phase of evolution \citep{karakas03b}. 
However, in massive stars ($>$\, 4\Msun) during the thermally pulsing AGB phase, lithium can be created by hot bottom burning 
\citep{boothroyd95,lattanzio96}. We searched for the presence of lithium in the stellar photospheres of 
both the post-AGB/RGB and YSO candidates by visually inspecting the spectra. We detected the presence of the 
LiI (6708 \AA) line in absorption in 7 out of the 63 post-AGB/RGB candidates. These 7 candidates with LiI detections are low luminosity post-RGB candidates. 
Current evolutionary models for these mass ranges do not predict an 
enhanced Li abundance.

We also detected the LiI (6708 \AA) line in absorption in 
3 of the 40 YSO candidates, indicating that these latter objects are 
probably early stage YSOs or massive YSOs. We note again that the 
low-resolution of the spectra could possibly affect the 
number of identifications.

\section{HR Diagrams}
\label{HR}

\begin{figure*}
\begin{center}
\begin{minipage}{230pt}
\resizebox{\hsize}{!}{\includegraphics[clip=true]{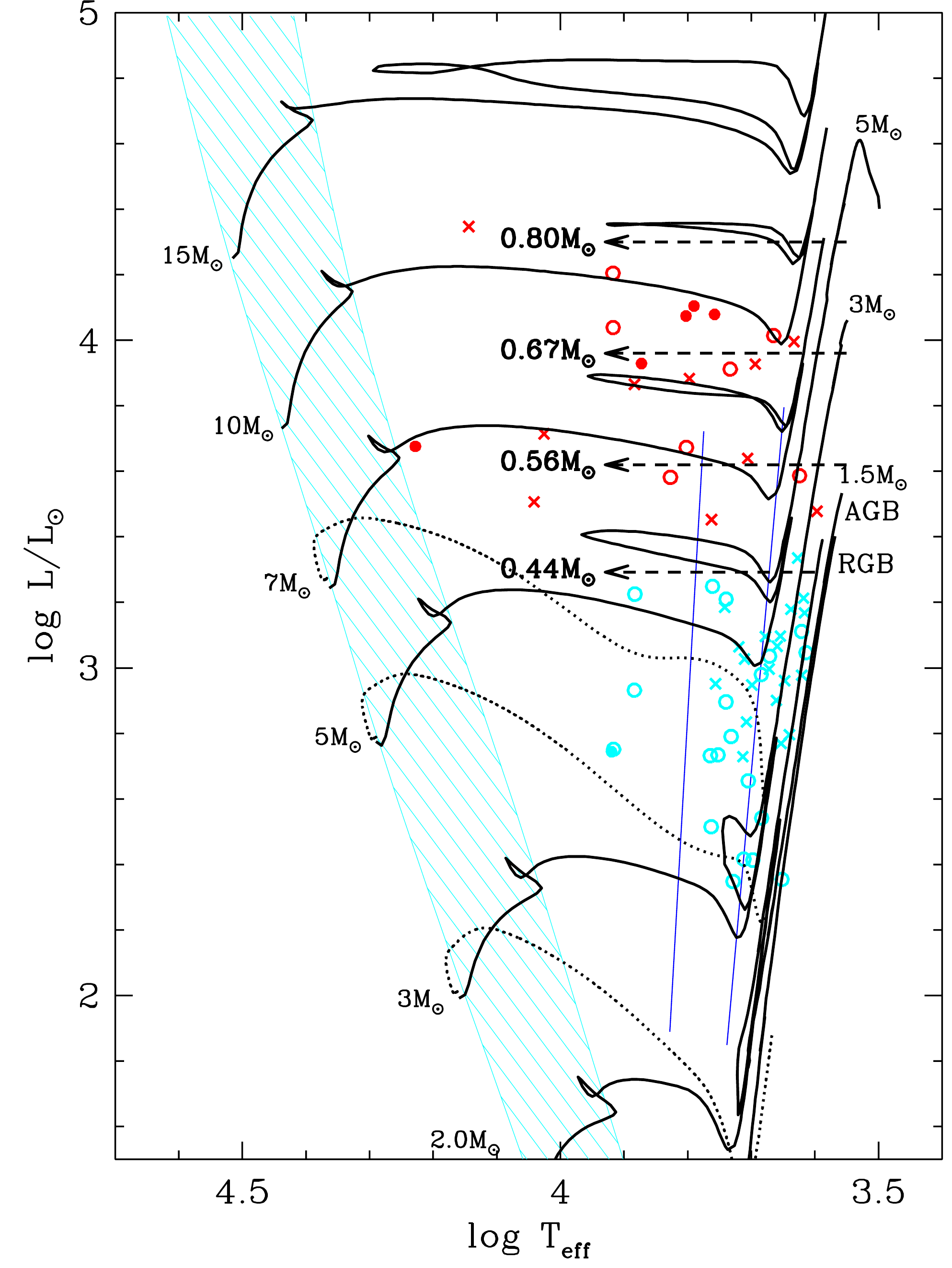}}
\end{minipage}
\hspace{0.1cm}
\begin{minipage}{230pt}
\resizebox{\hsize}{!}{\includegraphics[clip=true]{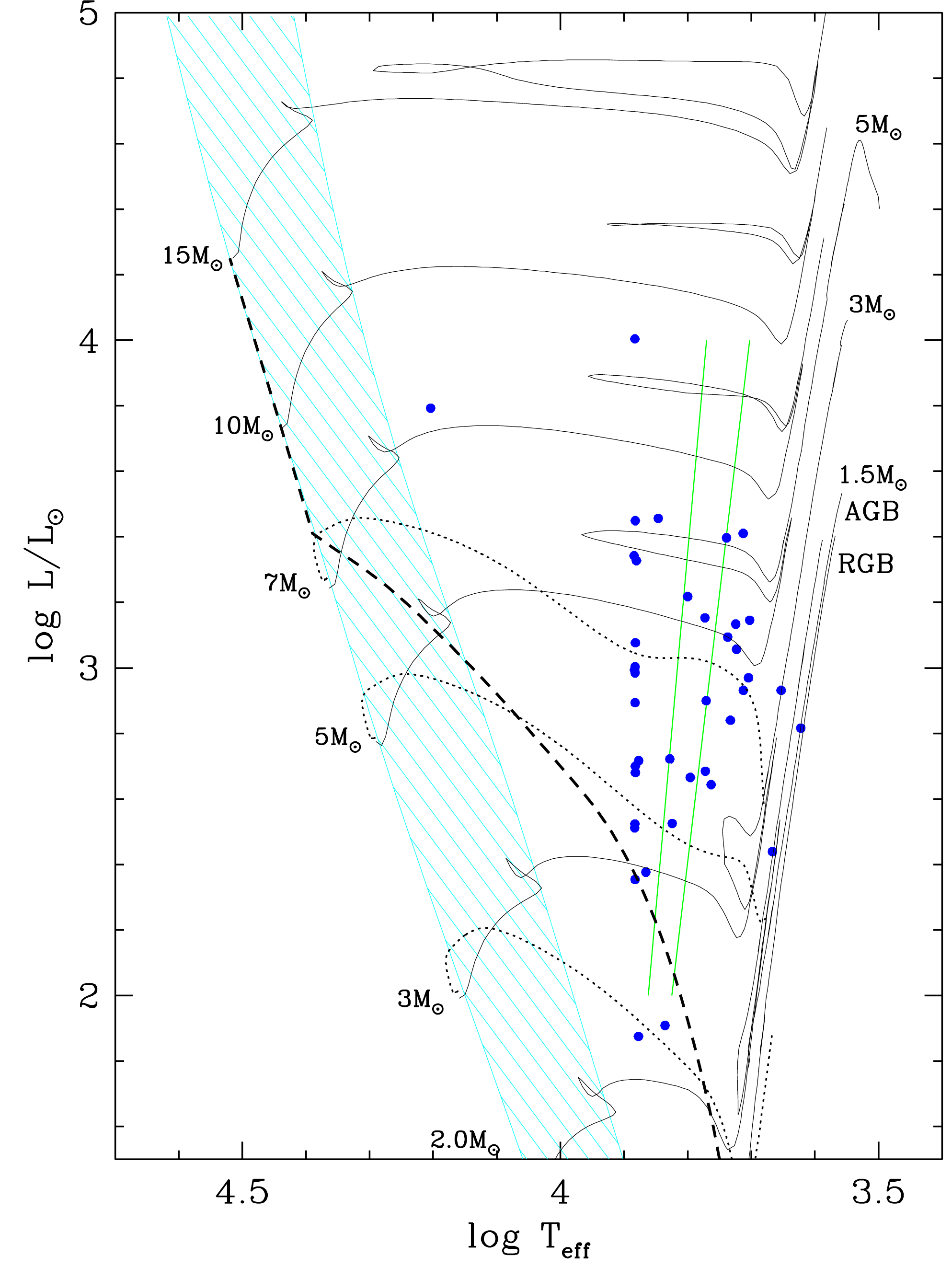}}
\end{minipage}
\end{center}
\caption{The HR diagram for the sample of the post-AGB/RGB candidates 
(left panel) and YSO candidates (right panel). In the left panel, 
the red symbols represent the post-AGB candidates and the cyan symbols represent the 
post-RGB candidates. The open circles represent disc sources and the filled circles 
represent the shell sources.  
The crosses represent those post-AGB/RGB candidates for which 
the nature of the SED is uncertain. In the right panel, the blue filled circles 
represent the YSO candidates. Each plot shows the main sequence as a cyan cross-hatched region. 
In both the plots the black solid lines represent evolutionary tracks starting from the main-sequence and 
the black dotted lines represent pre-main sequence evolutionary tracks. The black dashed arrows on 
the HR diagram for the 
post-AGB/RGB candidates schematically represents the post-AGB/RGB evolutionary tracks. Also shown 
on this plot is the empirical OGLE instability strip for the Population 
II Cepheids represented with blue vertical lines. 
In the right panel, the thick black dashed line in right panel is the 
birth-line and the green vertical lines represent the Cepheid instability strip. See text 
for further details.}
\label{hr}
\end{figure*}

To understand the evolutionary stage of the post-AGB/RGB and YSO candidates, we show their positions 
in the HR diagram in Figure~\ref{hr}. The left panel shows the post-AGB/RGB 
population. The post-AGB candidates are represented as red symbols and 
post-RGB candidates are represented as cyan symbols. The open circles represent the disc sources, 
the filled circles represent the shell sources and the crosses represent those sources classified as uncertain. 
The right panel shows the YSO population denoted using blue filled circles.
We note that the \teff\, values are those 
derived from the spectral fitting and the luminosities plotted are 
the photospheric luminosities ($L_{\rm phot}$).

Each plot shows the main sequence as a cyan cross-hatched region. Evolutionary 
tracks starting from the main sequence and continuing up to the AGB-tip according to the tracks of 
Bertelli et al. (2008,2009) are shown as black solid lines. Note that these tracks 
use a synthetic AGB calculation adopting unusual mass loss rates, and almost certainly terminate at too 
low a luminosity.  The plots also show the PISA pre-main sequence (PMS) 
evolutionary tracks \citep[black dotted lines:][]{tognelli11} 
up to the maximum 
computed mass of 7\Msun. A metallicity $Z$ = 0.004 was selected for both sets of evolutionary tracks. 
The masses of the evolutionary tracks are marked on the plots with the 
PMS and main-sequence masses marked on the left side of the plots and RGB-tip masses marked 
on the right side of the plots. The positions of the RGB and AGB are also marked.

In the figure showing the post-AGB/RGB candidates, post-AGB and post-RGB evolutionary 
tracks are shown schematically (black dashed arrows). The masses for the post-AGB 
evolutionary tracks are from \citet{vw94} for $Z$ = 0.004. The post-RGB evolutionary track masses 
are estimated from the RGB luminosity-core mass relation of the \citet{bertelli08} tracks 
with $Z$ = 0.004. 

In the HR diagram of the post-AGB/RGB candidates, the blue vertical lines shows the empirical 
OGLE instability strip for the Population 
II Cepheids \citep{soszynski08}, since post-AGB/RGB evolutionary tracks cross the Population 
II Cepheids instability strip.  In the HR diagram showing the YSO candidates, the green vertical 
lines on this plot denotes the Cepheid instability strip from \citet{chiosi93}. Also 
shown in the HR diagram showing the YSO candidates, is the birthline 
(thick black dashed line in right panel of Figure~\ref{hr}), which may be considered as the 
dividing line between the obscured protostellar and observable pre-main sequence stage of stellar evolution. 
The location of the birthline depends highly on the 
mass accretion rate, with higher accretion rates shifting the line to the right. A mass accretion rate of 10$^{-5} \Msun$/yr (used for the birthline in right panel of Figure~\ref{hr}) 
represents the typical value for stars in the 
mass range from few tenths of a solar 
mass to about 10\Msun\, \citep{stahler83,palla93}.

We find that most of the post-AGB/RGB and YSO candidates have \teff $<$ 10000K. 
The HR-diagram suggests that the post-AGB/RGB candidates are of $\sim$\,0.3\,$-$\,0.8\,\Msun\, and that the post-RGB candidates 
are mostly disc sources (and therefore inferred to be binaries). 

In the case of the YSO candidates, we find that the masses derived from the HR-diagram lie 
in the range $\sim$\,3\,$-$\,10\,\Msun. 
We also find that the majority of YSOs 
lie to the right of the birthline so they should not be visibly detectable. This discrepancy may be due 
to the assumption of symmetric and spherical dust 
shells in the birth line modelling (with asymmetries, it may be possible to see the central star through a region of low 
extinction), or too high and assumed accretion rate since the birthline depends on the mass-accretion rates \citep{palla93}. 
A low mass accretion rate could move the 
birthline to lower values of \teff\, so that our stars could become visible. A group of 
massive pre-main sequence stars 
similar to the Galactic Herbig AeBe stars was found in the LMC by \citet{lamers99} and these are also located 
above the traditional birth line used for the Galactic sources. \citet{lamers99} suggested that this could be due to either a shorter accretion 
timescale for Galactic Herbig AeBe stars due to lower metallicity in the LMC, or a lower dust-to-gas ratio in the LMC, 
again owing to the lower metallicity. Therefore for the SMC, a higher birth line for YSOs could be expected in the HR diagram. We note that the 
those candidates that were identified to have TiO emission features, within our survey,  also lie to the right of the birth line \citep{wood13} .

\section{Variability of post-AGB stars}
\label{lcanalysis}

\begin{figure*}
\centering
\includegraphics[bb = 4 1 554 746, width=14cm]{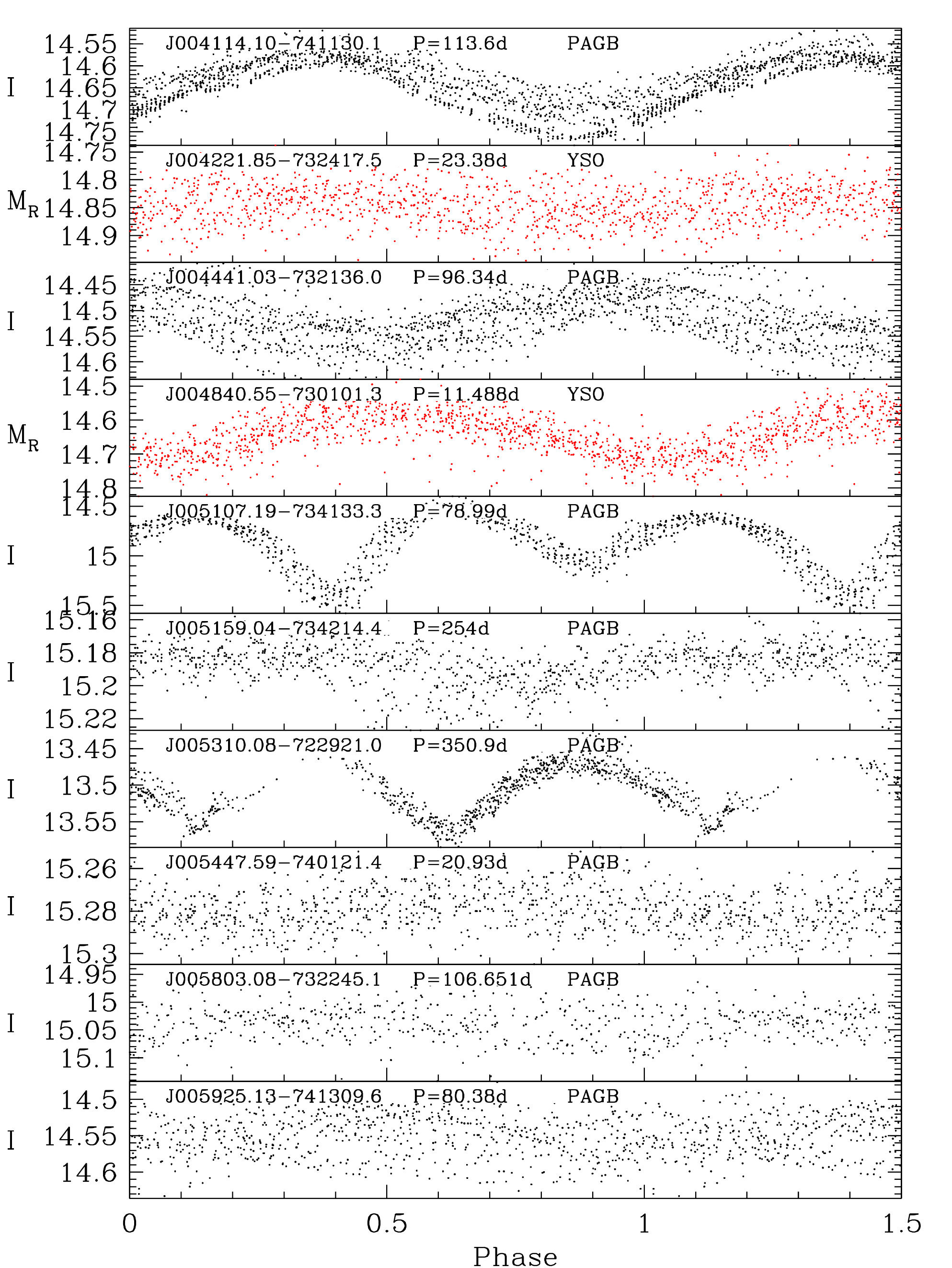}
\caption{The phased light curves for those stars that show a continuous periodic
variability. The object names, the periods and whether the object is
estimated to be a post-AGB/RGB star or YSO is indicated in each panel.
Phase zero is taken to be at JD=2448800.  Only MACHO or OGLE data
is used for a given object as indicated by the label on the magnitude
axis and point colour ($M_R$ and red points for MACHO data, $I$ and
black points for OGLE data). The RV Tauri star J005107.19-734133.3
is phased with the period between deep minima. The star
J005310.08-722921.0 is also phased using the period between alternate
minima although the evidence for an alternating depth of minima
is marginal (the star could be be an ellipsoidal binary).}
\label{phased}
\end{figure*}

The variability of the final sample of post-AGB/RGB and YSO candidates was examined by 
using the light curves from MACHO \citep{alcock92} and/or 
the OGLE\,II and OGLE\,III experiments \citep{udalski97,szymanski05,soszynski09,soszynski11}. Light curves exist for 38 of the 63 post-AGB/RGB 
candidates and 20 of the 40 YSO candidates (Figures~\ref{pagbq1lc} $-$~\ref{ysolc}). In Figure~\ref{phased} we show the 
phased light curves for those stars that show a continuous periodic variability. 

Of the 38 post-AGB/RGB candidates with light curves, 21 objects show no detectable periodicity. 
One of these 21 objects, J010623.71-724413.5, brightened 
by about 0.5 mags over a period of about 2500 days. This could be attributed to 
rapid changes in circumstellar dust obscuration. 

Thirteen stars, J004114.10-741130.1, J004441.03-732136.0, J004534.36-734811.8, J004614.67-723519.0, J004909.72-724745.4, J005113.04-722227.0, J005159.04-734214.4, J005310.8-722921.0, J005803.08-724405.1, J005447.59-740121.4, J005925.13-741309.6, J010021.78-730901.3, and J010254.90-722120.9 display semi-regular variability with 
periods from $\sim$\,20\,$-$\,500 days. Their periods are listed in Table~\ref{periods} when they could 
be determined. 

J005803.08-732245 
is also found to have a long secondary period (LSP) of 3700 days. LSPs are common in red giants 
\citep{wood99-pr,percy03,soszynski07,fraser08}. 
Two other candidates J004456.21-732256.6 and J003611.06-730447.0 exhibit LSPs of about 1800 and 1900 days, 
respectively. Stars with LSPs, also known as Sequence-D variables, are 
known to exhibit a mid-IR excess 
due to circumstellar dust \citep{wood99-pr,wood09}. The three stars: J005803.08-732245, J004456.21-732256.6, and J003611.06-730447.0, could be 
higher temperature analogues of the Sequence-D variables. 

J005107.19-734133.3 shows 
smooth oscillations. From the phased light curve for this star (see Figure~\ref{phased}), 
it is clear that 
these oscillations look RV-Tauri like with alternating deep and shallow minima and a period 
between alternating minima of 78.99 days. Based on the SED, J005107.19-734133.3 is classified as a disc source 
(see Section~\ref{sedanalysis}). RV-Tauri stars with circumstellar dust are mainly associated 
with circumbinary discs 
\citep[][]{vanwinckel99,gielen08}. Furthermore, visual inspection of the spectra of 
J005107.19-734133.3 indicates $s$-process enrichment 
(see Section~\ref{specanalysis}), similar to MACHO47.2496.8 which is a $s$-process rich RV-Tauri star in the 
LMC \citep{reyniers06}, suggesting that J005107.19-734133.3 is a 
newly discovered RV-Tauri star in the SMC. High resolution chemical abundance studies of J005107.19-734133.3 is 
required to confirm the $s$-process enrichment of this object.

The star J005310.08-722921.0 
displays a slow regular oscillations with a slight hint of alternation in minima and maxima as shown 
in the phased light curve of this 
object (see Figure~\ref{phased}). The time between alternate minima is about 350.9 days. 
For J005310.08-722921.0 the period seems too long to classify is as an RV Tauri star. 
The maximum period for RV Tauri stars in the SMC 
was found to be close to 100 days \citep{soszynski10}. The more likely scenario is that this 
star is in a binary system. 

J005311.41-740621.2 shows a fading in magnitude combined with a 
variation with a variable 
period between minima of 200 $-$ 400 days. The fading could be due to an LSP of $\sim$3000 days. 

Finally, J010342.34-721342.7 
is another star with a combination of a slow brightening and a long period of about 900 days. This 
may be attributed to slow changes in dust obscuration or changes in the accretion rate. 

Twenty of the YSO candidates have light curves. Amongst these candidates, six 
(J004301.63-732050.9, J004927.26-724738.4, J005159.81-723511.1,J005606.53-724722.7, J005800.62-721439.8 and J010634.50-721505.0) 
show erratic or secular long-term variations consistent with variations in dust obscuration by the circumstellar environment. 
J005318.28-733528.7 shows apparent variability which is probably not real but caused by an annual variation of about 365 days. 
Four stars, J004221.85-732417.5, J004451.87-725733.6, J004840.55-730101.3, and J005101.48-733100.4, show small oscillations with periods of 23.38, 22.5, 11.48, and 30 days, respectively. 
These small oscillations are similar to the ultra small amplitude oscillations displayed by stars that 
lie close to the Cepheid instability strip \citep{buchler09a}. This indicates that the 4 YSO candidates with small 
amplitude oscillations could be crossing the Cepheid instability strip on their way to the main sequence.

\begin{figure*}
\centering
\subfloat{\includegraphics[width=15cm]{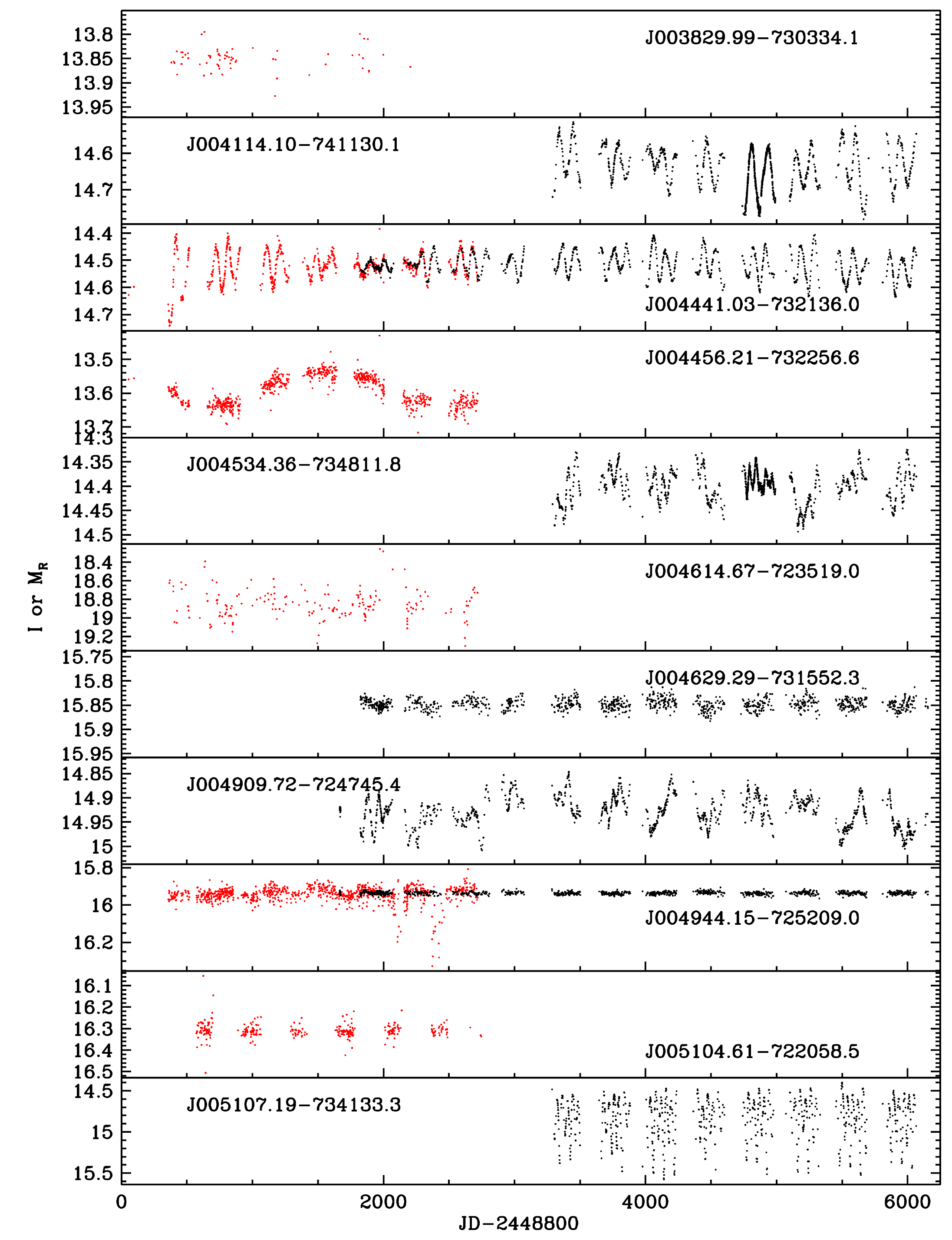}}\,
\caption{Light curves for the post-AGB/RGB Q1 candidates. The black light curves for dates later than $JD-$2448800 $>$ 3250 are 
from OGLE III, the black curves with 1700 $<$ JD$-$2448800 $<$ 3250 are from OGLE II while the red curves  with 0 $<$ JD$-$2448800 $<$ 2800
are MACHO red magnitudes normalised to the OGLE $I$ magnitudes over the interval 2000 $<$ JD$-$2448800 $<$ 3250. The light curves are ordered 
by RA.}
\label{pagbq1lc}
\end{figure*}
\begin{figure*}
\ContinuedFloat
\centering
\subfloat{\includegraphics[width=15cm]{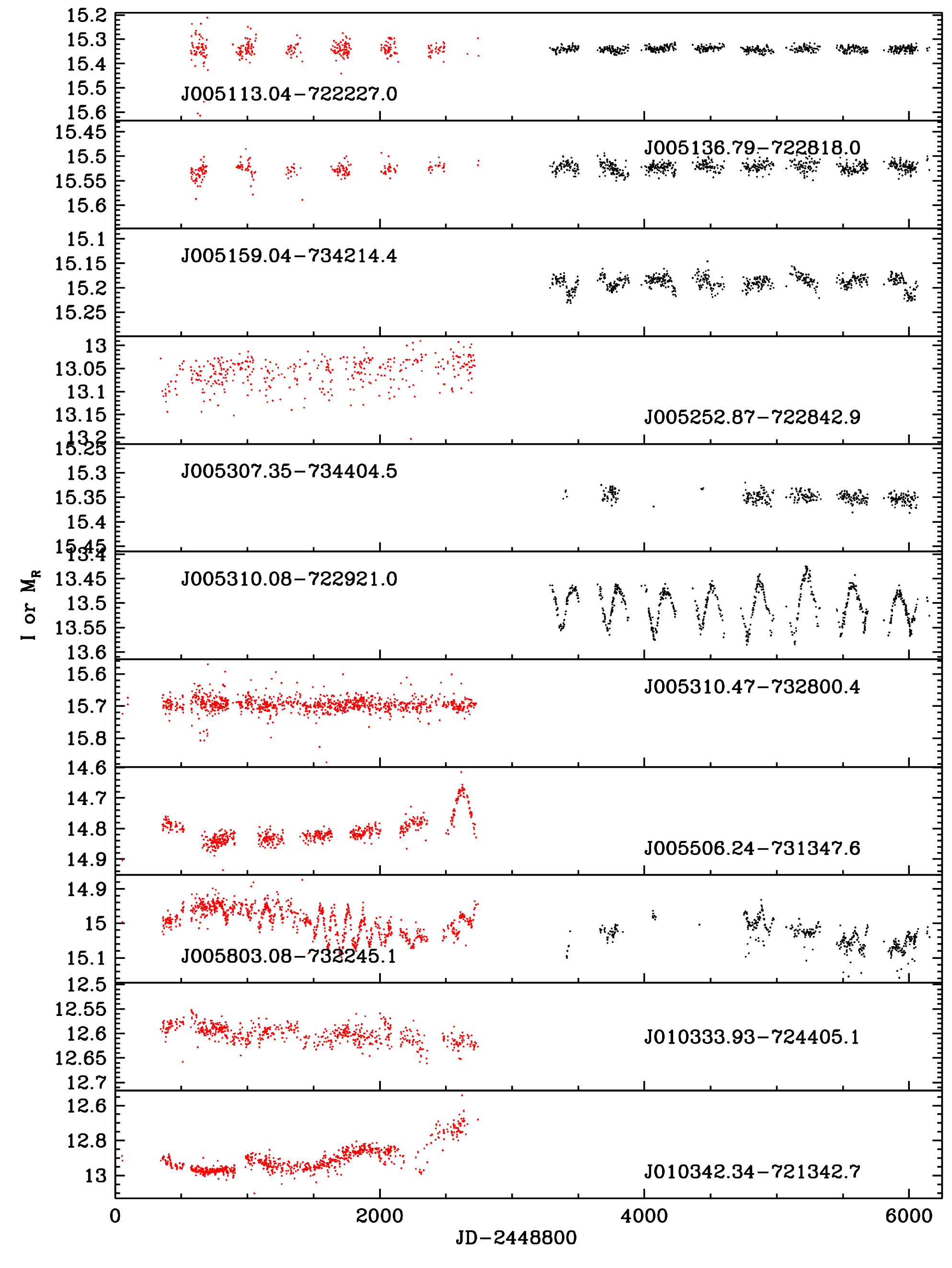}}\,
\caption{Figure~\ref{pagbq1lc} continued.}
\end{figure*}

\clearpage

\begin{figure*}
\centering
\includegraphics[bb = 18 11 587 772,width=13cm]{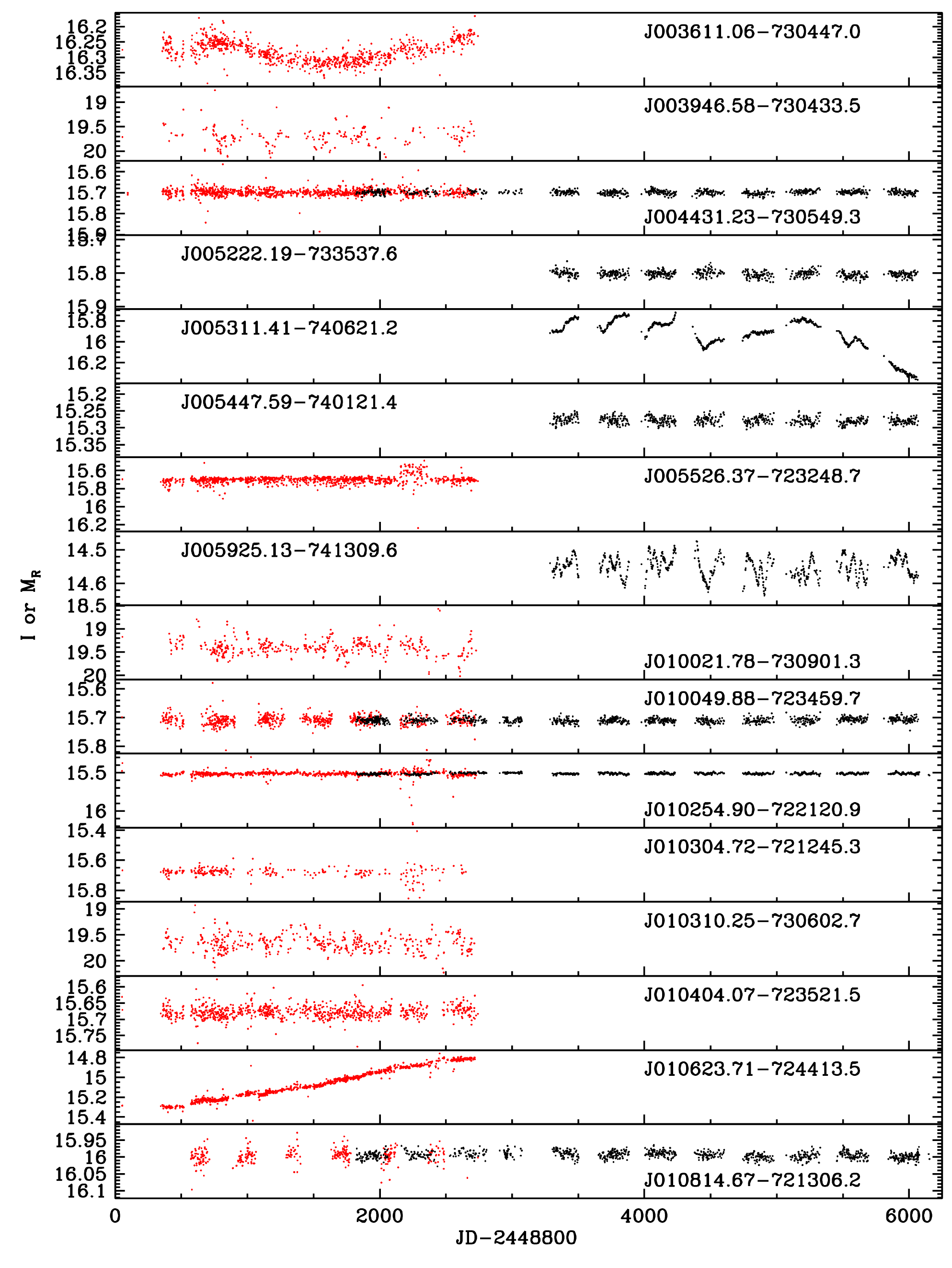}
\caption{Same as Figure~\ref{pagbq1lc}, but for the 
high probability Quality 2 post-AGB/RGB candidates.}
\label{pagbq2lc}
\end{figure*}
\clearpage

\begin{figure*}
\centering
\includegraphics[width=13cm]{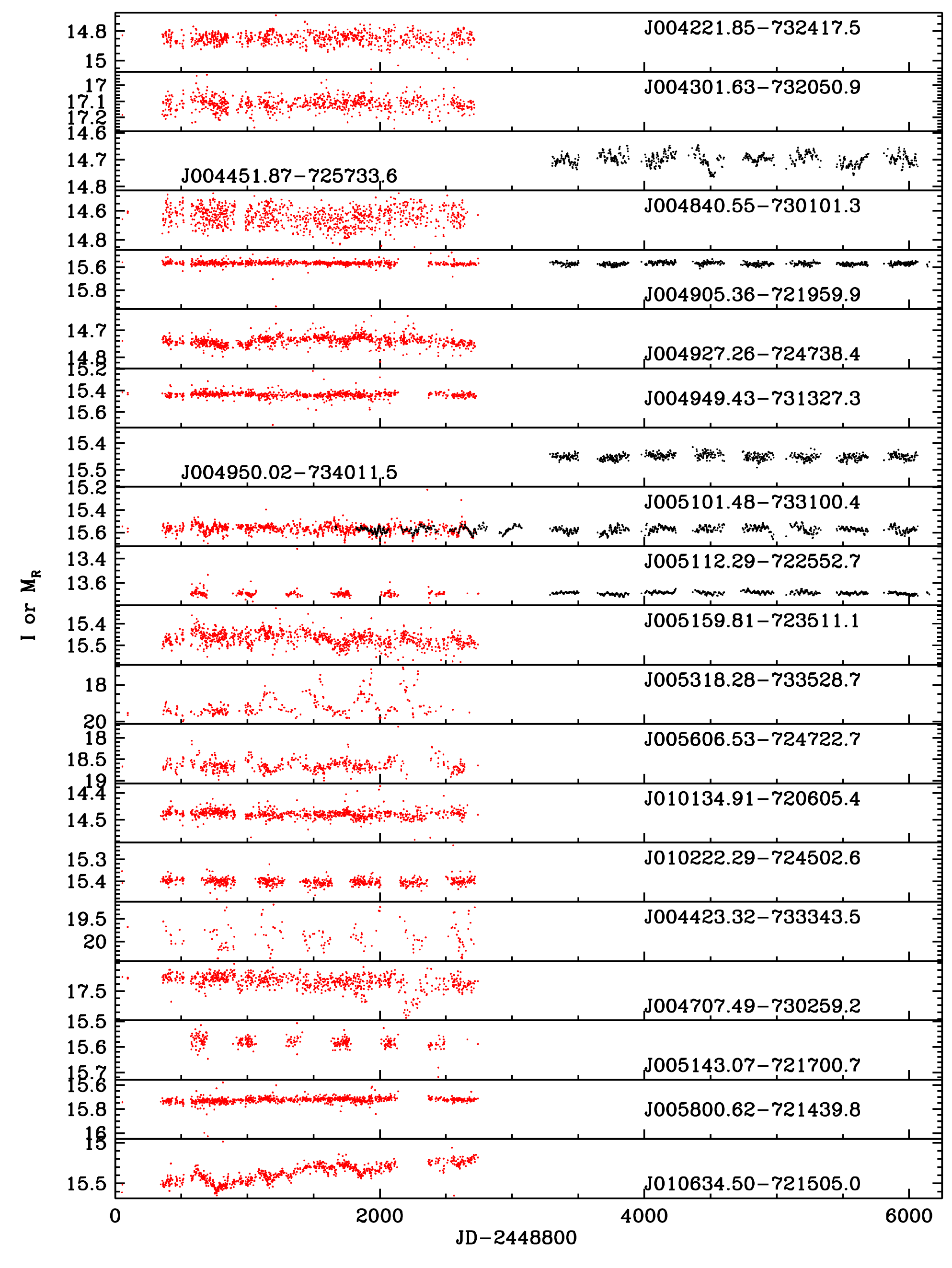}
\caption{Same as Figure~\ref{pagbq1lc} but for Quality 1 (top 15 light curves) and Quality 2 (bottom 5 light curves) 
YSO candidates.}
\label{ysolc}
\end{figure*}
\clearpage

\clearpage

\begin{table*}
\small
  \caption{The sample of post-AGB/RGB and YSO candidates with MACHO and/or OGLE light curves. 
Periods are given when they could be determined.}
\label{periods}
\vspace{-3mm}
\tabcolsep=1.75pt
   \centering
  \begin{tabular}{cccccccc}
  \hline
   Object & P (days) & Object & P (days) & Object & P (days) & Object & P (days)\\
  \hline
Post-AGB/RGB Q1 & & Post-AGB/RGB Q2 & & YSO Q1 & & YSO Q2\\
  \hline
J003829.99-730334.1 & -                           & J003611.06-730447.0 & 1900$\rightarrow$                 & J004221.85-732417.5 & 23.38    & J004423.32-733343.5 & - \\
J004114.10-741130.1 & 113.601             & J003946.58-730433.5 & -                                                 & J004301.63-732050.9 & -             & J004707.49-730259.2 & - \\
J004441.03-732136.0 & 96.338               & J004431.23-730549.3 & -                                                  & J004451.87-725733.6 & 22.5,       & J005143.07-721700.7 & - \\
J004456.21-732256.6 & 1800:                 & J005222.19-733537.6 & -                                                  & J004840.55-730101.3 & 11.488  & J005800.62-721439.8 & - \\
J004534.36-734811.8 & 59, 130              & J005311.41-740621.2 & 200-400, 2972$\rightarrow$ & J004905.36-721959.9 & -             & J010634.50-721505.0 & - \\
J004614.67-723519.0 & 87:                      & J005447.59-740121.4 & 20.926                                       & J004927.26-724738.4 & -             &                     & \\
J004629.29-731552.3 & -                          & J005925.13-741309.6 & 80.380                                       & J004949.43-731327.3  & -            &                     &  \\
J004909.72-724745.4 &  97                      & J010021.78-730901.3 & -                                                   & J004950.02-734011.5 & -            &                     &  \\
J004944.15-725209.0 &  -                         & J010049.88-723459.7 & -                                                   & J005101.48-733100.4 & 30:        &                     &  \\
J005104.61-722058.5 &  -                         & J010254.90-722120.9 & 40:                                               & J005112.29-722552.7 & -            &                     &  \\
J005107.19-734133.3 & 78.990              & J010304.72-721245.3 & -                                                    & J005159.81-723511.1 & -\\
J005113.04-722227.0 &  20                     & J010310.25-730602.7 & -                                                    & J005318.28-733528.7 & -\\
J005136.79-722818.0 & -                          & J010404.07-723521.5 & -                                                    & J005606.53-724722.7 & -\\
J005159.04-734214.4 & 254                    & J010623.71-724413.5 & -                                                    & J010134.91-720605.4 & -\\
J005252.87-722842.9 & -                          & J010814.67-721306.2 & -                                                    & J010222.29-724502.6 & -\\
J005307.35-734404.5 & -                          &     \\                   
J005310.08-722921.0 & 89, 120                     &     \\   
J005310.47-732800.4 & -                          &     \\  
J005506.24-731347.6 & -                          &     \\  
J005803.08-732245.1 &106.651, 3700  &     \\  
J010333.93-724405.1 & -                          &     \\  
J010342.34-721342.7 & 900:                   &      \\  
 \end{tabular}
\vspace{-3mm}
\begin{flushleft}
Note: Uncertain periods are denoted by ':'. $\rightarrow$ denotes that the true period is increasing from the estimated period with time.
\end{flushleft}
\normalsize
\end{table*}

\section{Completeness of the survey}
\label{completeness}

This study is aimed at identifying optically visible post-AGB and RGB candidates in the SMC. 
However, the survey obviously has its limitations and does not catalogue all the post-AGB/RGB candidates in the SMC. 
In this section we have listed the limitations that govern this survey and also provided the 
extent of the completeness of our survey.

The selection criteria used to identify the post-AGB/RGB candidates in this study 
requires a detection at 8$\mu$m. The presence of a valid 8$\mu$m detection and our selection 
criteria restricts our search to post-AGB/RGB candidates with an excess at 8$\mu$m. Older post-AGB/RGB 
stars with expanding shells for which the excess starts redder than 8$\mu$m will not be selected, 
for instance, the SMC counterparts of the hot Galactic post-AGB stars studied by 
\citet{gauba04} will not be selected if the photospheric 8$\mu$m detection is below the threshold. Furthermore, we require that all the 
selected candidates have a $V$\,$<$\,20, which only selects those post-AGB/RGB stars that are 
optically visible, therefore omitting those that are heavily dust enshrouded.

Based on our selection criteria, we selected a sample 
of 1194 objects, out of which 150 objects were given a priority 1, 303 were given a priority 2 and 
741 were given a priority 3. We note that the priorities were 
assigned based on the selection criteria used and the likelihood of finding post-AGB/RGB 
candidates with that selection 
criteria (see Section~\ref{intro}). We performed a low resolution optical survey that covered the SMC, as 
shown in Figure~\ref{fields}. From the initial selected sample of 1194 objects, we obtained spectra of 801 objects.  
Excluding the faint objects that 
had a low signal ($<$\,200 counts) and therefore a poor quality spectra we were left with a sample of 621 unique spectra.
In Table~\ref{samplevo} we give a summary of the evolution of the 
number of post-AGB/RGB stars and YSO candidates as the analysis has proceeded. In 
Table~\ref{samplevo2} we give a breakdown of the number of objects with respect to their assigned priorities 
as the analysis has proceeded. Taking into consideration the candidates for which we did not obtain 
the low resolution optical spectra and the objects that we rejected from the spectral analysis due to 
poor quality of their spectra, we estimate the completeness of the survey to be $\approx$\,50$\%$. However, when only 
considering the priority 1 objects, we find that we were able to study and assign candidature for 120 out of 
the 150 objects, resulting in $\approx$\,80$\%$ completeness. Similarly we estimate the completeness of the study of the priority 2 and priority 3 
objects to be $\approx$\,76$\%$ and $\approx$\,37$\%$, respectively.

We can provide a rough estimate 
of the number of post-AGB/RGB candidates we expect to find in the unstudied sample 
(the sample of objects whose spectra were not obtained and those objects that were rejected due to poor 
spectra, see Table 11). Nineteen out of the 120 priority 1 
objects turned out to be post-AGB/RGB candidates. So from the remaining unstudied sample of 30 priority 1 
post-AGB/RGB candidates, we can expect $\approx$\,5 more likely post-AGB/RGB candidates. 30 out of the 230 
priority 2 objects turned out to be post-AGB/RGB candidates. So from the remaining unstudied sample of 
73 priority 2
candidates, we can expect $\approx$\,10 more likely post-AGB/RGB candidates. Similarly, from the 
group of 469 unstudied priority 3 post-AGB/RGB candidates, we 
can expect $\approx$\,24 more likely post-AGB/RGB candidates. This implies that it might be possible to identify a 
further group of $\approx$\, 39 optically visible 
post-AGB/RGB candidates from the total unstudied group of initially selected 
candidates.  Based on the ratio of post-AGB to post-RGB candidates in our current final 
sample, we can expect to find around 13 additional post-AGB candidates and 26 additional post-RGB candidates. 

As part of the future work, we intend to complete the survey by obtaining the optical spectra for the 
objects that were not observed, and by re-observing those targets that had a poor quality spectra.

\begin{table*}
  \caption{A summary 
of the evolution of the number of post-AGB/RGB stars and YSO candidates as analysis has 
proceeded.}
  \centering
  \label{samplevo}
  \begin{tabular}{ll}
  \hline
   Stage & Numbers of objects at various stage analysis \\
  \hline
   Initial photometric selection & 1194 \\
   Objects with usable optical spectra & 621 \\
   Final sample of post-AGB/RGB and YSO candidates with confirmed  & 103 (63 post-AGB/RGB, 40 YSOs) \\
   SMC membership and with stellar parameters derived from their spectra & \\
  \hline
 \end{tabular}
\end{table*}

\begin{table*}
  \caption{A breakdown of the number of objects with respect to their assigned priorities, as analysis has 
proceeded.}
  \centering
  \label{samplevo2}
  \begin{tabular}{lcccc}
  \hline
   Stage & Total objects & Priority 1 objects & Priority 2 objects & Priority 3 objects \\
  \hline
   Initial photometric selection & 1194 & 150 & 303 & 741 \\
   Objects with optical spectra & 801 & 133 & 248 & 420 \\
   Objects rejected due to poor spectra & 180 & 13 & 19 & 148 \\
   Objects retained with good optical spectra & 621 & 120 & 229 & 272 \\
   Objects classified as M stars & 20 & 9 & 8 & 3 \\
   Objects classified as C stars & 140 & 12 & 126 & 2 \\
   Objects classified as PN & 46 & 24 & 10 & 12 \\
   Objects classified as red-shifted galaxies & 204 & 1 & 13 & 190 \\
   Objects classified as QSOs & 36 & 0 & 3 & 33 \\
   Objects with TiO in emission & 9 & 8 & 1 & 0 \\
   Objects with strong emission lines and an UV continuum  & 63 & 39 & 19 & 5 \\
   (probable hot post-AGB/RGB or luminous YSO candidates) & \\
   Sample of probable post-AGB/RGB and YSO candidates & 103 & 27 & 49 & 27 \\
   for which we carry out detailed spectral analysis & \\
   Final number of post-AGB/RGB candidates & 63 & 19 & 30 & 14\\
   Final number of YSO candidates & 40 & 8 & 18 &14 \\
  \hline
 \end{tabular}
\end{table*}

\section{Estimating Post-AGB/RGB Evolutionary Rates}
\label{evolrate}

The post-AGB/RGB phase of evolution is a very short lived phase. To be able to estimate the 
evolutionary rate, a complete sample of post-AGB stars is required. 
Furthermore, during the post-AGB/RGB phase, \teff\, is determined by the mass of the hydrogen 
envelope \citep{schoenberner81,vw94}. The rate of evolution in \teff\, is 
therefore determined by the luminosity (which determines the rate of consumption of 
the hydrogen envelope by nuclear burning) and by the mass loss rate 
(in the case of single post-AGB stars) and the mass re-accretion rate (in the case binary of post-AGB/RGB stars). 
This mass loss rate/mass accretion rate is essentially completely 
unknown. Based on the available sample of post-AGB/RGB candidates from this study, 
we now try to estimate this mass loss/accretion rate by 
determining the numbers of stars in a given \teff\, interval along the post-AGB/RGB track 
and comparing those numbers to the number of 
stars in the top magnitude of the RGB, where the duration of evolution is known.

To estimate the lifetime on the top one magnitude of the RGB, we used the 
\citet{bertelli08} evolutionary tracks. They show that 
stars in the mass range 1.0 to 1.8 \Msun\, and SMC-like 
metallicity (Z=0.004) take $\sim$\,3$\times$\,10$^{6}$ years to traverse the 
top magnitude of the RGB.  Subsequent AGB evolution through the 
same luminosity range takes $\sim$\,1$\times$\,10$^{6}$ years. Hence, the total 
time spent by a low mass star in the luminosity range corresponding 
to the top one magnitude of the RGB is $\sim$\,4$\times$\,10$^{6}$ years.  

In order to estimate the observed number of stars on the top 
magnitude of the RGB in the four fields SMC1-4, we followed the prescriptions in 
\citet{nie12}.  Stars in the
 SAGE-SMC catalog \citep{gordon11} were plotted in the 
$J$, $J-$[3.6] diagram and those in a parallelogram coinciding 
 with the top 1 magnitude of the RGB were selected. The 
parallelogram has sides $J$\,=\,13.9, $J$\,=\,14.9, 
$J-$[3.6]\,=\,3.25 $-$ 0.17$J$ and $J-$[3.6]\,=\,3.75 $-$ 0.17$J$.
We find the numbers of stars n1, n2, 
n3 and n4 in each of SMC1, SMC2, SMC3 and SMC4 are 6861, 9540, 5217, and 5343, 
respectively, with an error of approximately 5\% \citep{nie12}.
The total number of stars in the magnitude interval corresponding
to the top mag of the RGB in all 4 SMC fields is thus 26961.

Next we estimated the lifetimes of the stars in the early part of the post-AGB/RGB phase. 
If we assume that post-AGB stars, whether they leave the AGB by single 
 star mass loss or binary interaction, have all passed through the top 
 magnitude of the RGB then we can easily derive their average post-AGB lifetime. 
If there are N$_{\rm AGB}$ post-AGB stars $(\log L \ga 3.4)$ in all 4 SMC fields in a 
 certain evolutionary phase (say from the AGB to \teff = 6000\,K i.e. 
 $\sim$3.6 $<$ $\log$\teff $< \sim$3.8), then the lifetime of these stars in that 
 evolutionary phase is 4$\times$10$^{6}$*N$_{\rm AGB}$/26961 years. From our analysis, we find a 
total of N$_{\rm AGB}$ = 21 which results in a post-AGB lifetime of 3115 yrs. We note that 
the estimated lifetime assumes that the sample of in the fields SMC 1 $-$ 4 is complete.

For post-RGB stars, the calculation is not so easy. \citet{nie12}
 find that $\sim$4\% of red giants evolving up the RGB produce post-RGB stars
 when they fill their Roche lobes before reaching the RGB tip. The \citet{nie12} 
calculations were done for the LMC but we assume they also apply 
 for the SMC. Most of the post-RGB stars are produced during the top 
two magnitudes of the RGB where most of our post-RGB stars are observed. 
The median luminosity of these stars is $\log$\,L/\Lsun\,$\sim$\,2.8. In this case, 
 if N$_{\rm RGB}$ post-RGB stars are observed, then their average lifetime is 
 4$\times$10$^{6}*($N$_{\rm RGB}/0.04)/26961$ years. From our analysis, we find a 
total of N$_{\rm RGB}$ = 42 post-RGB candidates, which results in a post-RGB lifetime of 155780 yrs.

We need to compare the above lifetimes to those of post-AGB/RGB stars 
without external mass loss during the post-AGB life. The evolutionary track of \citet{bertelli08} with M = 1.2\Msun\, and metallicity Z=0.004 was used 
as the starting point.  The core mass at a number of luminosities on the RGB and AGB was extracted.  Then static models were made 
with varying envelope masses and hence \teff values in order to simulate stars 
that have left the RGB or AGB at these luminosities 
\citep[the static model code used was that of \citet{fox82}, with updated physics as described in][]{soszynski13}. 
Post-RGB and post-AGB stars evolve at constant luminosity to higher \teff\, values and, in the 
absence of mass loss, the rate of evolution is determined by the rate at which the hydrogen-rich envelope is consumed by the 
H-burning shell. Most of our observed post-RGB stars (which have $\log$ L/\Lsun $<$ 3.4) have $\log$\,\teff\,$<$ 3.8 
so we estimate the time $\Delta$t it takes for a post-RGB star to traverse from 
$\log$\,\teff(RGB)\,+\,0.05 to $\log$\,\teff\, = 3.8 by consuming the hydrogen-rich envelope. A helium 
mass fraction Y=0.25 was assumed and the H-burning shell was assumed to provide 
all the surface luminosity. Similarly, we compute the time $\Delta$t for a post-AGB star to 
traverse the interval 3.7 $< \log$\,\teff $<$ 3.9 where most of the observed post-AGB stars lie. 
These times are given in Table~\ref{table:lifetime}.

\begin{table}
   \centering
  \begin{tabular}{|cccc|}
  \hline
   Stage & $\log\,L$/\Lsun &  $M$$_{\rm core}$/\Msun  &  $\Delta$t\,(yr) \\    
  \hline
  RGB & 2.6 & 0.347 & 139000 \\
  RGB & 3.0 & 0.396 & 101000 \\
  RGB & 3.4 & 0.463 & 163000 \\
  AGB & 3.4 & 0.536 & 17400 \\
  AGB & 3.8 & 0.598 & 9300 \\
  \hline
 \end{tabular}
  \caption{Evolutionary rates for post-AGB/RGB stars of different core-mass ($M$$_{\rm core}$/\Msun).}
\label{table:lifetime}
\end{table}

We find that for the post-AGB stars, the observational lifetime = 3115, and the 
lifetime for post-AGB evolution without post-AGB mass loss from $\log$ \teff\,=\,3.7 to 3.9 
and for an intermediate luminosity star ($\log$L/\Lsun\,=\,3.8) is 9300 years. 
Formally, these numbers suggest a some mass loss is required to hasten the evolution. 
This mass loss rate is M$_{\odot}$ = 7\,$\times$\,10$^{-8}$ \Msun/yr but the uncertainties
 are very large. For the RGB stars, the observational lifetime = 155780, and the lifetime 
for post-AGB evolution without
 mass loss from $\log$ \teff(RGB)\,+\,0.05 to $\log$ \teff = 3.8 and for an intermediate luminosity
 star ($\log$L/\Lsun=3.0) is 101000 years. The agreement is good. Formally, the numbers suggest that a small amount of mass accretion is required in 
the post-RGB phase to match 
the observed and predicted numbers of post-RGB stars. Overall, 
these results suggest that the numbers of post-AGB and post-RGB stars that we have 
found are in reasonable agreement with stellar evolution models that have some mass loss in the 
post-AGB phase and a very low amount of re-accretion in the lower luminosity RGB phase. 

We note that 
the estimated lifetime assumes that the sample of stars in the fields SMC 1 $-$ 4 is complete. 
However, as mentioned in Section~\ref{completeness}, our survey is not entirely complete and 
we can expect to find an additional sample of 13 post-AGB candidates and 26 post-RGB candidates. 
To illustrate the impact of this addition, we recalculated the
lifetimes and we estimate a post-AGB lifetime of 5044 yrs 
and a post-RGB lifetime of 252216 yrs. These lifetimes are not greatly different from the lifetimes we have estimated from 
the current sample so that our conclusions on the estimated lifetimes for the post-AGB and post-RGB 
phases remain valid.

\section{Summary and Conclusions}

We have identified a sample of 63 high probability post-AGB/RGB candidates in the 
SMC with spectral types between A and K. Of these 63 objects, 42 are post-RGB 
candidates and 21 are post-AGB candidates. Being an evolved 
class of objects, they have a lower metallicity ([Fe/H] = -1.14) than the mean present-day SMC metallicity. 
The $J-$[3.6] vs [3.6]$-$[8]  colour-colour plot and the spectral energy distributions allowed us to distinguish between single (shell sources) and 
binary (mostly discs) post-AGB/RGB 
populations, resulting in 6 shell sources and 27 disc sources. For the remaining 30 sources, we were 
unable to establish their nature. However, majority of these sources are post-RGB candidates which 
are known to be binaries and therefore likely disc sources. Detailed 
studies are required to confirm the true nature of all these sources. The low resolution spectra of these objects 
revealed the definite presence of barium for 6 candidates and lithium for 7 candidates, both of which are expected products of the 
nucleosynthesis that during the AGB phase of evolution. Variability is displayed by 
38 of the 63 post-AGB/RGB candidates with the most common 
variability types being the Population II Cepheids (including RV-Tauri stars) and semi-regular variables. This 
study has resulted in the discovery of a new 
RV-Tauri star, J005107.19-734133.3, which shows signs of $s$-process enrichment, based on visual inspection of the 
low-resolution spectrum. We also used the reliable numbers of these objects, to study 
the evolutionary rates and mass loss/mass accretion rates. We found that the numbers 
of post-AGB require stellar evolution models with some mass loss and the number of RGB stars 
suggests a very small amount of re-accretion of gas.

This study has also resulted in a new sample of YSOs, since YSOs also display a large IR excess and 
are present in the luminosity range occupied by post-AGB stars. We identified a sample of 
40 high probability YSO candidates. The high probability population has temperatures ranging  
between 4000K and 9000K, high surface gravities, and a mean metallicity [Fe/H] = -0.62, which agrees well 
with the average present-day SMC metallicity. From the position of these YSO candidates on the HR diagram, we 
were able to infer that they have masses of $\sim$\,3$-$\,10\,\Msun. An interesting finding is that most of these YSO candidates lie to the right of the birthline where previous 
studies have shown that the objects are not visible. Both groups of 
YSO candidates showed H$\alpha$ emission and forbidden line emission indicative of disc 
accretion in YSOs. We were also able to identify the presence of Li in three candidates. 
Four of the YSO candidates with light curves show low amplitude periodicity which is 
probably associated with Cepheid-like pulsations as the stars cross the instability strip. 
Slow variations in the obscurations by circumstellar matter us seen in both post-AGB/RGB stars 
and YSOs but it is more common in the latter.

We have also identified a group of 63 hot objects whose spectra show emission lines and in some cases, a significant UV continuum. 
These objects are likely to be either hot post-AGB/RGB or luminous YSO candidates (presented in Appendix~\ref{YSO}). Based on a visual 
inspection of their spectra and SEDs, we were able to establish the most probable nature of the objects, resulting in 40 probable 
hot post-AGB/RGB candidates and 23 probable YSO candidates.

This study has also resulted in the discovery of a significant number contaminants. They are: M-stars, C-stars 
and PNe (presented in Appendix~\ref{PN_M_C}), a group of QSOs and 
red-shifted galaxies (to be presented in a following publications), 
and a group of stars with TiO band emission \citep{wood13}.

We note that, due to limitations introduced by the selection criteria, our study 
is restricted to optically visible post-AGB/RGB stars of spectral type A $-$ K, in the SMC. The completeness of this survey is 
$\approx$\,50$\%$ since we were not able to obtain spectra all of the candidates from within the initially selected sample of candidates and some 
of the candidates with optical spectra were rejected as their spectra were of poor quality due to the faintness of the targets combined with the low resolution 
of our spectra ($\approx$ 1300). Based on the current final sample of post-AGB/RGB candidates (of A $-$ K) in the SMC, we expect to find 
approximately an additional 39 such candidates (13 post-AGB and 26 post-RGB candidates).

\section*{Acknowledgments}
DK would like to thank Prof. Martin Asplund for his valuable discussions and advice throughout the project. DK would also like to 
thank George Zhou for his help while developing the spectral typing pipeline and Dr. Rob Sharp for his useful tips during data 
reduction of the AAOmega spectra

We thank the Australian Astronomical Observatory for allowing us to use the observatory facilities 
and our AAT support astronomer, 
Dr. Paul Dobbie, who was very helpful during our observing run. We thank the AAO Service 
Program, especially Dr. Sarah Brough and Dr. Daniel Zucker, our service observers, for 
observing one of the SMC fields. 

We would also like to thank the referee for his/her comments and suggestions. 

PRW was partly supported during this work by Australian Research Council Discovery Project grant DP1095368.

DK acknowledges support of the FWO grant G.OB86.13. HVW and DK acknowledges support of the 
KU Leuven contract GOA/13/012.

\bibliography{mnemonic,devlib}

\appendix
\section[]{Candidates with Strong Emission Features: Hot Post-AGBs }
\label{YSO}

In this section, we present the 63 objects that have strong emission features and a significant UV continuum. Based on their spectra, we find that, these objects can either be 
hot post-AGB candidates or YSO candidates with a strong UV continuum.  Table~\ref{tab:ysoemissionobjects} lists the important spectral features that we 
were able to identify. We also provide positional cross-matchings 
for those objects found in previous studies and the estimated 
observed luminosity ($L_{\rm obs}$) for each of the objects. $L_{\rm obs}$ is computed by 
integrating under the flux distribution at observed wavelengths, is a lower limit 
to the real luminosity 
of the object because of the large amount of flux beyond 24$\mu$m and for hot photospheres the luminosity contribution from the 
region blue-ward of the B filter. Figure~\ref{ysospecfull} shows the spectra of these 63 objects. Figure~\ref{ysospecpart} shows the 
same stars but the plots have been scaled to show the continuum. The observed SEDs of these 63 candidates are shown in Figure~\ref{fig:yso18_sed}. Based on a visual inspection 
of their spectra and SEDs we have tried to establish the most probable nature of the object (either hot post-AGB/RGB  or YSO). Hot-post AGB/RGB stars are likely to have an emission-line spectrum characterised by weak recombination lines of hydrogen and helium and various collisionally-excited forbidden lines of heavier elements \citep[e.g.,][]{vanwinckel03}. The spectra of YSO candidates are likely to show a broad H$\alpha$ line profile owing to the disc accretion  in YSOs \citep{natta02,jayawardhana02}. Furthermore the YSO objects show a flared SED peaking at longer wavelengths (mostly $>$100$\mu$m). This classification results in 40 
probable hot post-AGB/RGB candidates and 23 probable YSOs. In Table~\ref{tab:ysoemissionobjects} we list the probable nature of each of these objects.

\begin{table*}
\scriptsize{
{\renewcommand{\arraystretch}{0.8}
\caption{The list of candidates with emission lines and a UV continuum . In this table 'a' represents absorption, 
'e' represents emission, '0' indicates that the feature is not observed. ':' 
indicates that there is some line blending that has taken place or there is an absorption line with an emission 
core or the line indicates signs of strong winds and therefore mass-loss. '$?$' 
represents that the nature of the spectral line is uncertain. }
\medskip
\tabcolsep=1.5pt
\begin{tabular}{lllcccccccccccccccccc}
Name & Previous Identification & $L_{\rm obs}$/\Lsun & Type & H$\alpha$ & H$\beta$ & H$\gamma$ & [OIII] & [OIII] & HeI & HeI & [SII] & [SII] & [NII] & [NII] & CaII & CaII & CaII & Li & Ba & Pa \\ 
Wavelength (\AA) &  & minimum &  & 6563 & 4861 & 4341 & 4659 & 5007 & 4471 & 5876 & 6717 & 6731 & 6548 & 6584 & 8498 & 8542 & 8662 & 6708 & 4554 & - \\ 
\hline
\hline
J002810.39-725844.5 &IR$^{\rm 1}$,PN$^{\rm 2,3}$ & 15128 & YSO & e & e & e & 0  & e & 0 & 0 & e & e & 0 & 0 & 0 & 0 & 0 & 0 & 0 & 0 \\ 
 & I25$^{\rm  4}$,I100$^{\rm 5}$,FIR$^{\rm 6}$ \\
 J003543.47-732110.7 &I25$^{4}$,FIR$^{\rm 6},$I60$^{7}$ & 7009 & hot pA/R & e & e & e & 0 & 0 & 0 & 0 & e & e & 0 & 0 & e & e & e & a? & 0 & e \\ 
J003717.72-730020.8 &RGB$^{\rm 6}$ & 1135 &YSO & e & e & e & 0 & e & 0 & 0 & e & e & 0 & 0 & 0 & 0 & 0 & 0 & 0 & 0 \\ 
J004419.98-725205.8 & - & 1972 & hot pA/R &e: & e & e: & 0 & 0 & e: & 0 & 0 & 0 & 0 & 0 & a? & a & a & 0 & 0 & a? \\ 
J004535.80-731412.2 & - & 21252 & hot pA/R & 0 & a& a& 0& 0& a& a& 0& 0& 0& 0& 0& 0& 0& 0& 0& a \\
J004607.42-731124.6 & - & 2868& hot pA/R & e& e& a:& 0& e?& a& 0& 0& 0& 0& 0& a& a& a& 0& 0& 0 \\
J004654.98-730833.6 & FIR$^{\rm 6}$, SB$^{\rm 8,9}$ & 28202 & hot pA/R & e & e & e & e? & e? & 0 & 0 & e & e & 0 & 0 & 0 & 0 & 0 & 0 & 0 & e \\ 
J004655.70-733158.4 &- & 1161 & YSO & e & e & e & 0 & e & 0 & e? & e & e & 0 & 0 & 0 & 0 & 0 & 0 & 0 & 0 \\ 
J004706.19-730759.2 &- & 2277 & YSO & e & e & e & 0 & e & 0 & 0 & e & e & 0 & 0 & 0 & 0 & 0 & 0 & 0 & e \\ 
J004752.33-731711.5 &FIR$^{\rm 6}$ & 10649 & YSO & e & e & e & 0 & e & 0 & 0 & e & e & 0 & 0 & 0 & 0 & 0 & 0 & 0 & 0 \\ 
J004805.74-731743.3 &FIR$^{\rm 6}$ & 2553 &YSO& e & e & e & 0 & e & 0 & 0 & e & e & e? & e? & 0 & 0 & 0 & 0 & 0 & e \\ 
J004830.60-730353.1 &Em$^{\rm 8}$ & 1811 & hot pA/R & e & e & e & 0 & e & e? & e? & e & e & 0 & 0 & 0 & 0 & 0 & 0 & 0 & e \\ 
J004842.90-730311.0 & Em$^{\rm 8}$ & 1129 &  hot pA/R&e & e & e & 0 & e & e? & e? & e & e & 0 & 0 & 0 & 0 & 0 & 0 & 0 & e \\ 
J004849.01-731123.2 &- & 783 & hot pA/R&e & e & e: & 0 & e & 0 & e? & e & e & e? & e? & 0 & 0 & 0 & 0 & 0 & e \\ 
J004929.11-723532.2 &Em$^{\rm 10}$ & 3633 & hot pA/R &e & e & e & 0 & e & 0 & e? & e & e & 0 & e? & 0 & 0 & 0 & 0 & 0 & e \\ 
J004934.21-724855.1 &- & 667 & YSO &e & e & e: & 0 & e & 0 & e? & e & e & 0 & 0 & 0 & 0 & 0 & 0 & 0 & e \\ 
J005047.44-721019.6 & Em*$^{8,9}$& 2159& hot pA/R& e & e & a & 0 & 0 & a & a & e & e & 0 & 0 & a & a & a & a & 0 & a \\ 
J005131.58-730911.7 & -& 745 & hot pA/R & e & e & a & 0 & e & 0 & 0 & e & e & 0 & 0 & a & a & a & e? & 0 & 0 \\ 
J005157.72-731421.8 &Em$^{\rm 3,8,11}$ & 822 & hot pA/R&e & e & e & 0 & e & 0 & e? & e & e & 0 & 0 & 0 & 0 & 0 & 0 & 0 & e \\ 
J005246.33-724244.5 & Em*$^{8}$ & 8465 & hot pA/R & e & e & e: & 0 & e & a & e & e & e & 0 & 0 & e & e & e & 0 & 0 & e \\ 
J005252.48-731833.9 & Em*$^{8}$ & 3134 & hot pA/R & e & e & e & e & e & a & 0 & e & e & 0 & 0 & e & e & e & 0 & 0 & e \\ 
J005309.86-731141.9 & - & 3084 & hot pA/R & e: & a: & a & 0 & e & a & a & e & e & 0 & 0 & 0 & 0 & 0 & a & 0 & 0 \\ 
J005339.94-725218.6 &x-AGB$^{\rm 6}$ & 11425 &hot pA/R & e & e & e: & e? & e & e? & e & e & e & 0 & 0 & 0 & 0 & 0 & 0 & 0 & 0 \\ 
J005344.56-731237.1 &Em$^{\rm 8,9}$ & 6425 & hot pA/R& e & e & e & 0 & e & 0 & e? & 0 & 0 & 0 & 0 & 0 & e & e & a? & 0 & e \\ 
J005409.46-724143.1 &x-AGB$^{\rm 6}$,Be$^{\rm 8}$ & 25304 & hot pA/R& e & e & e & e & e & e & e & e & e & 0 & 0 & 0 & 0 & 0 & 0 & 0 & e \\ 
J005439.09-722923.1 & Em*$^{8}$ & 6891 & hot pA/R & e & e & e & e & e & 0 & e & 0 & 0 & 0 & 0 & e & e & e & 0 & 0 & e \\ 
J005444.94-724109.8 & - & 3671 & hot pA/R & e& e& e& 0& e& a& 0& e& e& 0& 0& 0& 0& 0& 0& 0& e \\
J005648.54-724820.4 & - & 2062 & hot pA/R & e& e& a:& 0& 0& a& a& e& e& 0& 0& 0& 0& 0& 0& 0& e? \\
J005734.23-722654.6 & - & 12117 & hot pA/R & a& a& a& 0& 0& a& ?& 0& 0& 0& 0& 0& 0& 0& 0& 0& a \\
J005753.72-723317.5 & - & 1878 & hot pA/R & e & e & e & 0 & e & a & 0 & 0 & 0 & 0 & 0 & e & e & e & 0 & 0 & e \\ 
J005809.94-721102.0 & - & 15088 & hot pA/R & e & e & 0 & 0 & e & a & 0 & e & e & 0 & 0 & e & e & e & 0 & 0 & e \\ 
J005929.09-720104.6 &Em$^{\rm 8,9}$ & 42499 & hot pA/R & e & e & e & 0 & e? & 0 & 0 & 0 & 0 & 0 & 0 & e & e & e & 0 & 0 & e \\ 
J005942.38-714445.6 &- & 1040 & YSO &e & e & e & 0 & e & 0 & e? & e & e & 0 & 0 & 0 & 0 & 0 & 0 & 0 & e \\ 
J010053.80-714649.1 &- & 1263 & YSO &e & e & e: & 0 & e & e? & 0 & e? & e? & 0 & 0 & 0 & 0 & 0 & 0 & 0 & 0 \\ 
J010138.43-715655.5 &HII$^{\rm 3,8,11}$ & 1194 & hot pA/R & e & e & e & e & e & e & e & e & e & 0 & 0 & 0 & e & e & 0 & 0 & e \\ 
J010229.46-720153.3 &RGB$^{\rm 6}$ & 495 & YSO &e & e & e & 0 & e & 0 & e? & e & e & 0 & 0 & 0 & 0 & 0 & 0 & 0 & e? \\ 
J010246.58-715127.5 &- & 404 & hot pA/R&e & e & e & e? & e & 0 & 0 & e & e & 0 & 0 & 0 & 0 & 0 & 0 & 0 & a? \\ 
J010250.06-724022.1 &Em$^{\rm 8}$ & 1623 & hot pA/R& e & e & e & 0 & e & 0 & 0 & e & e & 0 & 0 & e & e & e & 0 & 0 & e \\ 
J010318.52-721213.4 &FIR$^{\rm 6}$,I60$^{\rm 7}$ & 4081 &YSO& e & e & e: & 0 & e & 0 & 0 & e & e? & e? & 0 & 0 & 0 & 0 & 0 & 0 & e:? \\ 
J010505.75-715942.8 &G $^{\rm 12}$ & 745 & YSO&e & e & e & 0 & e & 0 & e? & e & e & 0 & e? & 0 & 0 & 0 & 0 & 0 & e? \\ 
J010525.79-715858.5 &- & 221 & hot pA/R &e & e & e & 0 & e & 0 & e? & e & e & e & e & 0 & 0 & 0 & 0 & 0 & 0 \\ 
J010528.61-715942.7 &- & 987 & YSO &e & e & e & e & e & e & e & e & e & e? & e? & 0 & 0 & 0 & 0 & 0 & e \\ 
J010546.40-714705.2 &21 $\mu$m source$^{13}$& 4106 & hot pA/R& e& e& e& e& e& e& e& e& e& 0& 0& e& e& e& 0& 0& e \\
J010603.22-724931.3 &- & 3485&hot pA/R& e& e& e&0 & 0& a?& 0& e?& e?& 0& 0& 0& 0& 0& 0& 0& e \\
J010619.56-715559.4 &- & 1396 & YSO & e & e & e: & 0 & e & 0 & 0 & e & e & 0 & 0 & 0 & 0 & 0 & 0 & 0 & 0 \\ 
J010640.30-731024.6 &Em $^{\rm 8,9}$& 14998& hot pA/R& e & e & a & 0 & 0 & a & 0 & 0 & 0 & 0 & 0 & e & e & e & 0 & 0 & e \\ 
J010710.99-723503.7 &- & 271 & hot pA/R & e & e & e & 0 & e & 0 & e? & e & e & 0 & 0 & 0 & 0 & 0 & 0 & 0 & 0 \\ 
J010722.82-723334.0 &Em$^{\rm 8}$ & 1267 & YSO &e & e & e & 0 & e? & 0 & 0 & e & e & 0 & 0 & 0 & 0 & 0 & 0 & 0 & e \\ 
J010832.86-715941.2 &Em$^{\rm 8}$ & 2716 & YSO & e & e & e & 0 & e & 0 & e? & e & e & 0 & 0 & 0 & 0 & 0 & 0 & 0 & e \\ 
J010834.02-715900.5 &Em$^{\rm 8}$ & 1503 & hot pA/R & e & e & e & e & e & e & e & e & e & 0 & 0 & 0 & e & 0 & 0 & 0 & e \\ 
J011029.10-725338.2 &Em$^{\rm 3,8}$ & 766 &hot pA/R & e & e & e & 0 & e & 0 & 0 & e & e & e & e & 0 & 0 & 0 & 0 & 0 & e \\ 
J011045.12-722137.5 &Em$^{\rm 8}$ & 3699 & hot pA/R & e & e & e & 0 & e & 0 & e & e? & e? & 0 & 0 & 0 & 0 & 0 & 0 & 0 & e \\ 
J011341.19-725049.8 &Em$^{\rm 8}$ & 3171 &hot pA/R& e & e & e & 0 & e? & 0 & 0 & 0 & 0 & 0 & 0 & 0 & 0 & 0 & 0 & 0 & e \\ 
J011404.66-731658.3 &FIR$^{\rm 6}$ & 20140 &YSO& e & e & e & 0 & e & 0 & 0 & e & e & e? & e? & 0 & 0 & 0 & 0 & 0 & e \\ 
J011417.81-731210.6 &- & 567 & hot pA/R & e & e & a:? & 0 & e & 0 & 0 & e & e & 0 & e? & 0 & 0 & 0 & 0 & 0 & e \\ 
J011542.86-730959.3 &FIR$^{\rm 6}$,Em$^{\rm 8}$ & 5125 & hot pA/R & e & e & e & 0 & e & 0 & 0 & e? & e? & 0 & 0 & 0 & 0 & 0 & 0 & 0 & e \\ 
J011545.85-732040.3 &- & 2241 & hot pA/R&e & e & e & 0 & e & 0 & e? & 0 & 0 & 0 & 0 & 0 & 0 & 0 & 0 & 0 & e \\ 
\hline
\multicolumn{20}{c}{Previously identified YSOs}\\ 
\hline
J004456.37-731010.8 &FIR$^{\rm 6}$,IR$^{12}$,YSO$^{\rm 14}$ & 28083 & YSO & e & e & e & e & e & e & e & e & e & e & e & o & 0 & 0 & 0 & 0 & e \\ 
 & I60$^{\rm 7}$, HII$^{\rm 8,12}$ \\
J005043.26-724656.0 &YSO$^{\rm 14}$ & 5634 &YSO& e & e & e & 0 & e & 0 & e & e & e & e & e & 0 & 0 &0 & 0 & 0 & e \\ 
J005058.10-730756.6 &Em$^{\rm 8}$,YSO$^{\rm 3}$ & 1936 &YSO& e & e & e: & 0 & e & 0 & 0 & e & e & 0 & 0 & 0 & 0 & 0 & 0 & 0 & e \\ 
J005238.82-732623.9 &YSO$^{\rm 14}$ & 6105 & YSO & e & e & e & 0 & e & 0 & e & e & e & e & e & 0 & 0 & 0 & 0 & 0 & e \\ 
J005606.40-722827.9 &Em$^{\rm 8}$,YSO$^{\rm 14}$ & 2267 &YSO & e & e & e & 0 & e & e? & e? & e? & e? & 0 & 0 & 0 & 0 & 0 & 0 & 0 & e \\ 
J010549.29-715948.4 &I25$^{\rm 4}$,FIR$^{\rm 6}$,YSO$^{\rm 14}$ & 11423 & YSO & e & e & e & 0 & e & 0 & 0 & e & e & e? & e? & 0 & 0 & 0 & 0 & 0 & e? \\ 
\hline
\label{tab:ysoemissionobjects}
\end{tabular}}}
\begin{flushleft}
Notes: In the table, 'hot pA/R' represents hot post-AGB/RGB candidates; $L_{\rm obs}$/\Lsun is computed by integrating under the flux distribution at observed 
wavelengths. A positional cross-matching was performed with all the catalogues mentioned in Table~\ref{tab:pagb1_param}. A positional matching was found 
with the following catalogues:
$^{1}$\citet{1997A&amp;AS..125..419L},
$^{2}$\citet{1985MNRAS.213..491M},
$^{3}$\citet{1995A&amp;AS..112..445M},
$^{4}$\citet{2003A&amp;A...401..873W} (25$\mu$m),
$^{5}$\citet{2003A&amp;A...401..873W} (100$\mu$m),
$^{6}$\citet{boyer11},
$^{7}$\citet{2003A&amp;A...401..873W} (60$\mu$m),
$^{8}$\citet{1993A&amp;AS..102..451M},
$^{9}$\citet{2000MNRAS.311..741M},
$^{10}$\citet{2007MNRAS.376.1270L},
$^{11}$\citet{2002AJ....123..269J},
$^{12}$\citet{2010AJ....139.1553V},
$^{13}$\citet{volk11},
$^{14}$\citet{2012MNRAS.tmp..229O}, 
Catalogue identifications: PN - Planetary nebula; 
Em*,EmO - object with emission features; G - Galaxy; HII - HII region; IR - Infrared source; 
I12 - IRAS 12$\mu$m source; I25 - IRAS 25$\mu$m source; I60 - IRAS 60$\mu$m source; I100 -IRAS 100$\mu$ source, YSO - 
Young stellar object. The following objects are defined in 
\citet{boyer11}: FIR - far-IR object, RGB - red giant branch star, x-AGB - dusty AGB star with superwind mass 
loss. \\
\end{flushleft}
\end{table*}

\clearpage

\begin{figure*}
\centering
\subfloat{\includegraphics[width=18.5cm]{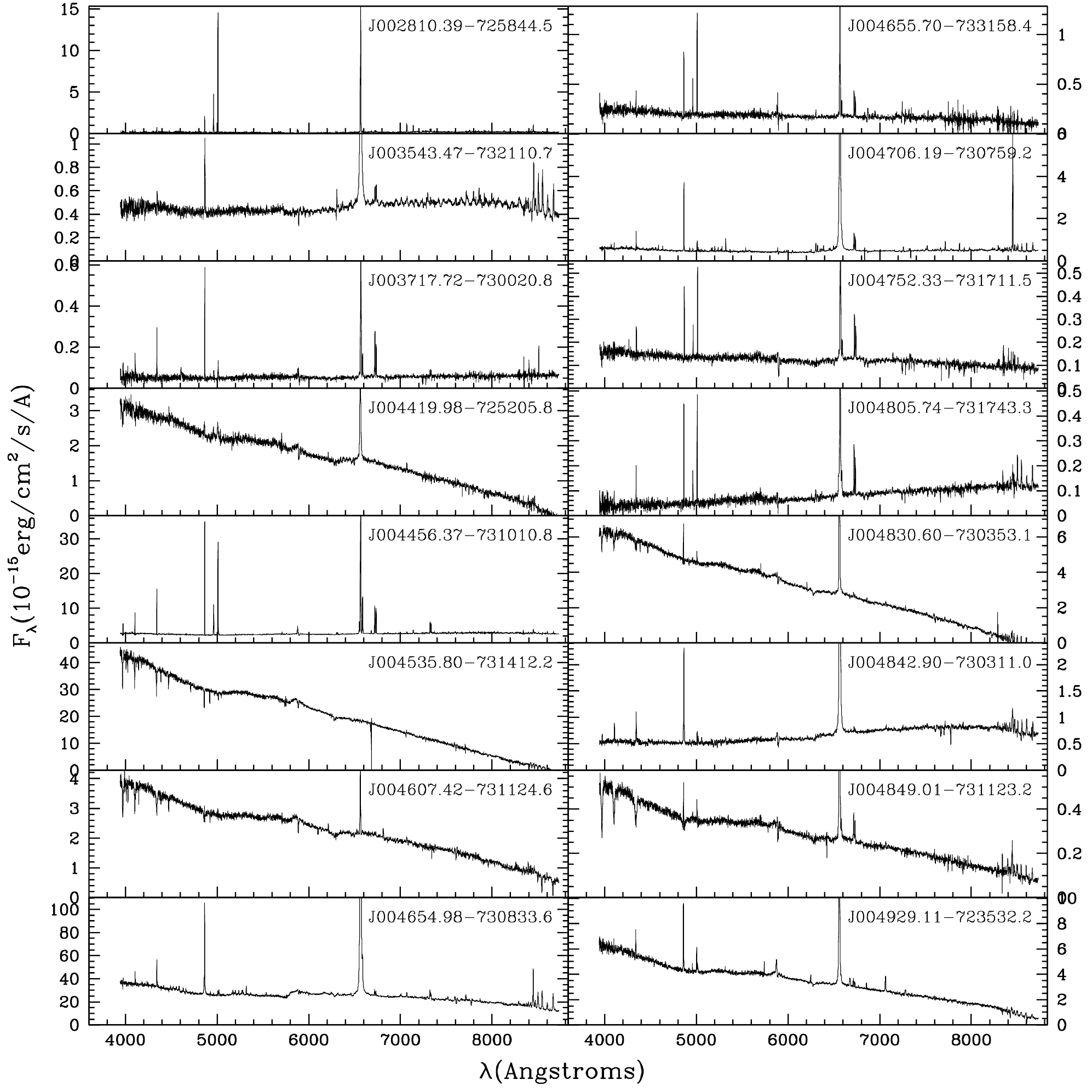}}
\caption{The low resolution AAOmega optical spectra for the 
sample of probable hot post-AGB/RGB and YSO candidates with emission features and a UV continuum. The spectra are ordered by RA.}
\label{ysospecfull}
\end{figure*}
\begin{figure*}
\ContinuedFloat
\centering
\subfloat{\includegraphics[width=18.5cm]{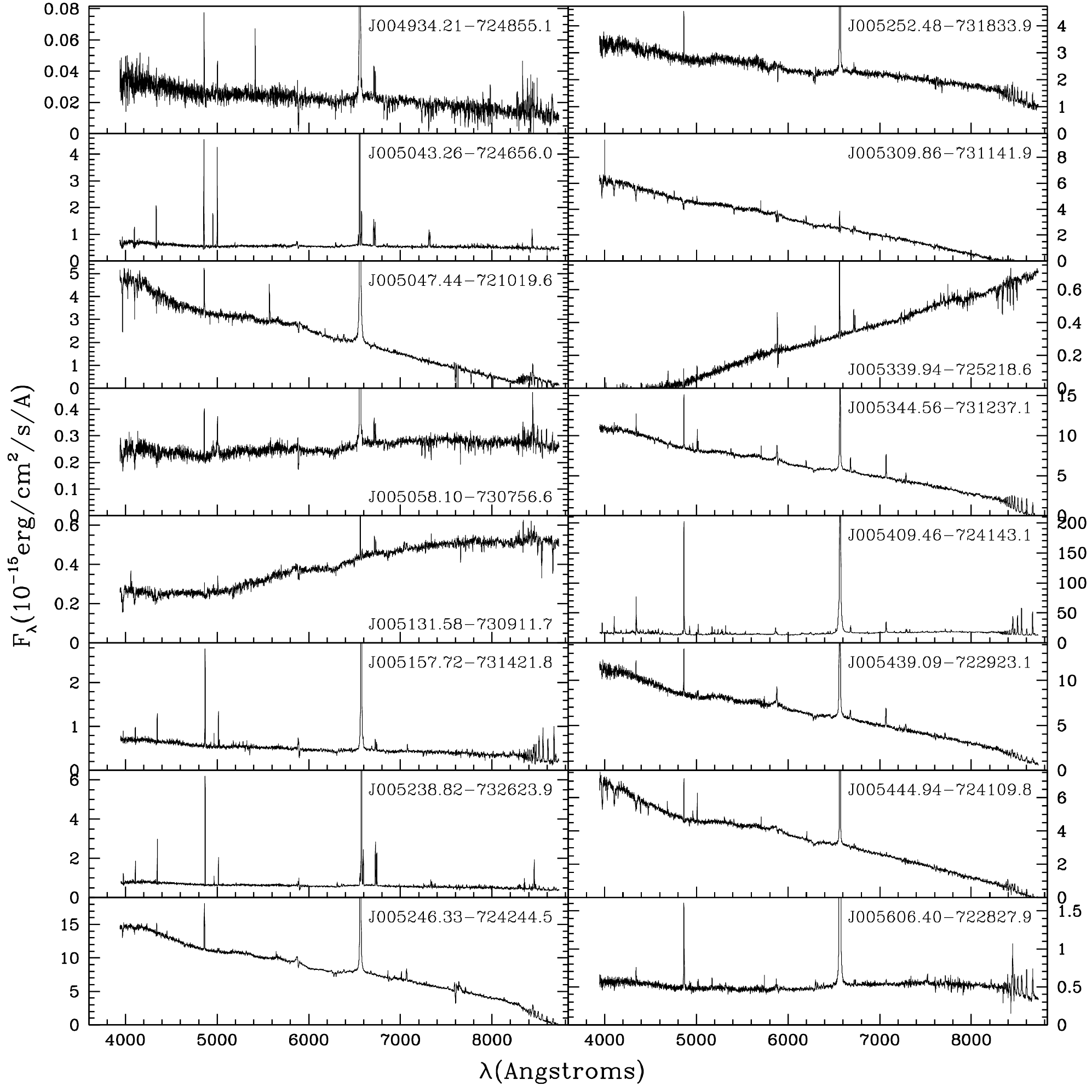}}
\caption{Figure~\ref{ysospecfull} continued.}
\end{figure*}
\begin{figure*}
\ContinuedFloat
\centering
\subfloat{\includegraphics[width=18.5cm]{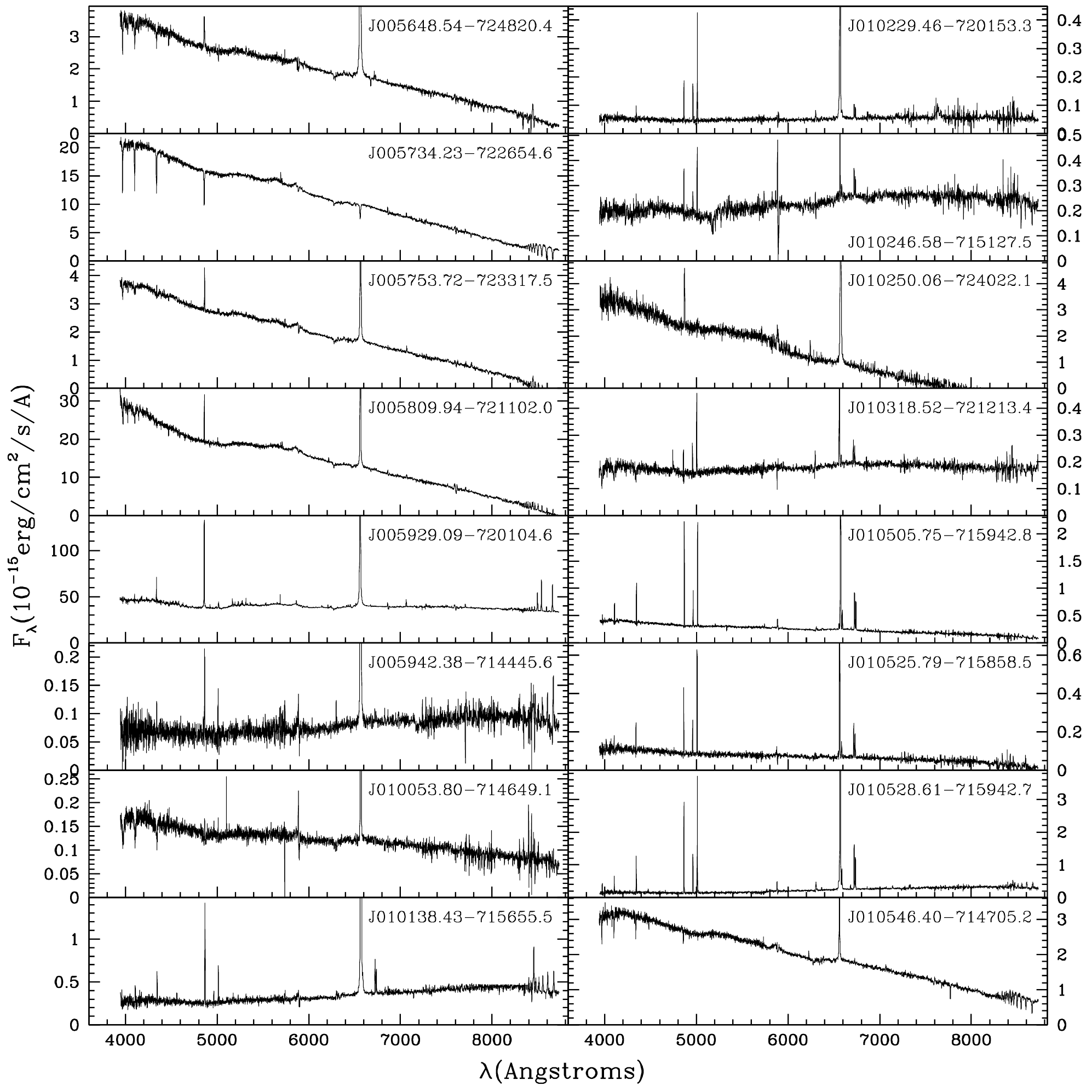}}
\caption{Figure~\ref{ysospecfull} continued.}
\end{figure*}
\begin{figure*}
\ContinuedFloat
\centering
\subfloat{\includegraphics[width=18.5cm]{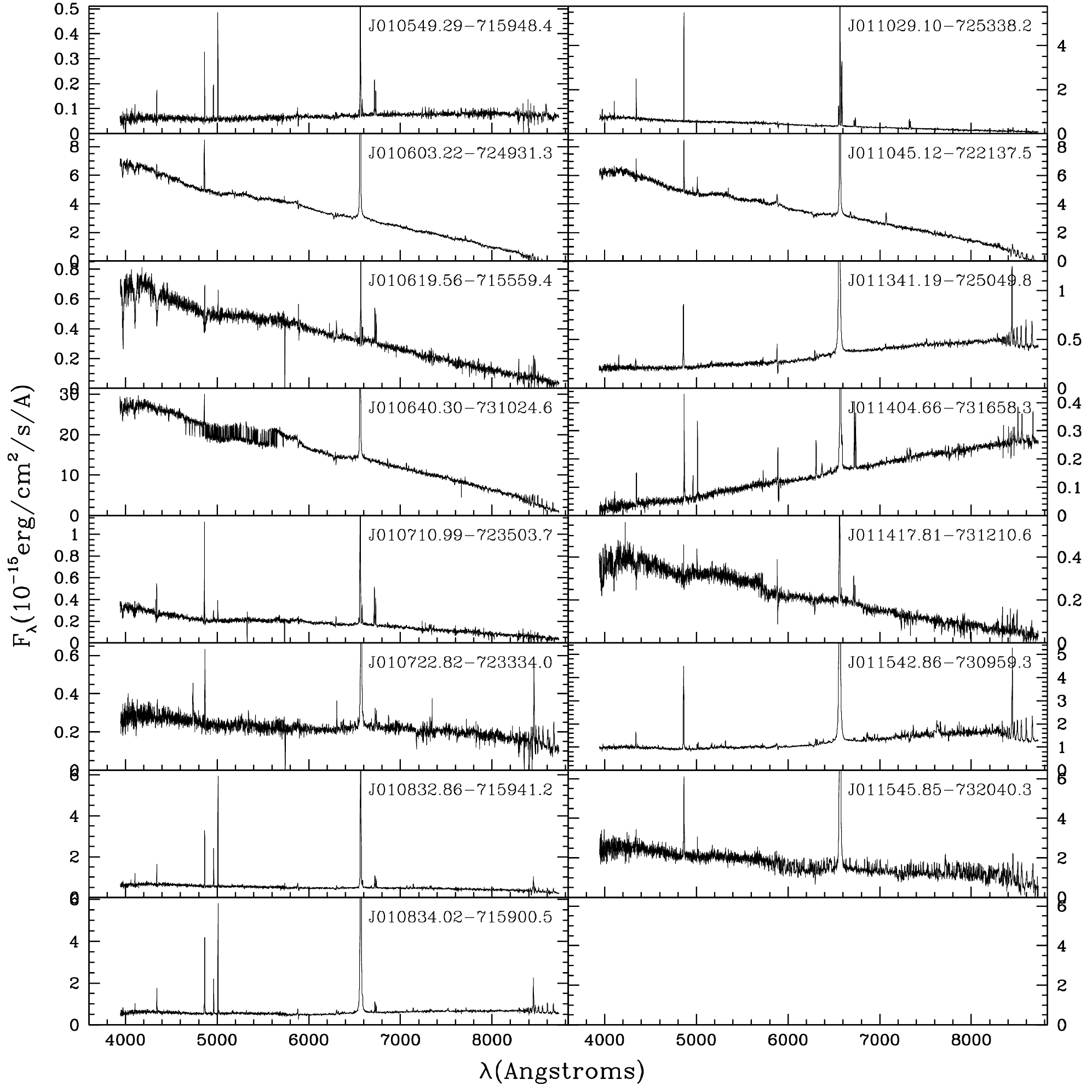}}
\caption{Figure~\ref{ysospecfull} continued.}
\end{figure*}
\clearpage

\begin{figure*}
\centering
\subfloat{\includegraphics[width=18.5cm]{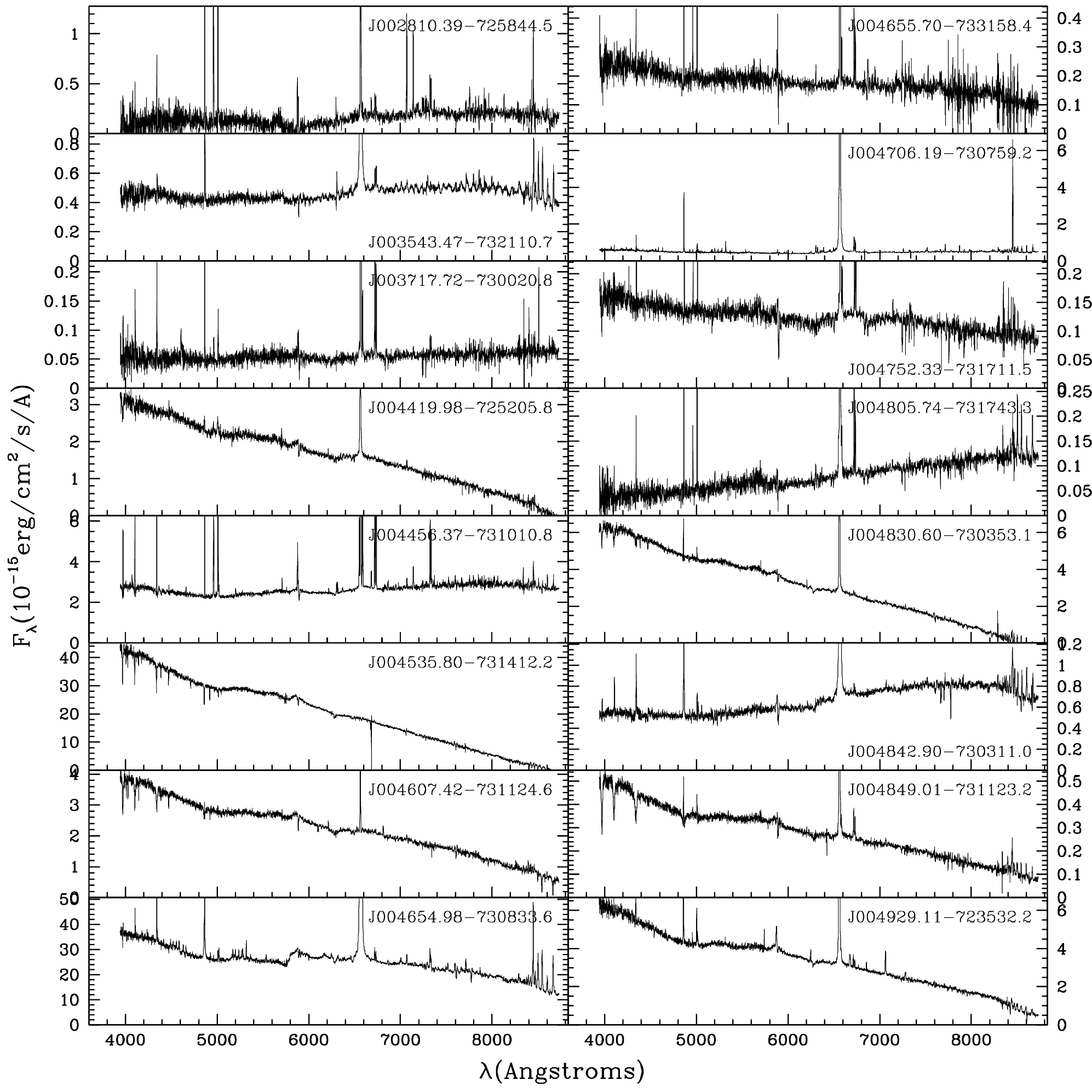}}
\caption{Same as Figure~\ref{ysospecfull} but scaled to show the continuum. 
Note that for some objects the spectra of the region $>$7000\AA\,\, is dominated by noise introduced 
during sky subtraction. The emission feature near the sodium doublet is an artefact of the 
data reduction process resulting from poor sky subtraction of the sodium doublet 
emission from the SMC, Galaxy and the night sky.}
\label{ysospecpart}
\end{figure*}
\begin{figure*}
\ContinuedFloat
\centering
\subfloat{\includegraphics[width=18.5cm]{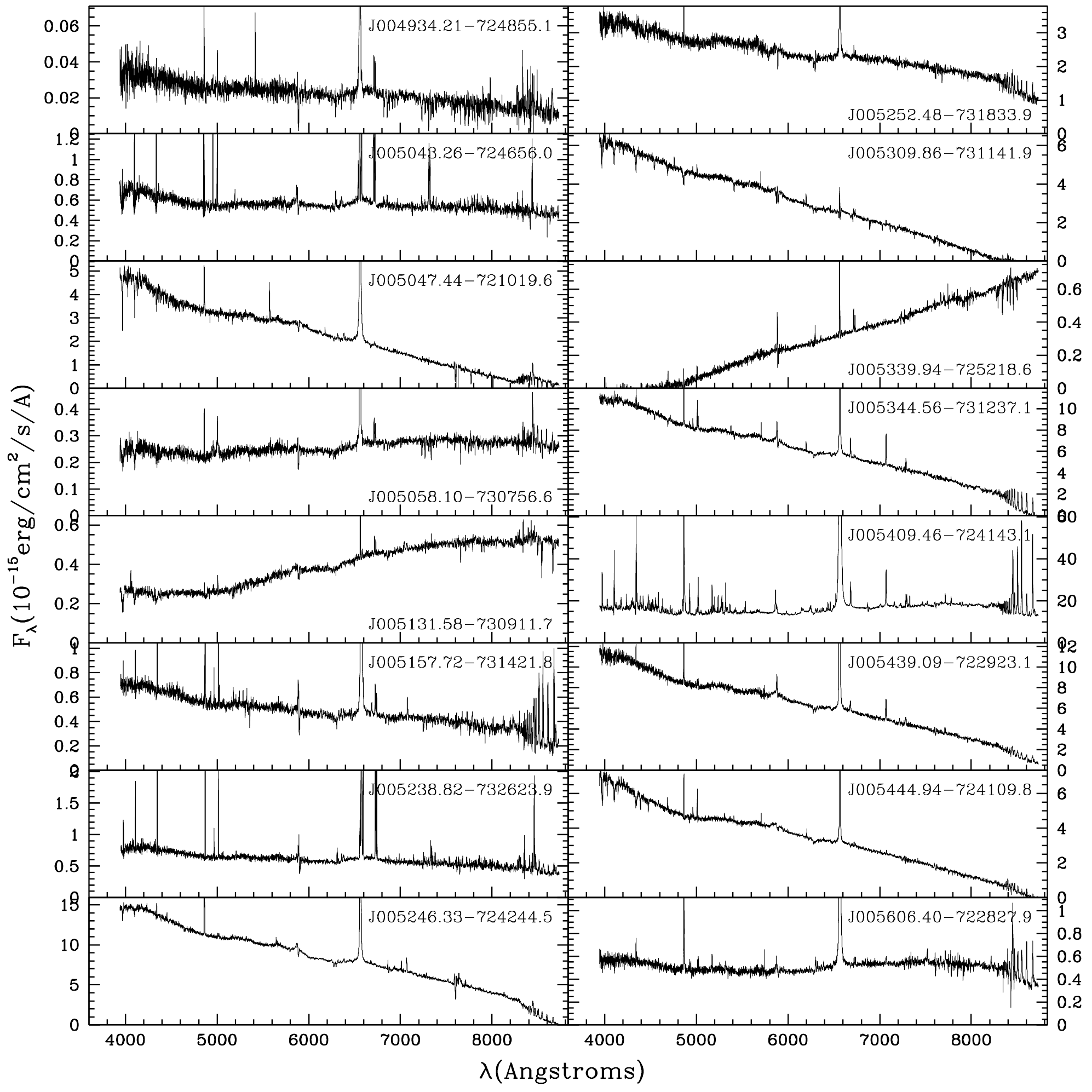}}
\caption{Figure~\ref{ysospecpart} continued.}
\end{figure*}
\begin{figure*}
\ContinuedFloat
\centering
\subfloat{\includegraphics[width=18.5cm]{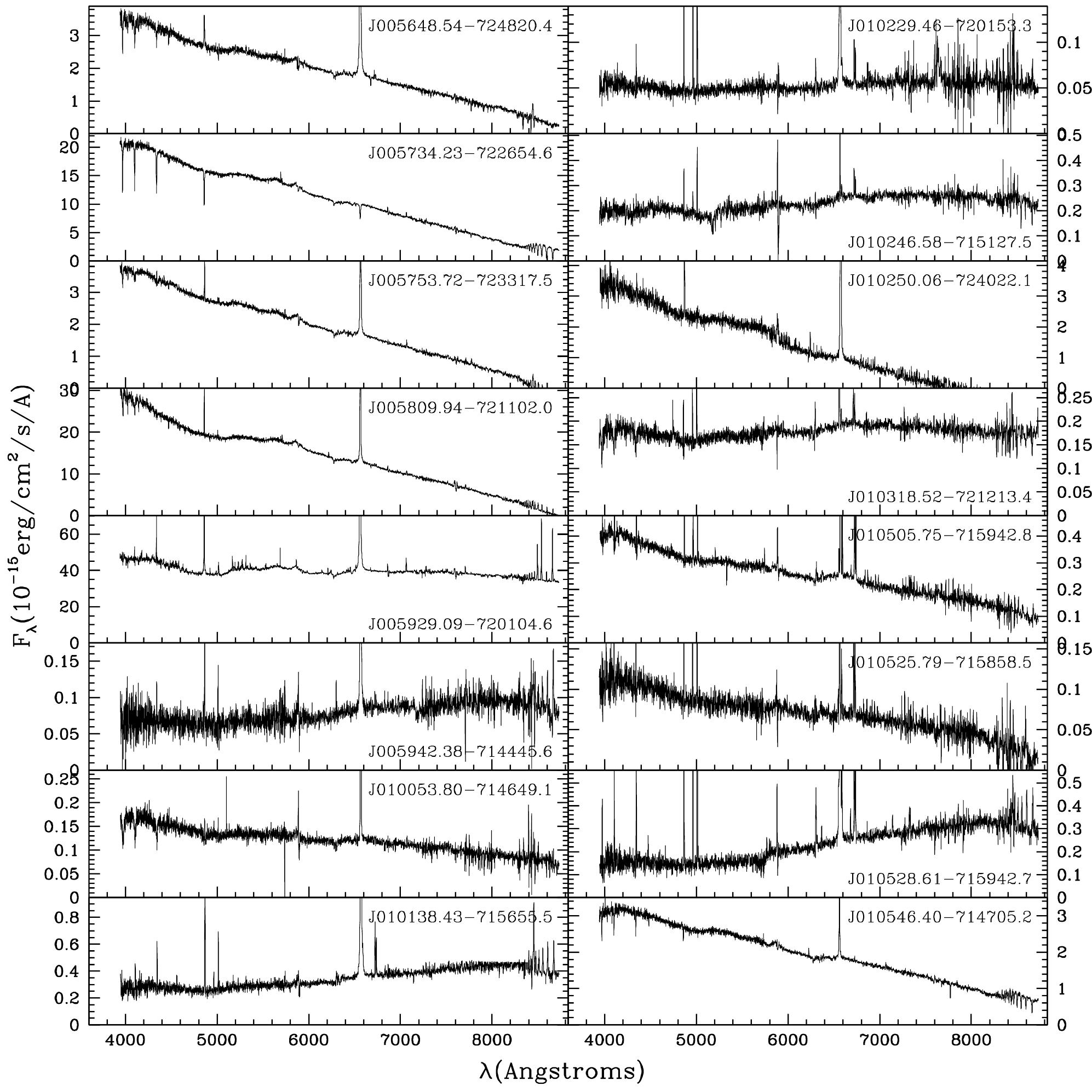}}
\caption{Figure~\ref{ysospecpart} continued.}
\end{figure*}
\begin{figure*}
\ContinuedFloat
\centering
\subfloat{\includegraphics[width=18.5cm]{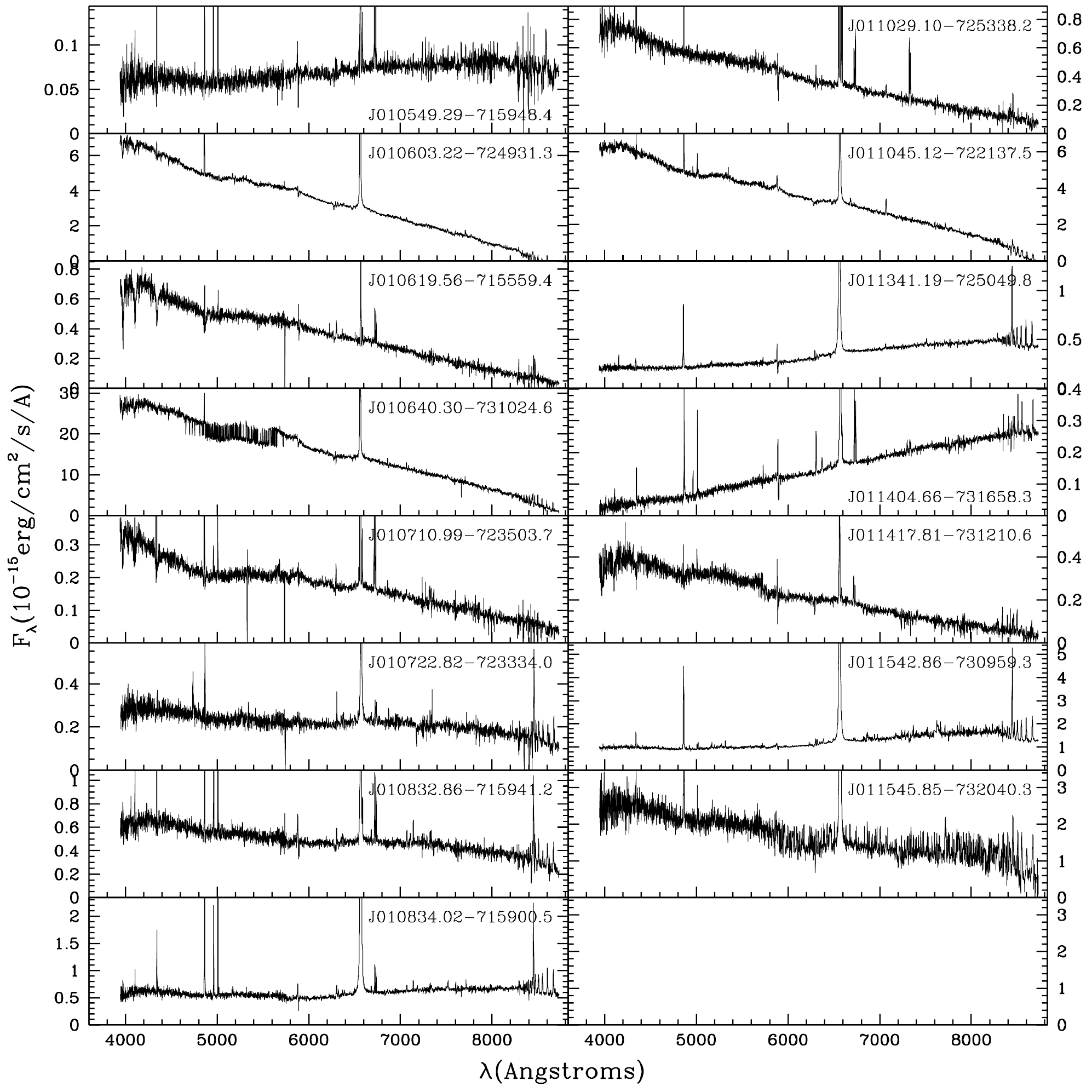}}
\caption{Figure~\ref{ysospecpart} continued.}
\end{figure*}
\clearpage

\begin{figure*}
\centering
\subfloat{\includegraphics[bb = 0 0 650 800,width=15cm]{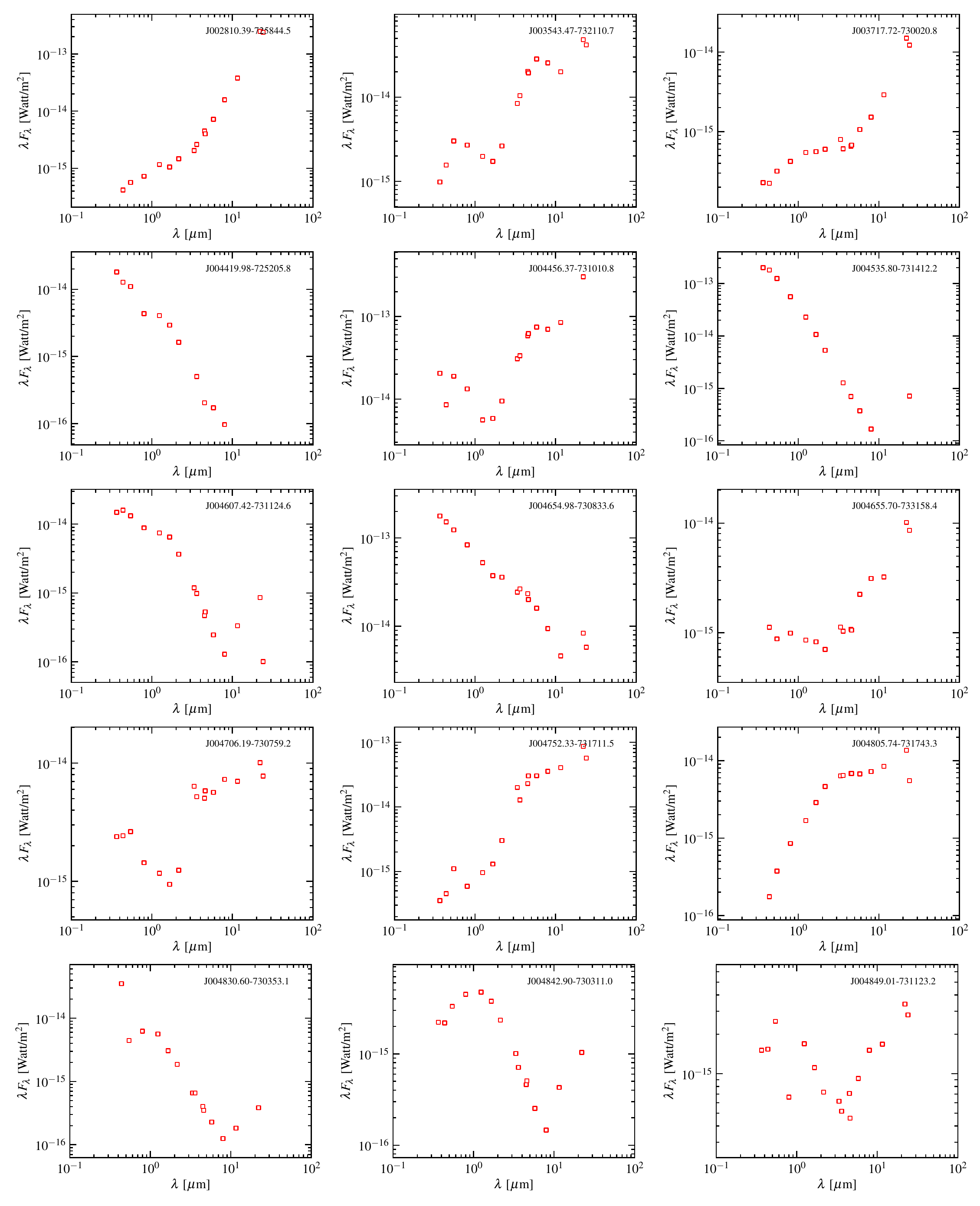}}\,
\caption{SEDs of the hot post-AGB/RGB and YSO candidates with strong emission lines and in some cases a significant UV continuum. 
The red open squares represent the original broadband 
photometry. The SED plots also show the name of the individual object. 
The SEDs are ordered by RA.}
\label{fig:yso18_sed}
\end{figure*}
\begin{figure*}
\ContinuedFloat
\centering
\subfloat{\includegraphics[bb = 0 0 650 800,width=15cm]{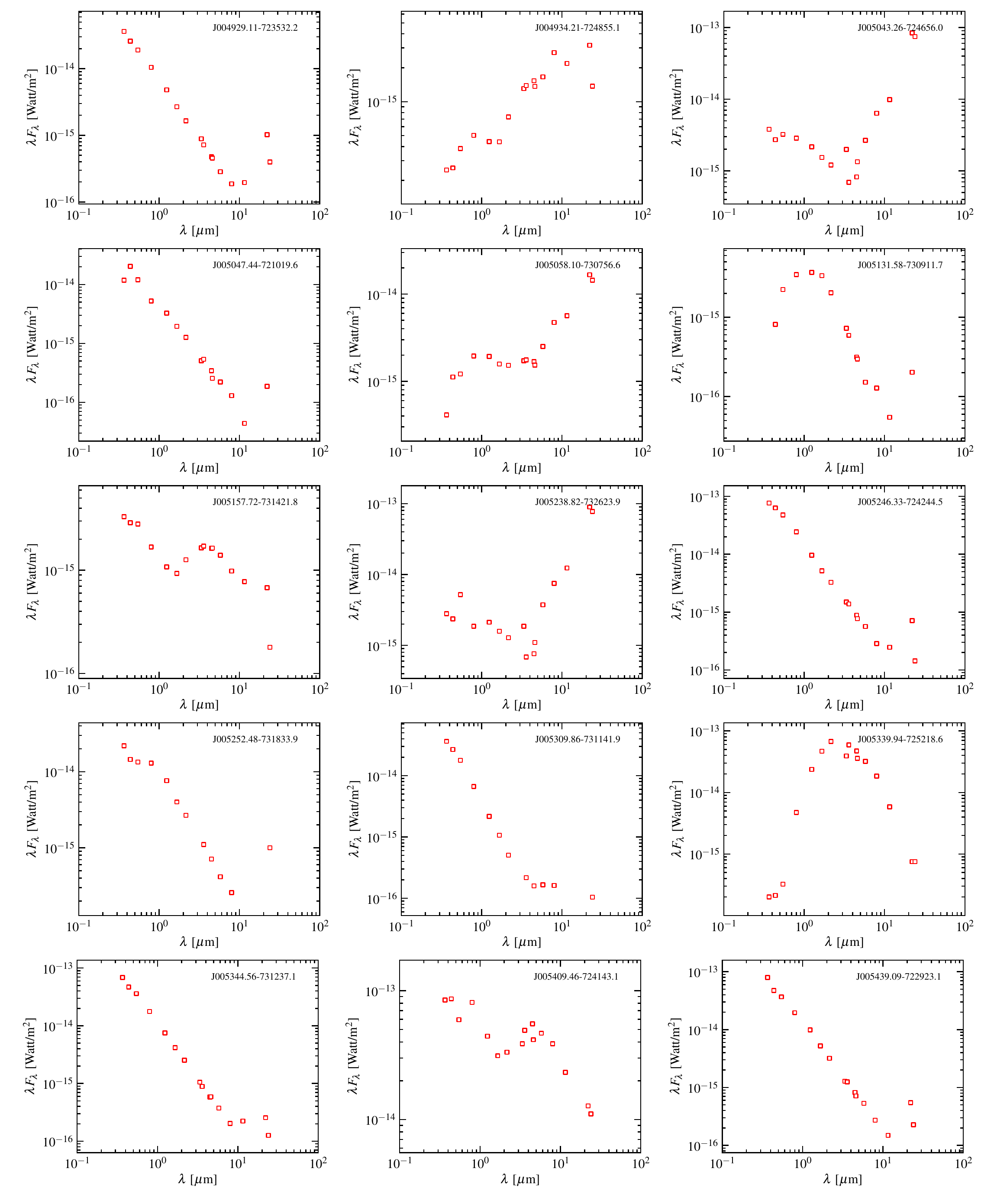}}\,
\caption{Figure~\ref{fig:yso18_sed} continued.}
\end{figure*}
\begin{figure*}
\ContinuedFloat
\centering
\subfloat{\includegraphics[bb = 0 0 650 800,width=15cm]{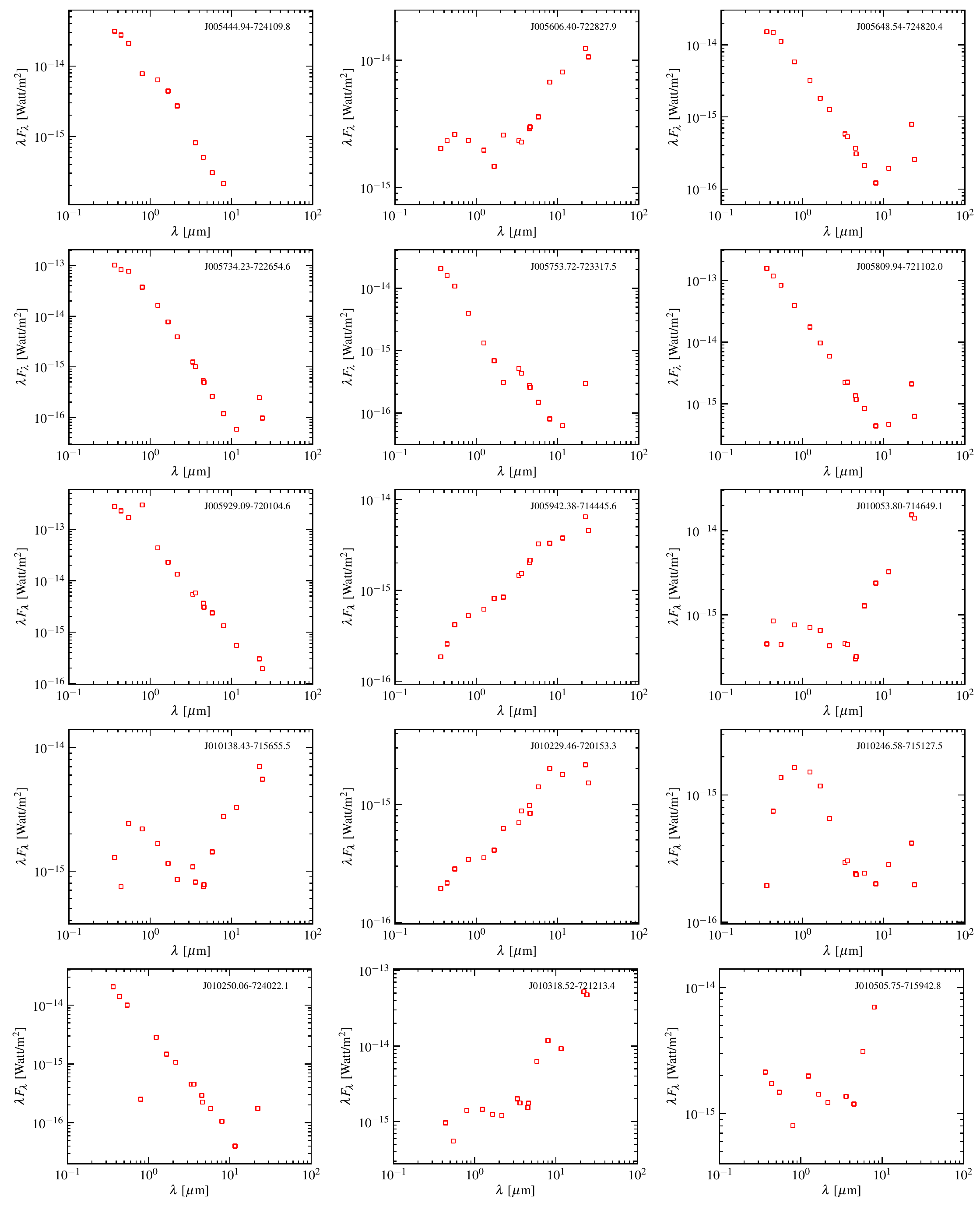}}\,
\caption{Figure~\ref{fig:yso18_sed} continued.}
\end{figure*}
\begin{figure*}
\ContinuedFloat
\centering
\subfloat{\includegraphics[bb = 0 0 650 800,width=15cm]{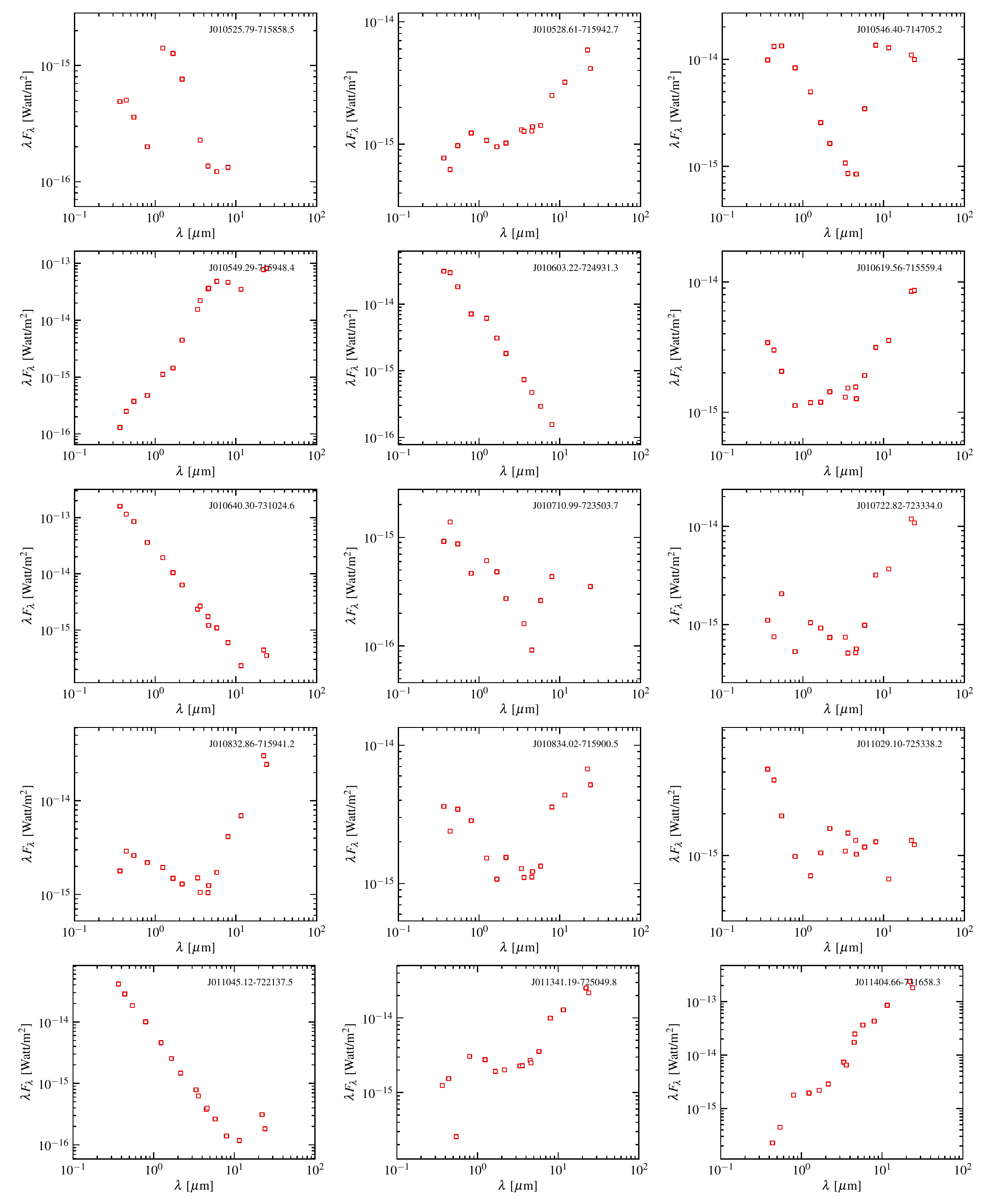}}\,
\caption{Figure~\ref{fig:yso18_sed} continued.}
\end{figure*}
\begin{figure*}
\ContinuedFloat
\centering
\subfloat{\includegraphics[bb = 0 0 288 216,width=5cm]{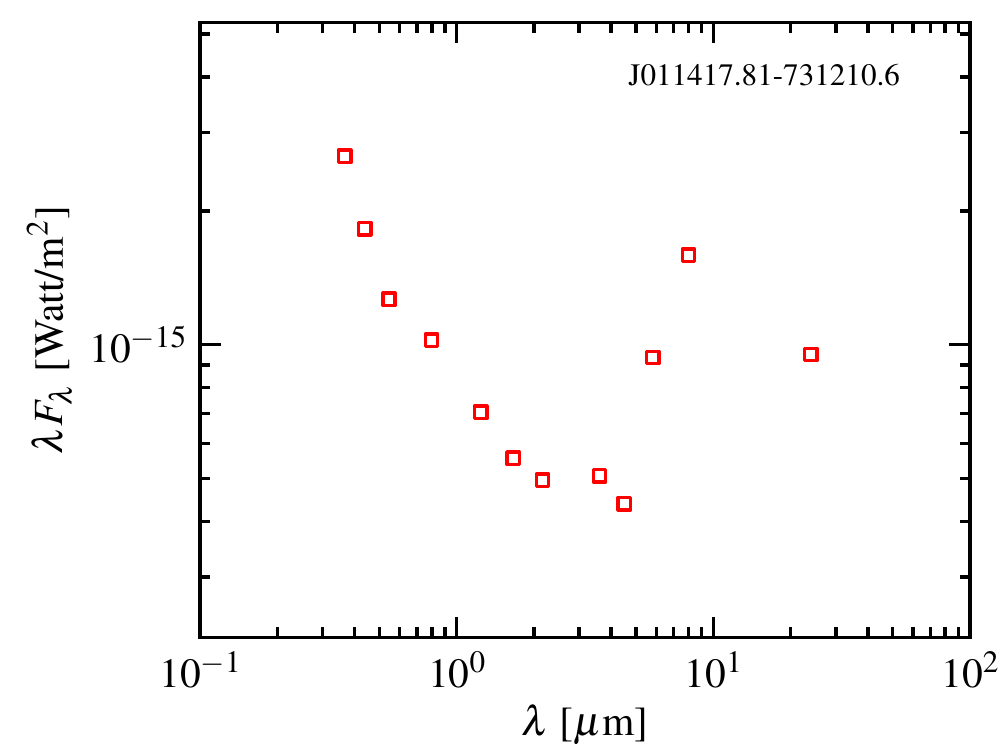}}\,
\subfloat{\includegraphics[bb = 0 0 288 216,width=5cm]{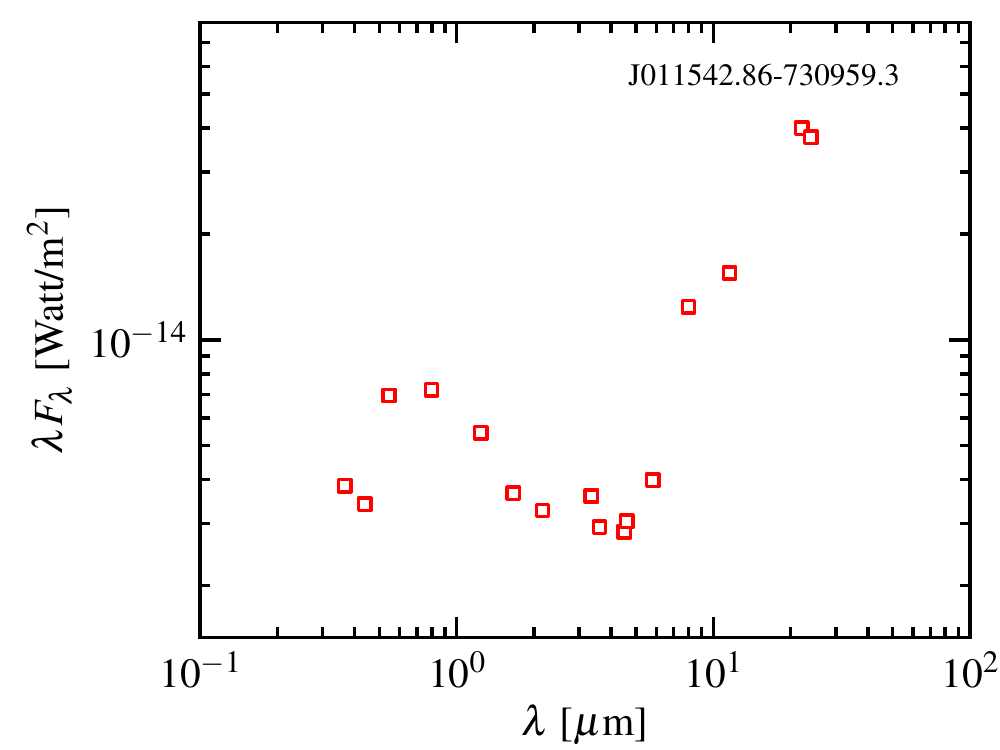}}\,
\subfloat{\includegraphics[bb = 0 0 288 216,width=5cm]{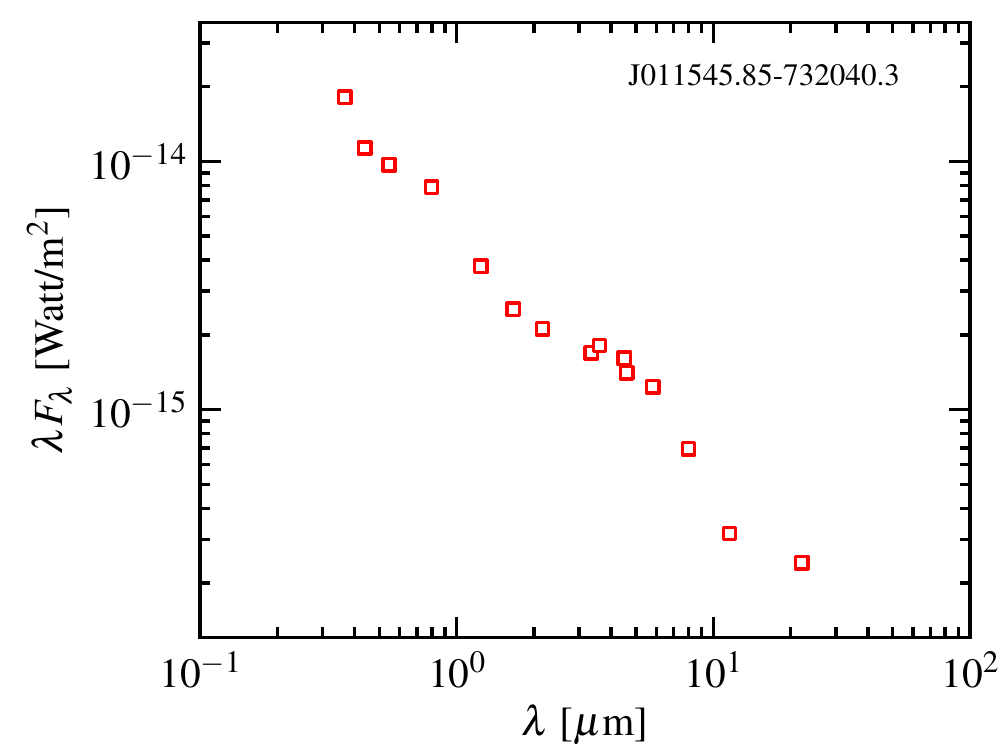}}\,
\caption{Figure~\ref{fig:yso18_sed} continued.}
\end{figure*}

\clearpage

\section[]{Tables of the PN, C-Stars, and M-Stars in our Sample}
\label{PN_M_C}

\begin{table*}
\scriptsize{
\caption{The M-stars in our sample.}
\begin{tabular}{cccc}
\hline
Name & Name & Name & Name \\ 
\hline
J003152.34-721801.0 & J003953.87-725124.2 & J004121.40-725015.1 & J004345.89-722337.5\\
J004348.97-733110.8 & J004451.17-715557.0 & J004535.62-721036.7 & J004546.33-733104.2\\
J004709.37-721050.5 & J004715.12-723338.8 & J005219.88-721046.2 & J005606.25-722452.4\\
J005754.95-731646.9 & J005841.03-742459.6 & J010411.22-731825.6 & J010808.85-735943.9\\
J010817.42-722024.7 & J010902.22-712410.2 & J011321.97-724520.1 & J011516.49-731143.4\\
\hline
\label{tab:M}
\end{tabular}}
\end{table*}

\begin{table*}
\scriptsize{
\caption{The C-stars in our sample}
\begin{tabular}{cccc}
\hline
Name & Name & Name & Name \\ 
J003552.00-725834.2 & J004702.94-740032.3 & J005442.40-742746.1 & J010146.23-731125.3  \\                  
J003633.21-735040.9 & J004705.26-723430.4 & J005448.70-720153.5 & J010149.96-734032.0  \\                  
J003635.58-733552.2 & J004713.71-732153.2 & J005449.07-723430.5 & J010225.78-735034.2  \\                  
J003647.96-731830.7 & J004714.54-732327.2 & J005454.11-725117.0 & J010233.40-735754.7  \\                  
J003744.45-741223.8 & J004718.08-723704.4 & J005459.64-733958.6 & J010241.01-723900.0  \\                  
J003755.00-721852.0 & J004744.15-731643.9 & J005505.65-711431.3 & J010252.47-725418.9  \\                  
J003819.73-735811.4 & J004816.89-724859.3 & J005515.82-725749.1 & J010308.49-731717.9  \\                  
J003830.89-733113.9 & J004822.93-734104.7 & J005526.57-724514.2 & J010315.06-732307.9  \\                  
J003905.60-724715.8 & J004859.45-733538.6 & J005530.74-732754.0 & J010337.61-714939.1  \\                  
J003913.22-735705.9 & J004903.79-730519.6 & J005544.42-725440.7 & J010404.96-715325.1  \\                  
J004010.18-730039.6 & J004931.46-730715.5 & J005559.90-723025.4 & J010419.07-734014.4  \\                  
J004024.50-742424.8 & J005016.58-732517.6 & J005606.14-720431.8 & J010442.50-720154.8  \\                  
J004035.42-741120.1 & J005023.29-740735.4 & J005617.49-722704.2 & J010449.69-723706.1  \\                  
J004100.00-722528.1 & J005031.25-722913.0 & J005635.44-713233.2 & J010525.11-743213.4  \\                  
J004152.00-730827.4 & J005044.35-723739.0 & J005645.11-712623.2 & J010532.11-720141.4  \\
J004156.75-730728.9 & J005101.97-725925.3 & J005705.78-741316.4* & J010613.22-742420.4  \\                  
J004210.62-735003.5 & J005108.11-731342.4 & J005710.92-723059.9 & J010617.23-722424.5  \\                  
J004216.95-733850.5 & J005109.23-731933.2 & J005720.49-731245.9 & J010653.02-734600.1  \\                  
J004221.72-722405.6 & J005113.84-721714.4 & J005732.74-722005.1 & J010716.65-731724.8  \\                  
J004239.07-724642.8 & J005116.18-713855.1 & J005742.47-723718.5 & J010720.13-724359.1  \\                  
J004248.70-734522.4 & J005222.26-730944.6 & J005748.96-730521.5 & J010740.35-721537.5  \\                  
J004308.69-734825.1 & J005229.38-722557.7 & J005758.16-725620.6 & J010812.94-725243.8  \\                  
J004325.12-721851.1 & J005233.40-725409.5 & J005809.47-725023.4 & J010828.25-723643.6  \\                  
J004339.51-731457.0 & J005234.70-720826.9 & J005835.16-725935.6 & J010828.79-720807.9  \\                  
J004345.74-742215.2 & J005235.32-731749.3 & J005900.39-730021.7 & J010925.78-722031.0  \\                  
J004417.52-733441.2 & J005241.15-731220.4 & J005936.60-722716.9 & J010928.89-722821.2  \\                  
J004456.97-730555.4 & J005310.07-721154.6 & J005941.63-731018.7 & J010935.05-731459.4  \\                  
J004522.31-741620.9 & J005313.97-731517.3 & J005951.31-732712.6 & J011014.44-731415.9  \\                  
J004534.30-722809.9 & J005338.82-733318.9 & J005958.75-720300.9 & J011053.17-721446.0  \\                  
J004608.33-734725.9 & J005353.50-733323.5 & J010010.43-713613.8 & J011123.43-724318.6  \\                  
J004614.25-732137.4 & J005353.54-734402.8 & J010020.83-730648.4 & J011136.36-724726.6  \\                  
J004617.88-720006.0 & J005354.98-722631.9 & J010021.17-741958.3 & J011140.46-731539.7  \\                  
J004626.21-730136.1 & J005402.65-725330.4 & J010024.23-714926.4 & J011246.96-720758.1  \\                  
J004641.36-725125.3 & J005408.46-721420.3 & J010111.31-722827.9 & J011451.00-730127.2  \\                  
J004658.55-724407.4 & J005410.74-730303.0 & J010144.14-731742.1 & J011914.67-733240.1  \\                  
\hline
\label{tab:C}
\end{tabular}}
\begin{flushleft}
Notes: *J005705.78-741316.4 has been previously identified as a 
symbiotic C-star by \citet{belczy00} which is confirmed by our 
low-resolution spectrum showing a photosphere of a C-star with 
strong emission lines.\\
\end{flushleft}
\end{table*}

\begin{table*}
\scriptsize{
{\renewcommand{\arraystretch}{0.8}
\caption{The new and previously known PN candidates in our sample. In this table 'a' represents absorption, 
'e' represents emission, '0' indicates that the feature is not observed. 
:' indicates that there is some line blending that has taken place or there is an absorption line with an emission 
core or the line indicates signs of strong winds and therefore mass-loss. '$?$' 
represents that the nature of the spectral line is uncertain. 'p' indicates a p-cygni line profile. Note: the 
low-resolution of the spectra could possibly affect the identification of a given line.}
\medskip
\tabcolsep=2pt
\begin{tabular}{llccccccccccccccccc}
\hline
Name & Previous Identification & H$\alpha$ & H$\beta$ & H$\gamma$ & [OIII] & [OIII] & HeI & HeI & [SII] & [SII] & [NII] & [NII] & CaII & CaII & CaII & Li & Ba & Pa \\ 
Wavelength (\AA) &  & 6563 & 4861 & 4341 & 4659 & 5007 & 4471 & 5876 & 6717 & 6731 & 6548 & 6584 & 8498 & 8542 & 8662 & 6708 & 4554 & - \\ 
\hline
\multicolumn{19}{c}{Previously uncatalogued planetary nebulae}\\ 
\hline
J004538.33-730438.0 &- & e & e & e & 0 & e & 0 & e & e & e & e & e & 0 & 0 & 0 & 0 & 0 & 0 \\ 
J004641.35-730613.1 & FIR$^{\rm 1}$ & e & e & e & 0 & e & 0 & e & e & e & e & e & 0 & 0 & 0 & 0 & 0 & e \\ 
J004825.75-730556.4 & YSO$^{\rm 3}$,Em$^{\rm 3,4}$ & e & e & e & 0 & e & 0 & e & e & e & e & e & 0 & 0 & 0 & 0 & 0 & e \\ 
J004836.45-725800.6 & FIR$^{\rm 1}$, I25$^{\rm 5}$, I60$^{\rm 6}$ & e & e & e & e & e & e & e & e & e & e & e & 0 & 0 & 0 & 0 & 0 & e \\ 
J005027.18-725254.6 & Em$^{\rm 7}$ & e & e & e & 0 & e & e & e & e & e & e & e & 0 & 0 & 0 & 0 & 0 & e? \\ 
J005058.49-735141.3 & I60$^{\rm 6}$,FIR$^{\rm 1}$ & e? & e? & e? & 0 & e? & e? & e & e & e & e & e & 0 & 0 & 0 & 0 & 0 & 0 \\ 
J005422.23-724329.7 & x-AGB$^{\rm 1}$ & e & e & e & 0 & e & 0 & e & e & e & e & e & 0 & 0 & 0 & 0 & 0 & 0 \\ 
J005610.87-721851.4 & -& e & e & e & 0 & e & 0 & e & e & e & e & e & 0 & 0 & 0 & 0 & 0 & 0 \\ 
J005617.17-721756.1 & - & e & e & e & 0 & e & e & e & e & e & e & e & 0 & 0 & 0 & 0 & 0 & e \\ 
J005706.16-733349.7 & - & e & e & e & 0 & e & e? & e & e & e & e & e & 0 & 0 & 0 & 0 & 0 & e? \\ 
J005856.88-720954.3 & YSOp$^{\rm 8}$ & e & e & e & 0 & e & e & e & e & e & 0 & 0 & e & e & e & 0 & 0 & 0 \\ 
J005905.38-721035.5 & FIR$^{\rm 1}$,I25$^{\rm 5}$ & e & e & e & e? & e & e & e & e & e & e & e &0 &0 & 0 & 0 & 0 & e \\ 
J005905.81-721127.1 & x-AGB$^{\rm 1}$,Em*$^{\rm 7}$ & e & e & e & e & e & e & e & e & e & e & e & 0 & 0 & 0 & 0 & 0 & e \\ 
J005912.20-720958.4 & RGB$^{\rm 1}$,YSOd?$^{\rm 8}$ & e & e & e & 0 & e & e & e & e & e & e & e & 0 & 0 & 0 & 0 & 0 & e \\ 
J010115.32-721637.4 & - & e & e & e & e & e & e & e & e & e & e & e & 0 & 0 & 0 & 0 & 0 & 0 \\ 
J010155.63-722948.0 & - & e & e & e & 0 & e & e? & e & e & e & e? & e? & 0 & 0 & 0 & 0 & 0 & e:? \\ 
J010248.21-720615.9 & - & e & e & e & 0 & e & e? & e & e & e & e & e & 0 & 0 & 0 & 0 & 0 & e \\ 
J010258.73-720347.5 & - & e & e & e & 0 & e & e & e & e & e & e & e & 0 & 0 & 0 & 0 & 0 & e? \\ 
J010307.48-720218.1 & - & e & e & e & 0 & e & e & e & e & e & e & e & 0 & 0 & 0 & 0 & 0 & 0 \\ 
J010322.32-720411.1$*$ & - & e & e & e & 0 & e & e & e & e & e & e & e & a & a & a & 0 & 0 & 0 \\ 
J010336.26-720404.2$*$ & - & e & e & e & 0 & e & e? & e & e & e & e? & e? & a? & a? & a? & 0 & 0 & 0 \\ 
J010405.71-720700.3$*$ & - & e & e & e & 0 & e & 0 & e & e & e & e & e & a? & a? & a? & 0 & 0 & 0 \\ 
J010442.52-721007.4$*$ & - & e & e & e & e & 0 & e & 0 & e & e & e? & e & a? & a? & a? & 0 & 0 & 0 \\ 
J010455.13-720055.0 & - & e & e & e & e & e & e & e & e & e & e & e & a? & a? & a? & 0 & 0 & 0 \\ 
J010458.98-715429.9 & - & e & e & e & e & e & e & e & e & e & e & e & 0 & 0 & 0 & 0 & 0 & 0 \\ 
J010529.27-720830.6 & - & e & e & e & 0 & e & 0 & e & e & e & e & e & 0 & 0 & 0 & 0 & 0 & 0 \\ 
J010659.66-725042.8 & RGB,FIR$^{\rm 1}$, YSO$^{\rm 2,9}$, Em$^{\rm 7}$ & e & e & e & 0 & e? & e? & 0 & e & e & e? & e? & 0 & 0 & 0 & 0 & 0 & e \\ 
J011347.56-731710.1 & - & e & e & e & e & e & e & e & e & e & e & e & 0 & 0 & 0 & 0 & 0 & e \\ 
J011358.02-731747.4 & FIR$^{\rm 1}$,Em*$^{\rm 7}$ & e & e & e& 0 & e & 0 & e & e & e & 0 & 0 & 0 & 0 & 0 & 0 & 0 & e \\ 
J011447.02-732058.8 & RSG$^{\rm 1}$ & e & e & e & e? & e & e? & e? & e & e & e? & e? & 0 & 0 & 0 & 0 & 0 & e? \\ 
\hline
\multicolumn{19}{c}{Previously identified planetary nebulae}\\ 
\hline
J003238.86-714159.5 & PN$^{\rm 10,11,12}$ &e & e & e & 0 & e & e & e & e & e & e & e & 0 & 0 & 0 & 0 & 0 & e \\ 
J003421.94-731321.5 & PN$^{\rm 10,12}$ & e & e & e & e & e & e & e & e & e & e & e & 0 & 0 & 0 & 0 & 0 & e \\ 
J004121.62-724516.4 & PN$^{\rm 7,10,12}$ & e & e & e & e & e & e & e & e & e & e & e & 0 & 0 & 0 & 0 & 0 & e \\ 
J004127.73-734706.5 & FIR$^{\rm 1}$, PN$^{\rm 7,10,12,13}$ & e & e & e & e & e & e & e & e & e & e & e & 0 & 0 & 0 & 0 & 0 & e \\ 
J004325.27-723818.4 & PN$^{\rm 7,10,12,14}$ & e & e & e & e & e & e & e & e & e & e & e & 0 & 0 & 0 & 0 & 0 & e \\ 
J004353.84-725514.1 & PN$^{\rm 7,11}$ & e & e & e & 0 & e & e & e & e & e & e & e & 0 & 0 & 0 & 0 & 0 & e \\ 
J004947.48-741440.0 & PN$^{\rm 11}$ & e & e & e & 0 & e & 0 & e & e & e & e? & e & 0 & 0 & 0 & 0 & 0 & e? \\ 
J005035.04-734257.9 & PN$^{\rm 7,10,14}$ & e & e & e & e & e & e & e & e & e & e & e & 0 & 0 & 0 & 0 & 0 & e \\ 
J005127.12-722611.6 & PN$^{\rm 3,7,10,12,14}$ & e & e & e & e & e & e & e & e & e & e & e & 0 & 0 & 0 & 0 & 0 & e \\ 
J005136.55-732016.9 & PN$^{\rm 3,7,11,12,14}$ & e & e & e & e & e & e & e & e & e & 0 & e? & 0 & 0 & 0 & 0 & 0 & e \\ 
J005142.17-725027.4$*$ & PN$^{\rm 14}$ & e & e & e & e & e & e & e & e & e & e & e & a & a & a & 0 & 0 & 0 \\ 
J005156.30-712444.3 & PN$^{\rm 7,10,12}$ & e & e & e & e & e & e & e & e & e & e & e & 0 & 0 & 0 & 0 & 0 & e \\ 
J005311.04-724507.4 & PN$^{\rm 3,7,10,12,13,14}$ & e & e & e & e & e & e & e & e & e & e & e & 0 & 0 & 0 & 0 & 0 & e \\ 
J005619.40-720658.3 & PN$^{\rm 5,7,11}$,I25$^{\rm 15}$ & e & e & e & e & e & e & e & e & e & e & e & 0 & 0 & 0 & 0 & 0 & e \\ 
J005639.30-723907.1 & PN$^{\rm 3,7,11,14}$ & e & e & 0 & 0 & e & 0 & 0 & e & e & 0 & 0 & 0 &  0 & 0 & 0 & 0 & e \\ 
J005842.84-722716.6 & PN$^{\rm 7,12,14,15}$ & e & e & e & e & e & e & e & e & e & e & e & 0 & 0 & 0 & 0 & 0 & e \\ 
\hline
\label{tab:PN}
\end{tabular}}}
\begin{flushleft}
Notes: '*' - indicates that these objects are likely to be symbiotic stars as their spectra show a 
photosphere of a cool star with a red continuum and strong emission
lines.\\
A positional cross-matching was performed with all the catalogues mentioned in Table~\ref{tab:pagb1_param}. A positional matching was found 
with the following catalogues:
$^{1}$\citet{boyer11},
$^{2}$\citet{2012MNRAS.tmp..229O}, 
$^{3}$\citet{1980ApJS...42....1J},
$^{4}$\citet{1981PASP...93..431S},
$^{5}$\citet{2003A&amp;A...401..873W} (25$\mu$m),
$^{6}$\citet{2003A&amp;A...401..873W} (60$\mu$m),
$^{7}$\citet{1993A&amp;AS..102..451M},
$^{8}$\citet{2007MNRAS.376.1270L},
$^{9}$\citet{2010AJ....139.1553V},
$^{10}$\citet{1978ApJ...221..586S},
$^{11}$\citet{1995A&amp;AS..112..445M},
$^{12}$\citet{2000MNRAS.311..741M},
$^{13}$\citet{1997A&amp;AS..125..419L},
$^{14}$\citet{2002AJ....123..269J},
$^{15}$\citet{1985MNRAS.213..491M}.
Previous identifications: YSOd - Definite YSO; YSOp- Probable YSO;  Em*,EmO - object with emission features; I25 - I 25$\mu$m source; I60 - I 60$\mu$m source; PN - Planetary nebula; 
FIR - far-IR object; RGB - red giant branch star; RSG - red supergiant, x-AGB - dusty AGB star with superwind mass loss (defined in \citet{boyer11}).\\
The positional cross matching was done with the catalogues 
mentioned in Table~\ref{tab:pagb1_param}. 
\end{flushleft}
\end{table*}

\clearpage

\section[]{The Final Sample of High Probability Post-AGB/RGB and YSO Candidates}
\label{specstuff}

Figures~\ref{fig:pagb1_sed}\,$-$~\ref{fig:yso2_sed} show the SEDs of the objects 
before and after de-reddening. Figures~\ref{pagb1spec}\,$-$\,~\ref{yso2spec} show the optical spectra of the sample of 
Q1 and Q2 post-AGB/RGB and YSO candidates.  We have summarised 
some of the most prominent features in Tables~\ref{tab:pagb1lines}\,$-$\,~\ref{tab:yso2lines}. 

\begin{figure*}
\centering
\subfloat{\includegraphics[bb = 0 0 661 801,width=15cm]{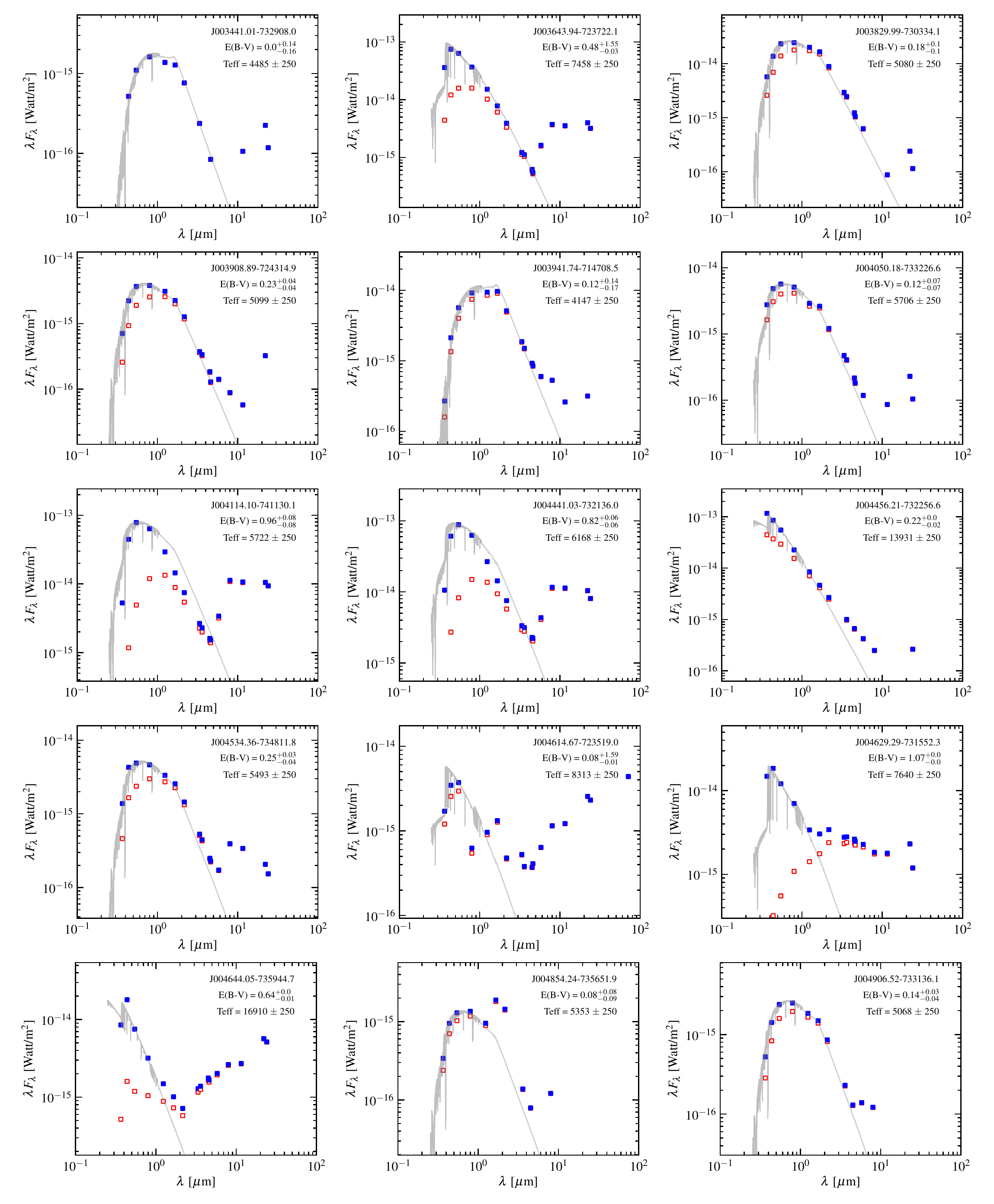}}\,
\caption{SEDs of Q1 post-AGB/RGB candidates. The red open squares represent the original broadband 
photometry. The blue filled squares represent the dereddened broadband 
photometry. Up to a wavelegth of 10500\AA, we over-plot the Munari synthetic spectrum which is estimated to have the 
best-fit to the observed spectra. The SED distribution in the IR is 
represented with the corresponding Kurucz atmospheric model take from 
\citet{castelli03}. The SED plots also show the name of the individual object, the estimated E(B-V) 
value with error bars (see 
Section~\ref{reddening}) and the estimated \teff\, value (see Section~\ref{STP}).}
\label{fig:pagb1_sed}
%
\end{figure*}
\begin{figure*}
\ContinuedFloat
\centering
\subfloat{\includegraphics[bb = 0 0 652 804,width=15cm]{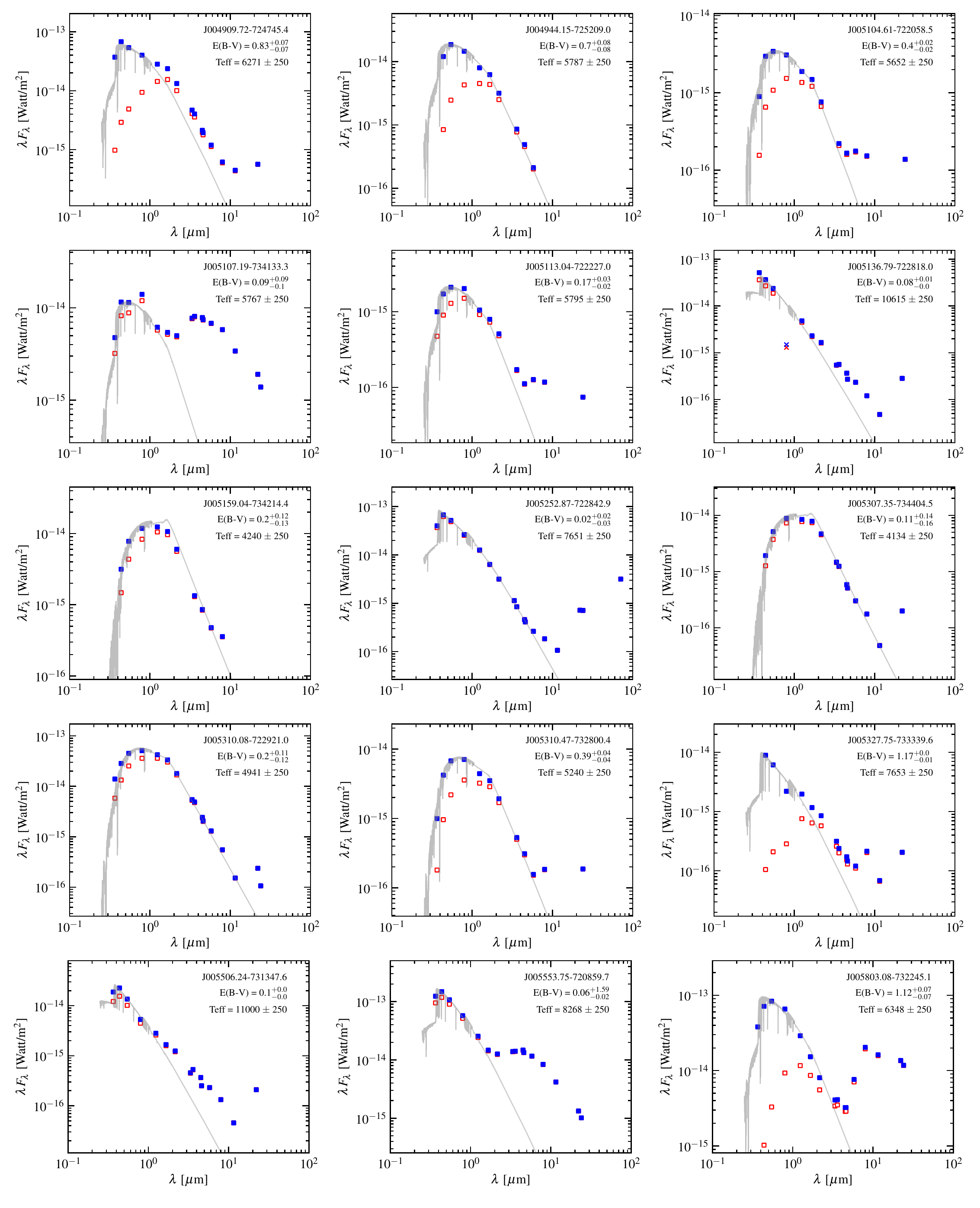}}\,
\caption{Figure~\ref{fig:pagb1_sed} continued.}
\end{figure*}
\begin{figure*}
\ContinuedFloat
\centering
\subfloat{\includegraphics[bb = 0 0 641 477,width=15cm]{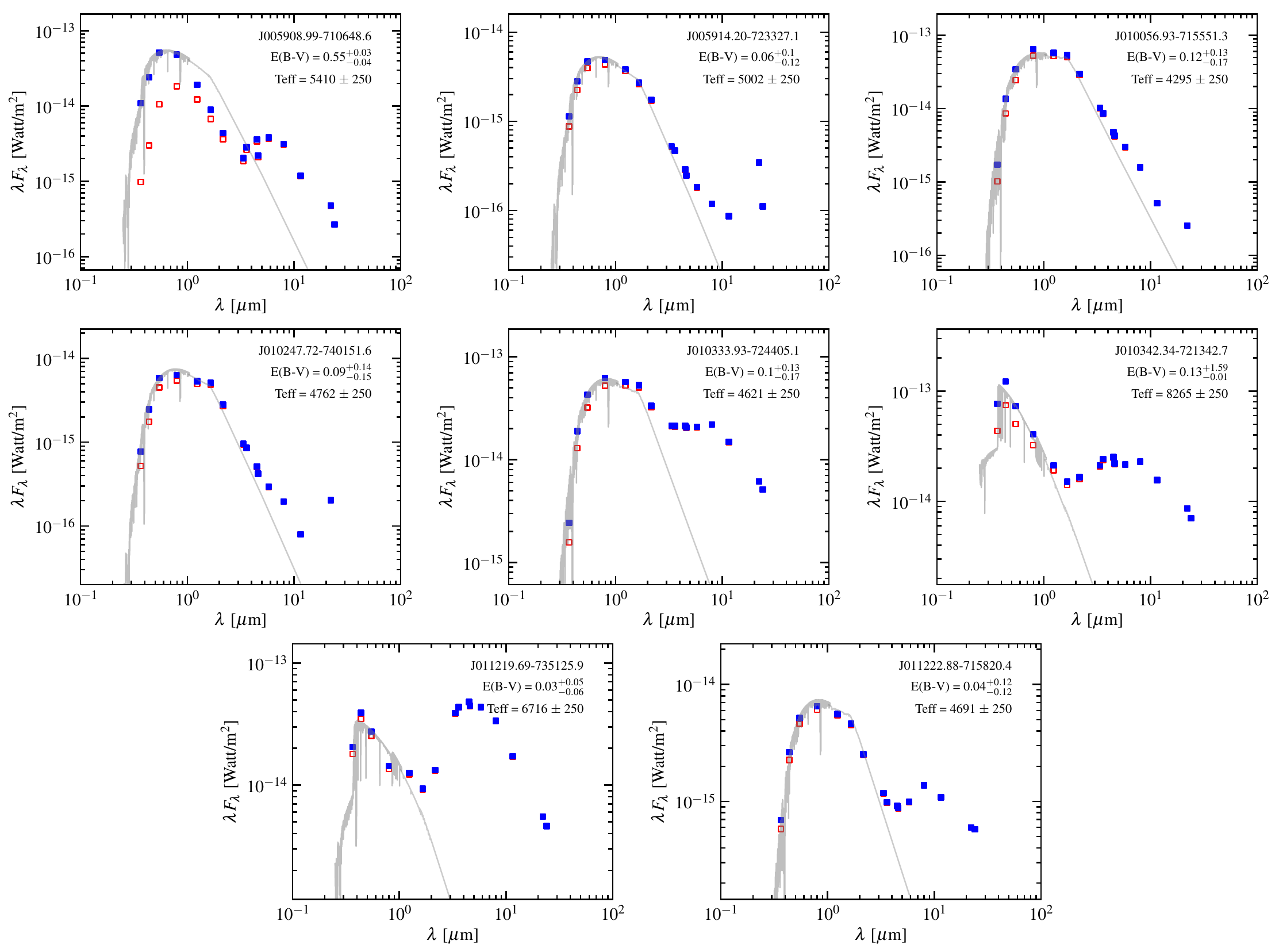}}\,
\caption{Figure~\ref{fig:pagb1_sed} continued.}
\end{figure*}

\begin{figure*}
\centering
\subfloat{\includegraphics[bb = 0 0 650 790,width=15cm]{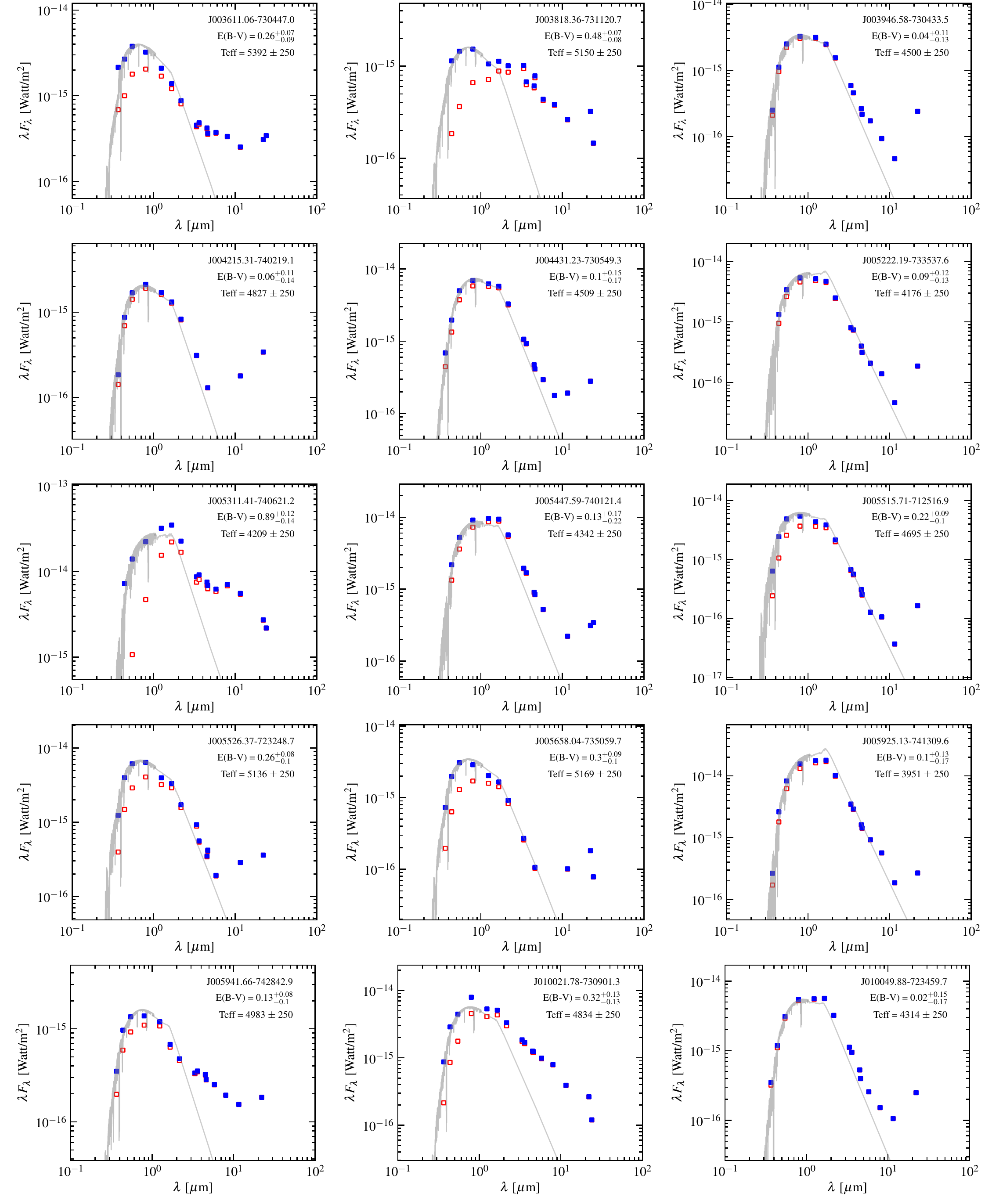}}\,
\caption{Same as Figure ~\ref{fig:pagb1_sed}, but for the 
Q2 post-AGB/RGB candidates.}
\label{fig:pagb2_sed}
\end{figure*}
\begin{figure*}
\ContinuedFloat
\centering
\subfloat{\includegraphics[bb = 0 0 643 640,width=15cm]{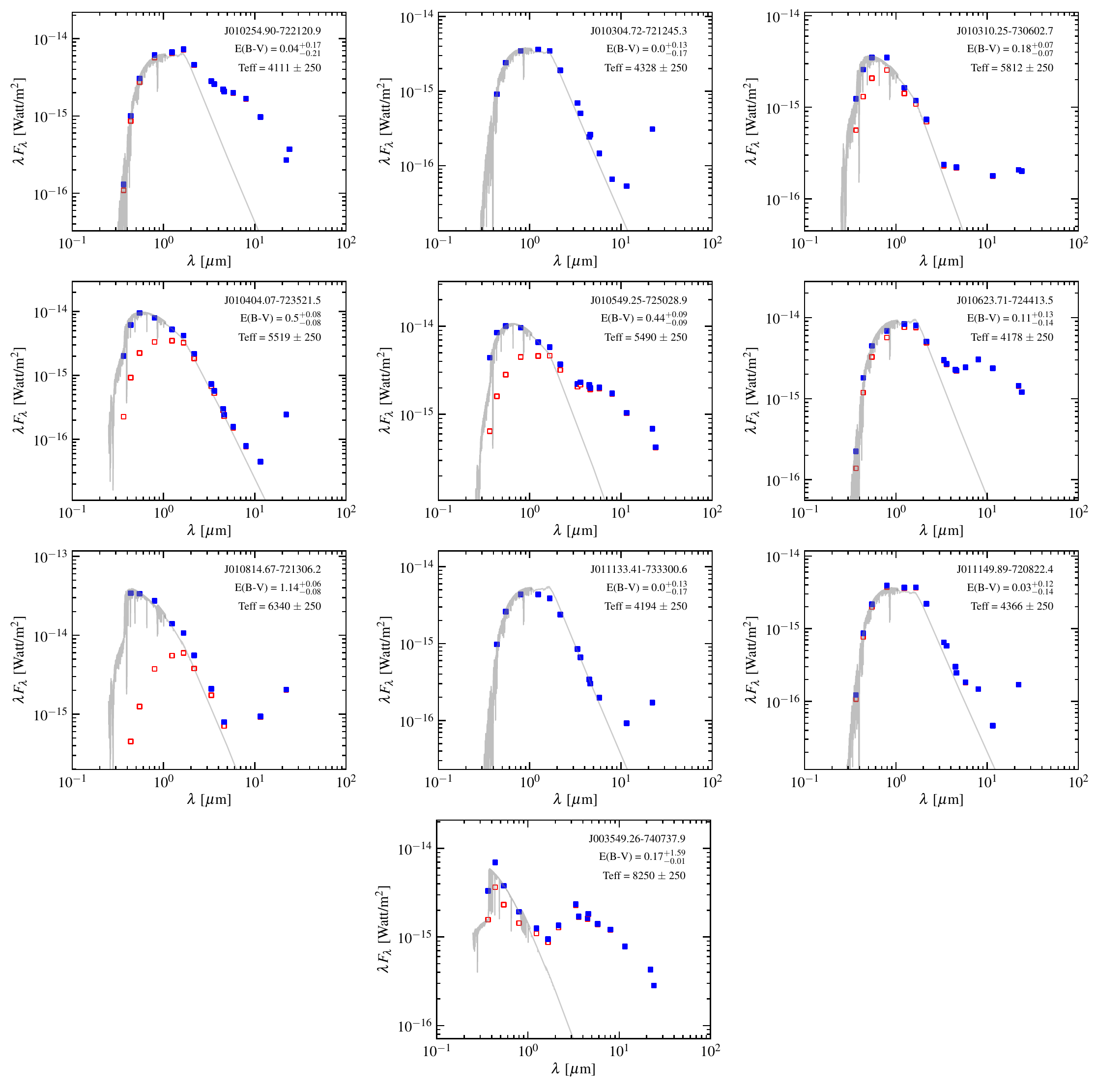}}\,
\caption{Figure ~\ref{fig:pagb2_sed} continued.}
\end{figure*}

\begin{figure*}
\centering
\subfloat{\includegraphics[bb = 0 0 648 795,width=15cm]{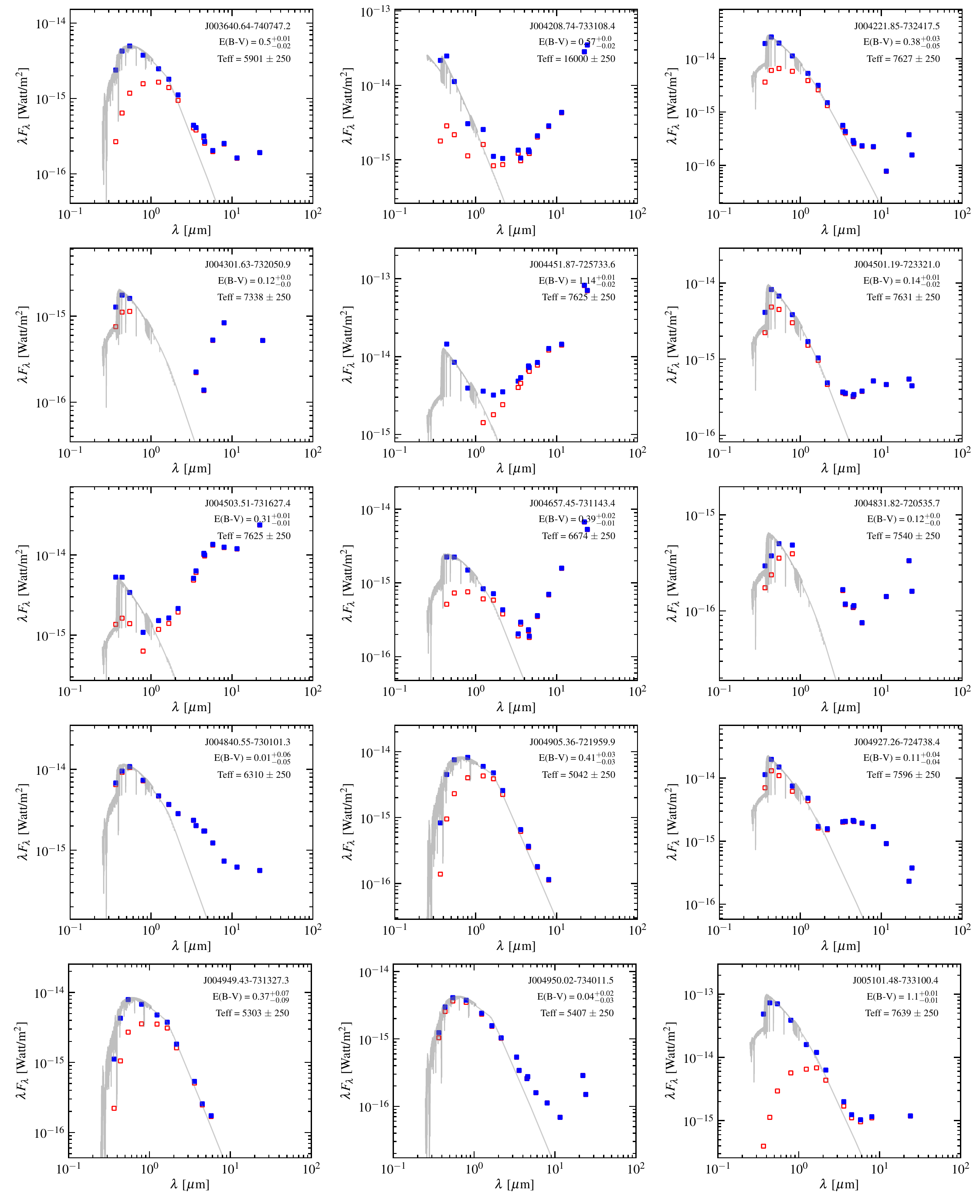}}\,
\caption{Same as Figure ~\ref{fig:pagb1_sed}, but for the Q1 
YSO candidates}
\label{fig:yso1_sed}
\end{figure*}
\begin{figure*}
\ContinuedFloat
\centering
\subfloat{\includegraphics[bb = 0 0 652 637,width=15cm]{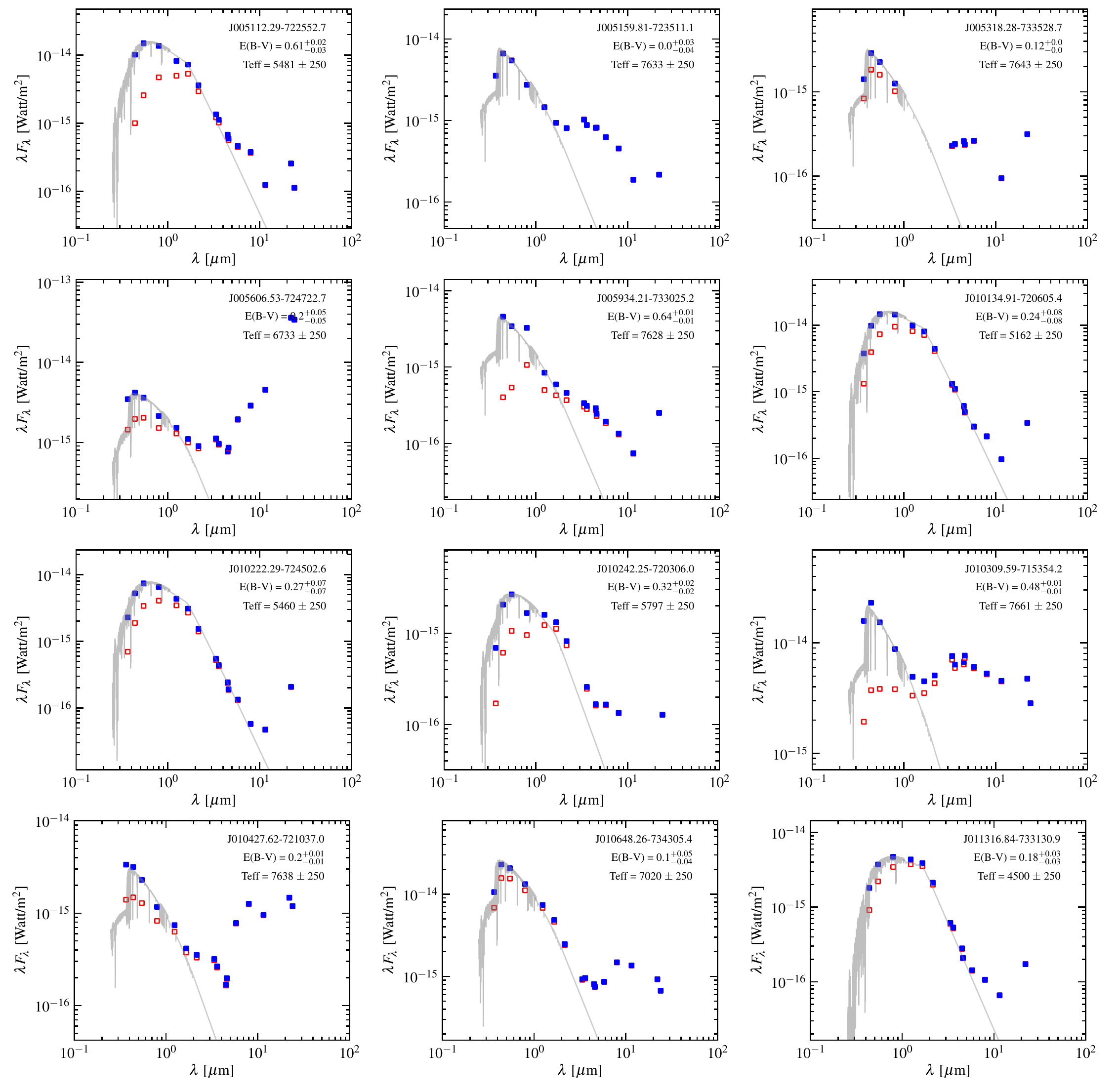}}\,
\caption{Figure~\ref{fig:yso1_sed} continued.}
\end{figure*}

\begin{figure*}
\centering
\subfloat{\includegraphics[bb = 0 0 647 790,width=15cm]{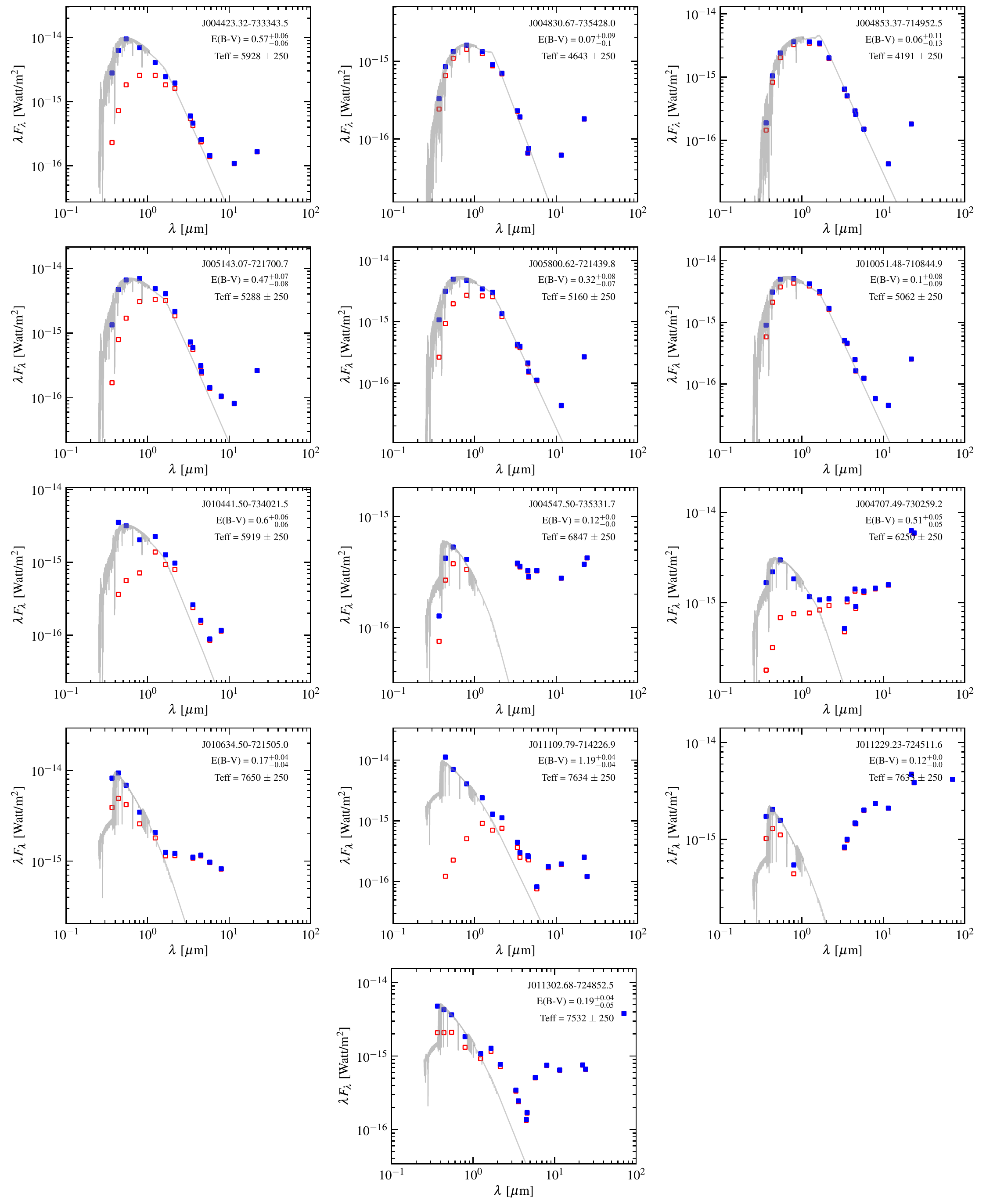}}\,
\caption{Same as Figure ~\ref{fig:pagb1_sed}, but for the Q2 
YSO candidates.}
\label{fig:yso2_sed}
\end{figure*}

\begin{table*}
{\renewcommand{\arraystretch}{0.8}
\caption{Features in the spectra of the sample of Q1 
post-AGB/RGB candidates. In this table 'a' represents absorption, 
'e' represents emission, '0' indicates that the feature is not observed. ':' 
indicates that there is some line blending that has taken place or there is an absorption line with an emission 
core or the line indicates signs of strong winds and therefore mass-loss. '$?$' 
represents that the nature of the spectral line is uncertain. 'p' indicates a p-cygni line profile. Note: the 
low-resolution of the spectra could possibly affect the identification of a given line.}
\medskip
\tabcolsep=2pt
\begin{tabular}{lrrrrrrrrrrrrrrrrr}
\hline
Name & H$\alpha$ & H$\beta$ & H$\gamma$ & [OIII] & [OIII] & HeI & HeI & [SII] & [SII] & [NII] & [NII] & CaII & CaII & CaII & Li & Ba & Pa \\ 
Wavelength (\AA) & 6563 & 4861 & 4341 & 4659 & 5007 & 4471 & 5876 & 6717 & 6731 & 6548 & 6584 & 8498 & 8542 & 8662 & 6708 & 4554 & - \\ 
\hline
\multicolumn{18}{c}{Candidates with [Fe/H] estimates from spectra}\\ 
\hline
J003441.01-732908.0 & a & 0 & 0 & 0 & 0 & 0 & 0 & 0 & 0 & 0 & 0 & a & a & a & 0 & 0 & 0 \\ 
J003643.94-723722.1 & a & a & a & 0 & 0 & 0 & 0 & 0 & 0 & 0 & 0 & a & a & a & 0 & a & a \\ 
J003829.99-730334.1 & a & a & a & 0 & 0 & 0 & 0 & 0 & 0 & 0 & 0 & a & a & a & 0 & 0 & 0 \\ 
J003908.89-724314.9 & a & a & ? & 0 & 0 & 0 & 0 & 0 & 0 & 0 & 0 & a & a & a & 0 & 0 & 0 \\
J003941.74-714708.5 & a & a? & 0 & 0 & 0 & 0 & 0 & 0 & 0 & 0 & 0 & a & a & a & a? & 0 & 0 \\ 
J004050.18-733226.6 & a & a & a & 0 & 0 & e & 0 & 0 & 0 & 0 & 0 & a & a & a & 0 & 0 & 0 \\ 
J004114.10-741130.1 & a & a & 0 & 0 & 0 & 0 & 0 & 0 & 0 & 0 & 0 & a & a & a & 0 & a & a \\ 
J004441.03-732136.0 & 0 & a: & a: & 0 & ? & 0 & 0 & 0 & 0 & 0 & 0 & a & a & a & 0 & a & a \\ 
J004906.52-733136.1 & e & e  & e  & 0 & 0 & e & e & e & e & e? & e? & a & a & a & a? & 0& 0 \\
J004909.72-724745.4 & e: & a: & a: & 0 & e & 0 & 0 & e & e & a: & 0 & a & a & a & a & 0 & a \\ 
J004944.15-725209.0 & e & e & a & 0 & e & 0 & 0 & e & e & 0 & 0 & a & a & a & a & 0 & 0 \\ 
J005107.19-734133.3 & e & a: & a & 0 & 0 & 0 & 0 & 0 & 0 & 0 & 0 & a & a & a & 0 & a & a \\ 
J005159.04-734214.4 & a & a & 0 & 0 & 0 & 0 & 0 & 0 & 0 & 0 & 0 & a & a & a & a? & 0 & 0 \\ 
J005252.87-722842.9 & a & a & a & 0 & 0 & 0 & 0 & e & e & 0 & 0 & a & a & a & 0 & 0 & a \\ 
J005307.35-734404.5 & a & a & 0 & 0 & 0 & 0 & 0 & 0 & 0 & 0 & 0 & a & a & a & 0 & 0 & 0 \\ 
J005310.08-722921.0 & e & a & a & 0 & 0 & 0 & 0 & 0 & 0 & 0 & 0 & a & a & a & 0 & 0 & 0 \\ 
J005803.08-732245.1 & a & a: & a: & 0 & 0 & 0 & 0 & 0 & 0 & 0 & 0 & a & a & a & 0 & 0 & a \\ 
J005914.20-723327.1 & e & e & 0 & 0 & e & 0 & 0 & e & e & 0 & 0 & a & a & a & 0 & 0 & 0 \\ 
J010056.93-715551.3 & a & a & a & 0 & 0 & 0 & 0 & 0 & 0 & 0 & 0 & a & a & a & a & 0 & 0 \\ 
J010247.72-740151.6 & a & a & 0 & 0 & 0 & 0 & 0 & 0 & 0 & 0 & 0 & a & a & a & 0 & a & 0 \\ 
J010333.93-724405.1 & a & a & a & e & 0 & 0 & 0 & 0 & 0 & 0 & 0 & a & a & a & 0 & 0 & 0 \\ 
J011219.69-735125.9 & e & e & a & 0 & 0 & a & a & 0 & 0 & 0 & 0 & a & a & a & 0 & 0 & a \\ 
J011222.88-715820.4 & a & a? & a? & 0 & 0 & 0 & 0 & 0 & 0 & 0 & 0 & a & a & a & a? & 0 & 0 \\ 
\hline
\multicolumn{18}{c}{Candidates with assumed [Fe/H] = -1.00}\\ 
\hline
J004534.36-734811.8 & e & a & a & 0 & 0 & 0 & 0 & 0 & 0 & 0 & 0 & 0& 0& 0& 0& 0& a \\
J004456.21-732256.6 & e & e & e: & 0 & a?: & a: & 0 & e & 0 & 0 & 0 & e & e & e & 0 & 0 & e \\ 
J004614.67-723519.0 & e & e & 0 & 0 & 0 & 0 & 0 & 0 & 0 & 0 & 0 & 0 & 0 & 0 & 0 & 0 & 0 \\ 
J004629.29-731552.3 & e & e & e: & 0 & e & 0 & 0 & e & e & 0 & 0 & 0 & 0 & 0 & a? & 0 & e \\
J004644.05-735944.7 & e & e & a: & 0 & 0 & 0 & 0 & 0 & 0 & 0 & 0 & e & e & e & 0 & 0 & 0 \\ 
J004854.24-735651.9 & a: & a: & 0 & 0 & 0 & 0 & 0 & 0 & 0 & 0 & 0 & a & a & a & 0 & 0 & 0 \\ 
J005104.61-722058.5 & e & e: & ? & 0 & e? & 0 & 0 & e & e & e? & e? & a?& a? & a? & 0& 0& 0 \\
J005113.04-722227.0 & e & e & ? & 0 & e & 0 & 0 & e & e & 0 & 0 & 0 & 0& 0& ? & 0& e:? \\
J005136.79-722818.0 & e & e & e & 0 & e & 0 & 0 & 0 & 0 & 0 & 0 & 0& 0 & 0 & 0& 0& e \\
J005310.47-732800.4 & e & e & ? & 0 & e & e? & ? & ? & ? & ? & ? & a? & a? & a? & 0& 0& ? \\
J005327.75-733339.6 & e: & a & a &  0 & 0 & 0 & 0 & 0 & 0 & 0 & 0 & ? & ? & ? & 0& 0& 0 \\
J005506.24-731347.6 & e & e & a: & 0 & 0 & a & 0 & 0 & 0 & 0 & 0 & 0 & 0& 0& 0 & ? & a \\
J005553.75-720859.7 & e & e & a: & 0 & 0 & 0 & 0 & 0 & 0 & 0 & 0 & e & e & e & 0 & 0 & 0 \\ 
J005908.99-710648.6 & e?& ? & ? & 0 & 0 & 0 & 0 & 0 & 0 & 0 & 0 & a & a & a & 0& a & 0 \\
J010342.34-721342.7 & e & e & a & 0 & e & a & 0 & 0 & 0 & 0 & 0 & e & e & e & 0 & 0 & 0 \\ 
\hline
\label{tab:pagb1lines}
\end{tabular}}
\end{table*}

\begin{table*}
{\renewcommand{\arraystretch}{0.8}
\caption{Same as Table ~\ref{tab:pagb1lines}, for the 
Q2 post-AGB/RGB candidates.}
\medskip
\tabcolsep=2pt
\begin{tabular}{lrrrrrrrrrrrrrrrrr}
\hline
Name & H$\alpha$ & H$\beta$ & H$\gamma$ & [OIII] & [OIII] & HeI & HeI & [SII] & [SII] & [NII] & [NII] & CaII & CaII & CaII & Li & Ba & Pa \\ 
Wavelength (\AA) & 6563 & 4861 & 4341 & 4659 & 5007 & 4471 & 5876 & 6717 & 6731 & 6548 & 6584 & 8498 & 8542 & 8662 & 6708 & 4554 & - \\ 
\hline
\multicolumn{18}{c}{Candidates with [Fe/H] estimates from spectra}\\ 
\hline
J003611.06-730447.0 & a & a? & 0 & 0 & 0 & 0 & 0 & 0 & 0 & 0 & 0 & 0 & a & 0 & 0 & 0 & 0 \\ 
J003818.36-731120.7 & e & 0 & 0 & 0 & 0 & 0 & 0 & 0 & 0 & 0 & 0 & e? & a & a & 0 & 0 & 0 \\ 
J003946.58-730433.5 & a & a? & 0 & 0 & 0 & 0 & 0 & 0 & 0 & 0 & 0 & a & a & a & 0 & 0 & 0 \\ 
J004215.31-740219.1 & a & a & 0 & 0 & 0 & 0 & 0 & 0 & 0 & 0 & 0 & a & a & a & 0 & 0 & 0 \\ 
J004431.23-730549.3 & e & 0 & 0 & 0 & e & 0 & e & 0 & 0 & 0 & 0 & a & a & a & 0 & 0 & 0 \\ 
J005222.19-733537.6 & 0 & 0 & 0 & 0 & 0 & 0 & 0 & 0 & 0 & 0 & 0 & a & a & a & a? & 0 & 0 \\ 
J005311.41-740621.2 & 0 & 0 & 0 & 0 & 0 & 0 & 0 & a? & a? & 0 & 0 & a & a & a & 0 & 0 & 0 \\ 
J005447.59-740121.4 & a & a & 0 & 0 & 0 & 0 & 0 & e & e & 0 & 0 & a & a & a & 0 & 0 & 0 \\ 
J005515.71-712516.9 & a & a & 0 & 0 & 0 & 0 & 0 & 0 & 0 & 0 & 0 & a & a & a & 0 & 0 & 0 \\ 
J005526.37-723248.7 & e & e & 0 & 0 & e & 0 & 0 & e & e & 0 & 0 & a & a & a & 0 & 0 & 0 \\ 
J005658.04-735059.7 & a & 0 & 0 & 0 & 0 & 0 & 0 & 0 & 0 & 0 & 0 & a & a & a & 0 & 0 & 0 \\ 
J005925.13-741309.6 & a & a & a & 0 & 0 & e & 0 & 0 & 0 & 0 & 0 & a & a & a & a & 0 & 0 \\ 
J005941.66-742842.9 & e & a & 0 & 0 & 0 & 0 & 0 & 0 & 0 & 0 & 0 & a & a & 0 & 0 & a? & 0 \\ 
J010021.78-730901.3 & e & a & 0 & 0 & 0 & 0 & 0 & 0 & e & 0 & 0 & a & a & a & a & 0 & 0 \\ 
J010049.88-723459.7 & e & e & 0 & e & e & 0 & 0 & e & e & 0 & 0 & a & a & a & a & 0 & 0 \\ 
J010254.90-722120.9 & e & e & 0 & 0 & e & 0 & 0 & e? & e? & 0 & 0 & a & a & a & a? & 0 & 0 \\ 
J010304.72-721245.3 & e & e & e & 0 & e & 0 & e: & e & e & e & e & a & a & a & 0 & 0 & 0 \\ 
J010310.25-730602.7 & a: & a & 0 & 0 & 0 & 0 & 0 & 0 & 0 & 0 & 0 & a & a & a & 0 & 0 & 0 \\ 
J010404.07-723521.5 & e: & e & 0 & 0 & 0 & 0 & 0 & 0 & 0 & 0 & 0 & a & a & a & 0 & 0 & 0 \\ 
J010549.25-725028.9 & e & e & a & 0 & e & 0 & 0 & 0 & e & 0 & 0 & a & a & a & a? & 0 & 0 \\ 
J010623.71-724413.5 & a & 0 & a & e & 0 & 0 & 0 & 0 & 0 & 0 & 0 & a & a & a & a & 0 & 0 \\ 
J010814.67-721306.2 & e & 0 & a & 0 & 0 & 0 & 0 & 0 & 0 & 0 & 0 & a & a & a & a? & 0 & 0 \\ 
J011133.41-733300.6 & a & a & 0 & 0 & 0 & 0 & 0 & 0 & 0 & 0 & 0 & a & a & a & 0 & 0 & 0 \\ 
J011149.89-720822.4 & a & a & a & e & 0 & a & e & 0 & 0 & 0 & 0 & a & a & a & a? & 0 & 0 \\ 
\hline
\multicolumn{18}{c}{Candidates with assumed [Fe/H] = -1.00}\\ 
\hline
J003549.26-740737.9 & e & e & 0 & 0 & 0 & 0 & 0 & 0 & 0 & 0 & 0 & e & e & e & 0 & 0 & 0 \\ 
\hline
\label{tab:pagb2lines}
\end{tabular}}
\end{table*}

\begin{table*}
{\renewcommand{\arraystretch}{0.8}
\caption{Same as Table ~\ref{tab:pagb1lines}, but for the Q1 YSO candidates.}
\medskip
\tabcolsep=2pt
\begin{tabular}{lrrrrrrrrrrrrrrrrr}
\hline
Name & H$\alpha$ & H$\beta$ & H$\gamma$ & [OIII] & [OIII] & HeI & HeI & [SII] & [SII] & [NII] & [NII] & CaII & CaII & CaII & Li & Ba & Pa \\ 
Wavelength (\AA) & 6563 & 4861 & 4341 & 4659 & 5007 & 4471 & 5876 & 6717 & 6731 & 6548 & 6584 & 8498 & 8542 & 8662 & 6708 & 4554 & - \\ 
\hline
\multicolumn{18}{c}{Candidates with [Fe/H] estimates from spectra}\\ 
\hline
J004927.26-724738.4 & e & a & a & 0 & e & 0 & 0 & e & e & 0 & 0 & a & a & a & a? & 0 & a \\ 
J004949.43-731327.3 & e & 0 & 0 & e? & e & 0 & 0 & e & e & 0 & 0 & a & a & a & a & 0 & 0 \\ 
J010134.91-720605.4 & 0 & a & 0 & 0 & e & 0 & 0 & 0 & 0 & 0 & 0 & a & a & a & 0 & 0 & 0 \\ 
J010222.29-724502.6 & 0 & 0 & 0 & 0 & 0 & 0 & 0 & e & e & 0 & 0 & a & a & a & 0 & 0 & 0 \\ 
J010648.26-734305.4 & e & a & a & 0 & 0 & 0 & 0 & 0 & 0 & 0 & 0 & a & a & a & 0 & 0 & a \\ 
J011316.84-733130.9& ? & ? & ?& 0 & 0 & 0 & 0 & 0 & 0 & 0 & 0 & a& a& a& 0& 0& 0 \\
\hline
\multicolumn{18}{c}{Candidates with assumed [Fe/H] = -1.00}\\ 
\hline
J003640.64-740747.2& e& e:?& ?& 0&0& 0& 0& 0& 0& e?& e?& ?& a& a& 0& 0& 0 \\
J004208.74-733108.4 & e & e: & a: & e & e & a & 0 & e & e & 0 & 0 & e & e? & e? & 0 & 0 & 0 \\ 
J004221.85-732417.5 & e & a & a & 0 & 0 & 0 & 0 & 0 & 0 & 0 & 0 & a & a & a & 0 & 0 & a \\ 
J004301.63-732050.9 & e & e: & e: & 0 & 0 & e? & 0 & e & e & e? & 0 & 0 & 0 & 0 & 0 & 0 & 0 \\ 
J004451.87-725733.6 & e & e & e:& 0 & 0 & 0 & 0 & e& e& 0& 0& 0 & 0& 0& 0& 0& e \\
J004501.19-723321.0& e& a: & a: & 0 & 0 & 0 & 0 & 0 & 0 & 0 & 0 & 0& 0& 0& 0& 0& a? \\
J004503.51-731627.4& e & e & e: & 0  & 0 & 0 & 0 & e & e & 0& 0& 0 & 0 & 0 & 0& 0& e \\
J004657.45-731143.4 & e & e & e & 0 & e & 0 & e & e & e & e? & e? & 0 & 0& 0& 0& 0& ? \\
J004831.82-720535.7 & e & a & a & 0 & 0 & 0 & 0 & e & e & 0 & 0 & a & a & a & a & 0 & 0 \\ 
J004840.55-730101.3 & e & e & 0 & 0 & e? & 0 & e & e & e & 0 & 0 & e & e & e & 0 & 0 & e \\ 
J004905.36-721959.9 & ? & ? & ? & 0 & ?   & 0 & 0 & 0 & 0 & 0 & 0 & a & a & a & 0 & 0 & 0\\
J004950.02-734011.5 & a & a & a:? & 0 & 0e? & 0 & 0 & 0 & 0 & 0 & 0 & a & a & a& 0& 0& 0 \\
J005101.48-733100.4* & e & e & ? & e & e & ? & ? & e & e& e?& e?& a& a& a& 0& a?& 0 \\
J005112.29-722552.7 & e & e & ? & e & e & e? & 0 & e & e & 0 & 0 & a & a & a & a? & 0& 0 \\
J005159.81-723511.1 & e & e: & a: & 0 & e & 0 & 0 & e & e & 0 & 0 & e & e & e & 0 & 0 & 0 \\ 
J005318.28-733528.7 & e & a: & a: & 0 & e & 0 & 0 & e & e & 0 & 0 & 0 & 0 & 0 & 0 & 0 & 0 \\ 
J005606.53-724722.7 & e & e & e & 0 & e & a: & 0 & e & e & 0 & 0 & 0 & 0 & 0 & 0 & 0 & 0 \\ 
J005934.21-733025.2 & e & a:? & a:? & 0 & 0 & 0 & 0 & 0 & 0 & 0 & 0 & 0& 0& 0& 0& 0& 0 \\
J010242.25-720306.0 & e & e & e: & 0 & e & 0 & 0 & e& e& 0& 0& 0& 0& 0& 0& 0& e \\
J010309.59-715354.2 & e& ep& ep?& e?& e& ?& 0& e& e& e& e & e& e& e& 0& 0& 0 \\
J010427.62-721037.0 & e & e & e: & 0 & e & e? & 0 & e& e& e& e & ?& ? & ? & 0& 0& ? \\
\hline
\label{tab:yso1lines}
\end{tabular}}
\end{table*}

\begin{table*}
{\renewcommand{\arraystretch}{0.8}
\caption{Same as Table ~\ref{tab:pagb1lines}, but for the 
Q2 YSO candidates.}
\medskip
\tabcolsep=2pt
\begin{tabular}{lrrrrrrrrrrrrrrrrr}
\hline
Name & H$\alpha$ & H$\beta$ & H$\gamma$ & [OIII] & [OIII] & HeI & HeI & [SII] & [SII] & [NII] & [NII] & CaII & CaII & CaII & Li & Ba & Pa \\ 
Wavelength (\AA) & 6563 & 4861 & 4341 & 4659 & 5007 & 4471 & 5876 & 6717 & 6731 & 6548 & 6584 & 8498 & 8542 & 8662 & 6708 & 4554 & - \\ 
\hline
\multicolumn{18}{c}{Candidates with [Fe/H] estimates from spectra}\\ 
\hline
J004423.32-733343.5 & e & 0 & 0 & 0 & e & 0 & 0 & e & e & 0 & 0 & a & a & a & a & 0 & 0 \\ 
J004830.67-735428.0 & a & 0 & 0 & 0 & 0 & 0 & 0 & 0 & 0 & 0 & 0 & a & a & a & 0 & 0 & 0 \\ 
J004853.37-714952.5 & a & a & 0 & 0 & 0 & 0 & 0 & 0 & 0 & 0 & 0 & a & a & a & 0 & 0 & 0 \\ 
J005143.07-721700.7 & e & e & e & 0 & e & 0 & e & e & e & 0 & 0 & a & a & a & 0 & 0 & 0 \\ 
J005800.62-721439.8 & e & e & e & 0 & e & 0 & 0 & e & e & 0 & 0 & a & a & a & 0 & 0 & 0 \\ 
J010051.48-710844.9 & a & a & 0 & 0 & 0 & 0 & 0 & 0 & 0 & 0 & 0 & a & a & a & 0 & 0 & 0 \\ 
J010441.50-734021.5 & a & 0 & 0 & 0 & e & 0 & 0 & 0 & 0 & 0 & 0 & a & a & a & 0 & 0 & 0 \\ 
\hline
\multicolumn{18}{c}{Candidates with assumed [Fe/H] = -1.00}\\ 
\hline
J004547.50-735331.7 & e & a: & a: & 0 & 0 & 0 & 0 & 0 & e: & 0 & 0 & 0 & 0 & 0 & 0 & 0 & 0 \\ 
J004707.49-730259.2 & e & e & 0 & 0 & e & 0 & 0 & e & e & 0 & 0 & 0 & 0 & 0 & a? & 0 & 0 \\ 
J010634.50-721505.0 & e & e & e & 0 & 0 & 0 & 0 & e & e & 0 & 0 & e & e & e & 0 & 0 & 0 \\ 
J011109.79-714226.9 & a: & a: & a: & 0 & 0 & 0 & 0 & 0 & 0 & 0 & 0 & 0 & 0 & 0 & 0 & 0 & 0 \\ 
J011229.23-724511.6 & e & e & a & 0 & 0 & 0 & 0 & 0 & 0 & 0 & 0 & e & e & e & 0 & 0 & 0 \\ 
J011302.68-724852.5 & e & a: & a: & 0 & 0 & 0 & 0 & 0 & 0 & 0 & 0 & e & 0 & 0 & 0 & 0 & 0 \\ 
\hline
\label{tab:yso2lines}
\end{tabular}}
\end{table*}

\begin{figure*}
\centering
\subfloat{\includegraphics[width=18.5cm]{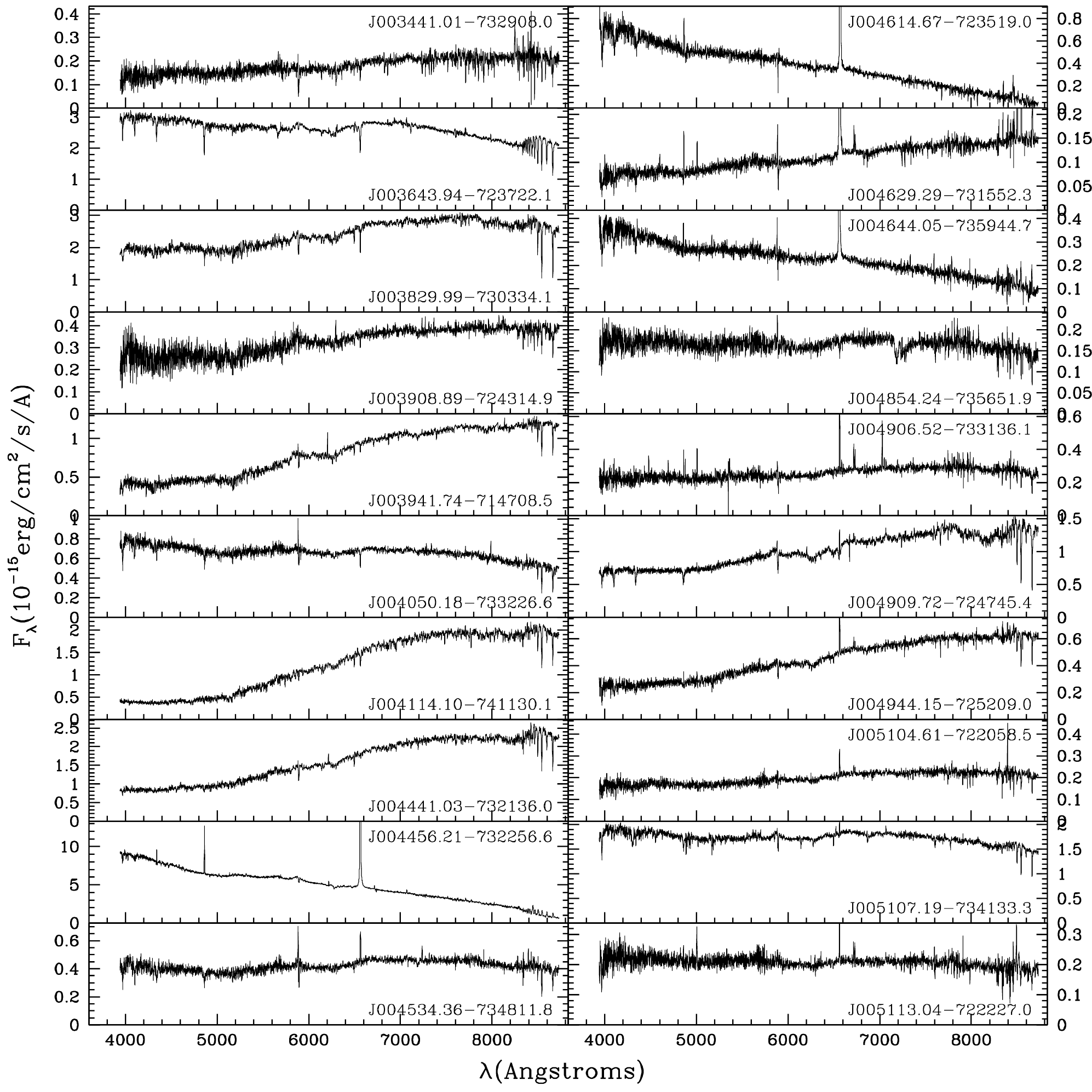}}
\caption{The low resolution AAOmega optical spectra for the 
sample of high probability Quality 1 post-AGB/RGB 
candidates. Occasionally, the spectra can have fluxes less than zero at the end due to poor sky subtraction. 
Note that for some objects the spectra of the region $>$7000\AA\,\, is dominated by noise introduced 
during sky subtraction. The emission feature near the sodium doublet is an artefact of the data reduction process resulting from poor sky subtraction 
of the sodium doublet emission from the SMC, Galaxy and the night sky. We note that, occasionally, the spectra can have 
fluxes less than zero at the end due to poor sky subtraction. The spectra are ordered by RA.}
\label{pagb1spec}
\end{figure*}
\clearpage

\begin{figure*}
\ContinuedFloat
\centering
\subfloat{\includegraphics[width=18.5cm]{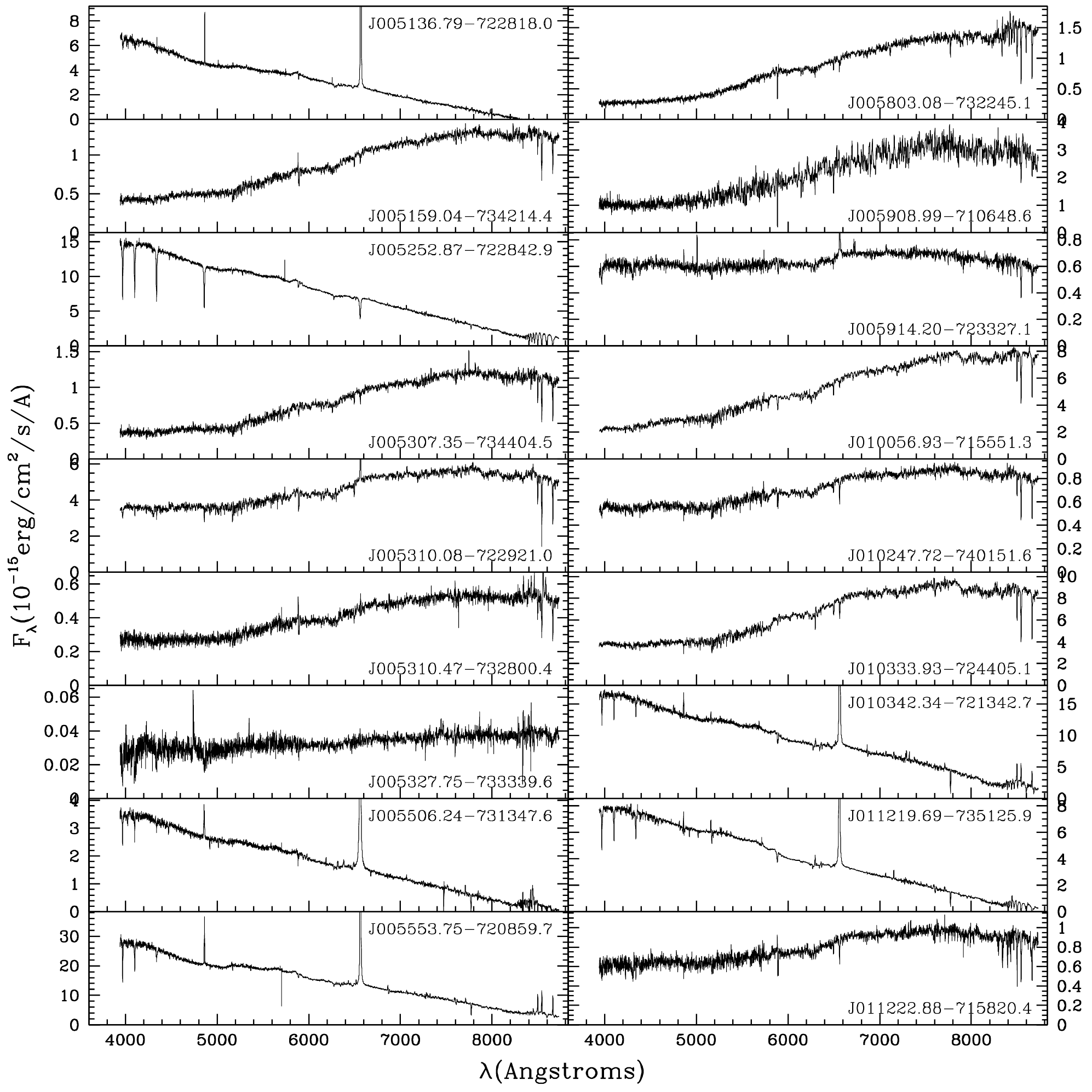}}
\caption{Figure~\ref{pagb1spec} continued.}
\end{figure*}
\clearpage

\begin{figure*}
\centering
\subfloat{\includegraphics[width=18.5cm]{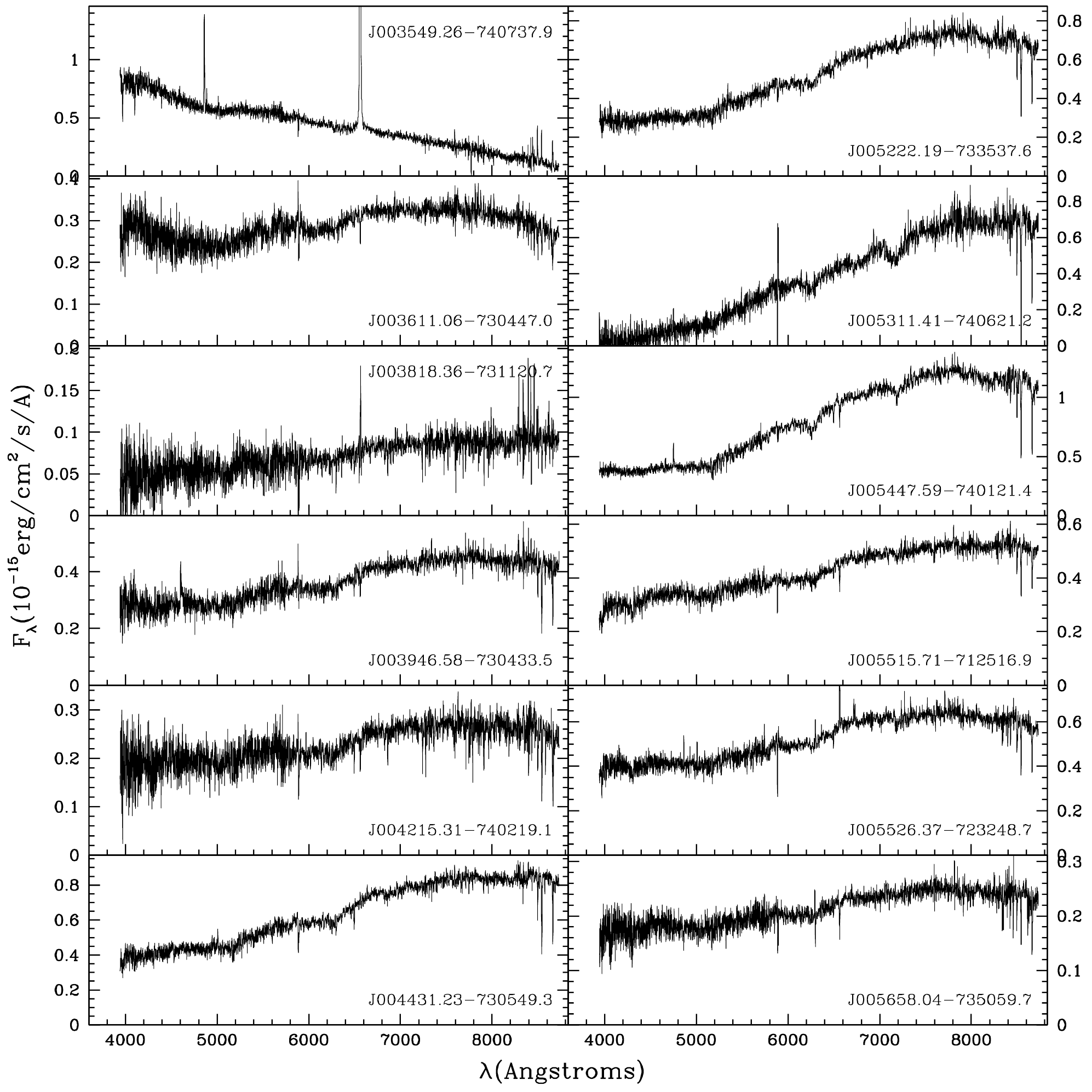}}
\caption{Same as Figure~\ref{pagb1spec}, but for the 
high probability Quality 2 post-AGB/RGB candidates.}
\label{pagb2spec}
\end{figure*}
\clearpage

\begin{figure*}
\ContinuedFloat
\centering
\subfloat{\includegraphics[width=18.5cm]{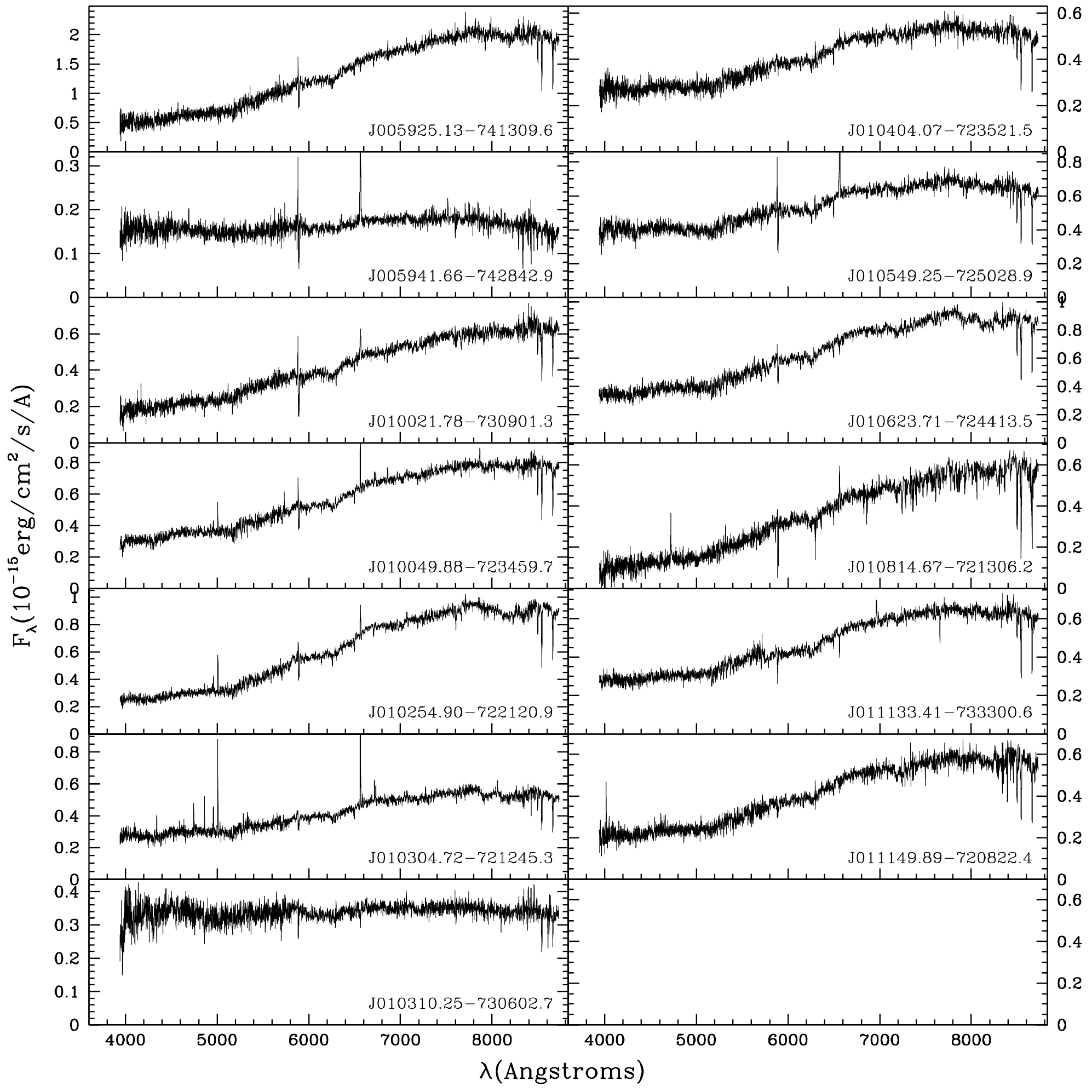}}
\caption{Figure~\ref{pagb2spec} continued.}
\end{figure*}
\clearpage

\begin{figure*}
\centering
\subfloat{\includegraphics[width=18.5cm]{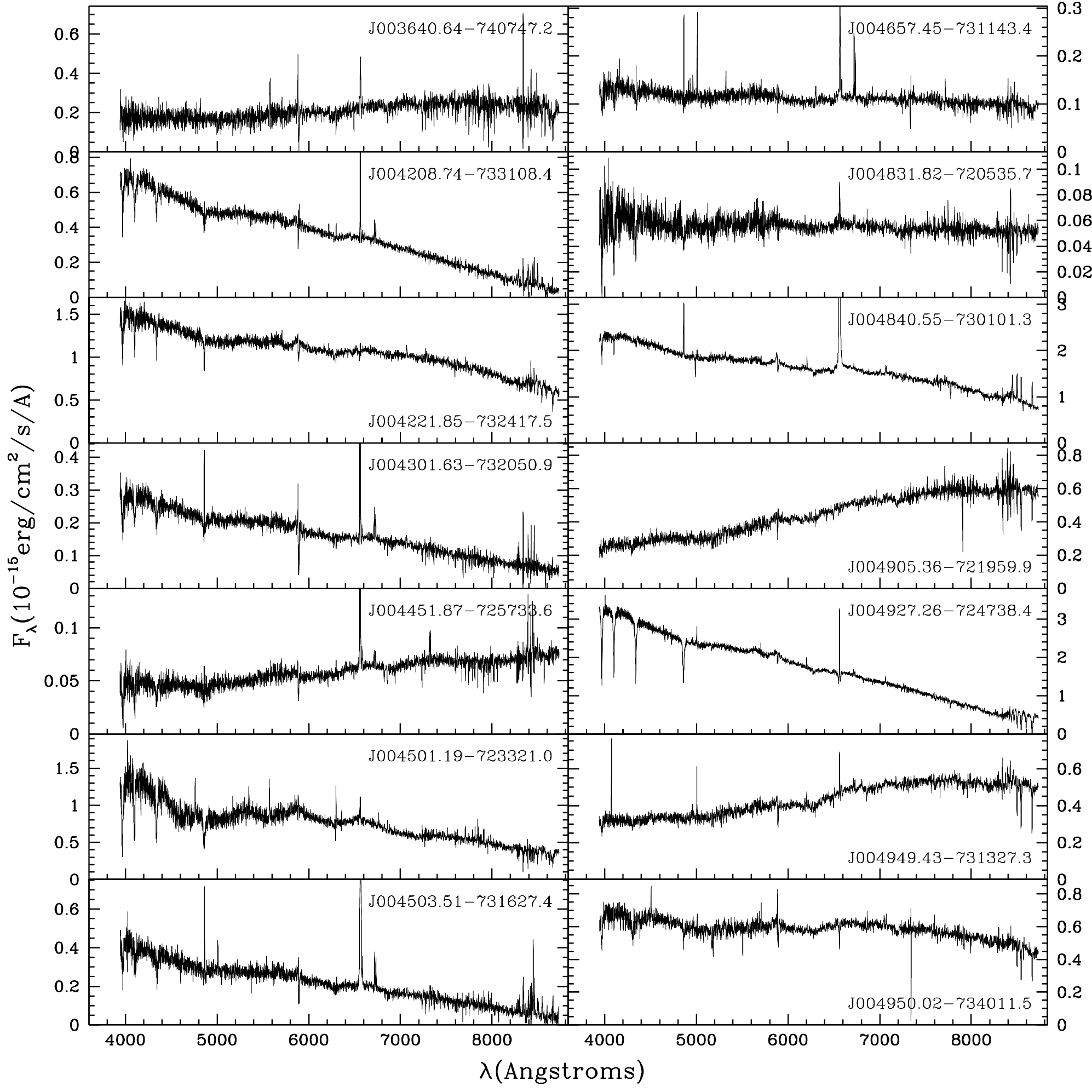}}
\caption{Same as Figure~\ref{pagb1spec} but for the 
high probability Quality 1 YSO candidates.}
\label{yso1spec}
\end{figure*}
\clearpage

\begin{figure*}
\ContinuedFloat
\centering
\subfloat{\includegraphics[width=18.5cm]{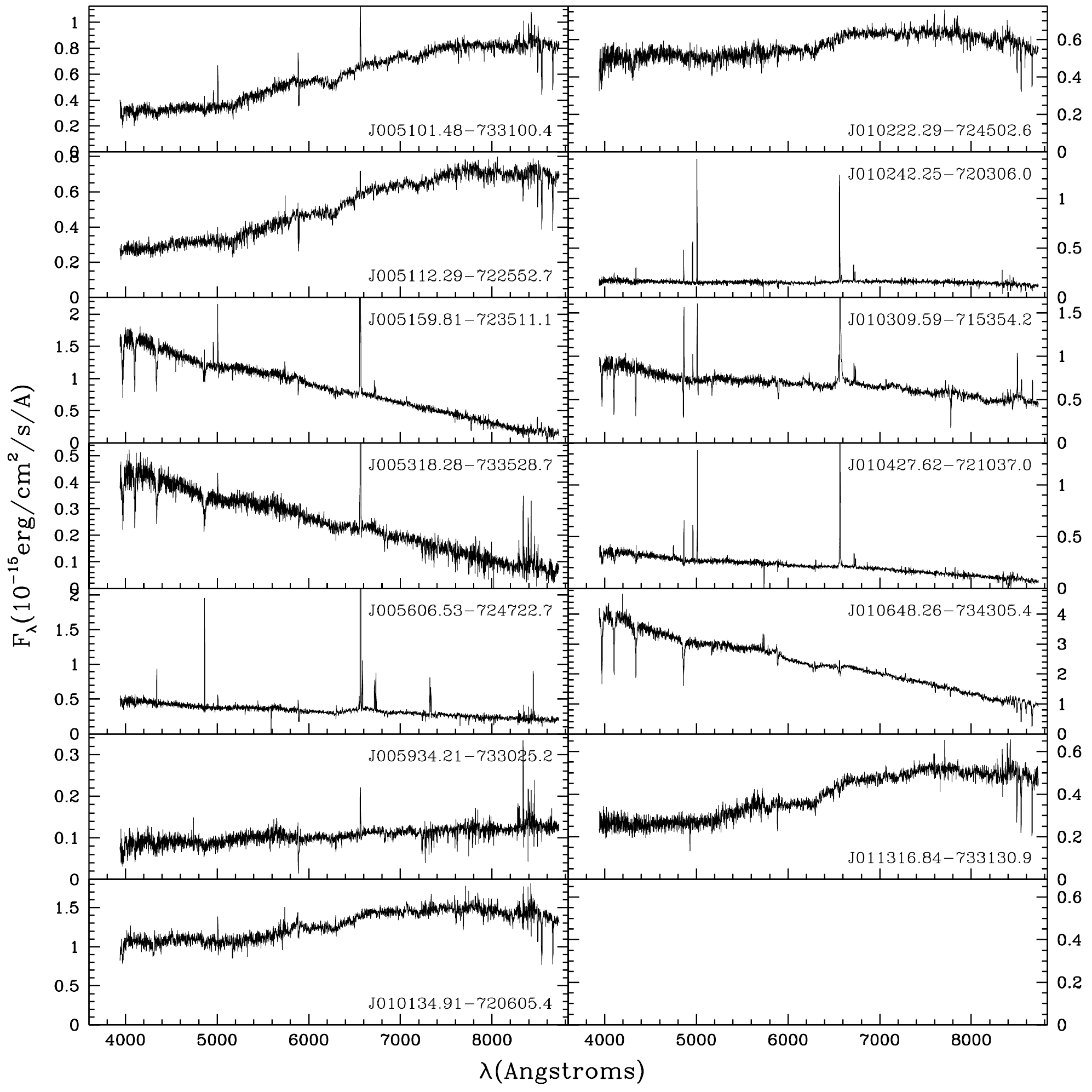}}
\caption{Figure~\ref{yso1spec} continued.}
\end{figure*}
\clearpage

\begin{figure*}
\centering
\subfloat{\includegraphics[width=18.5cm]{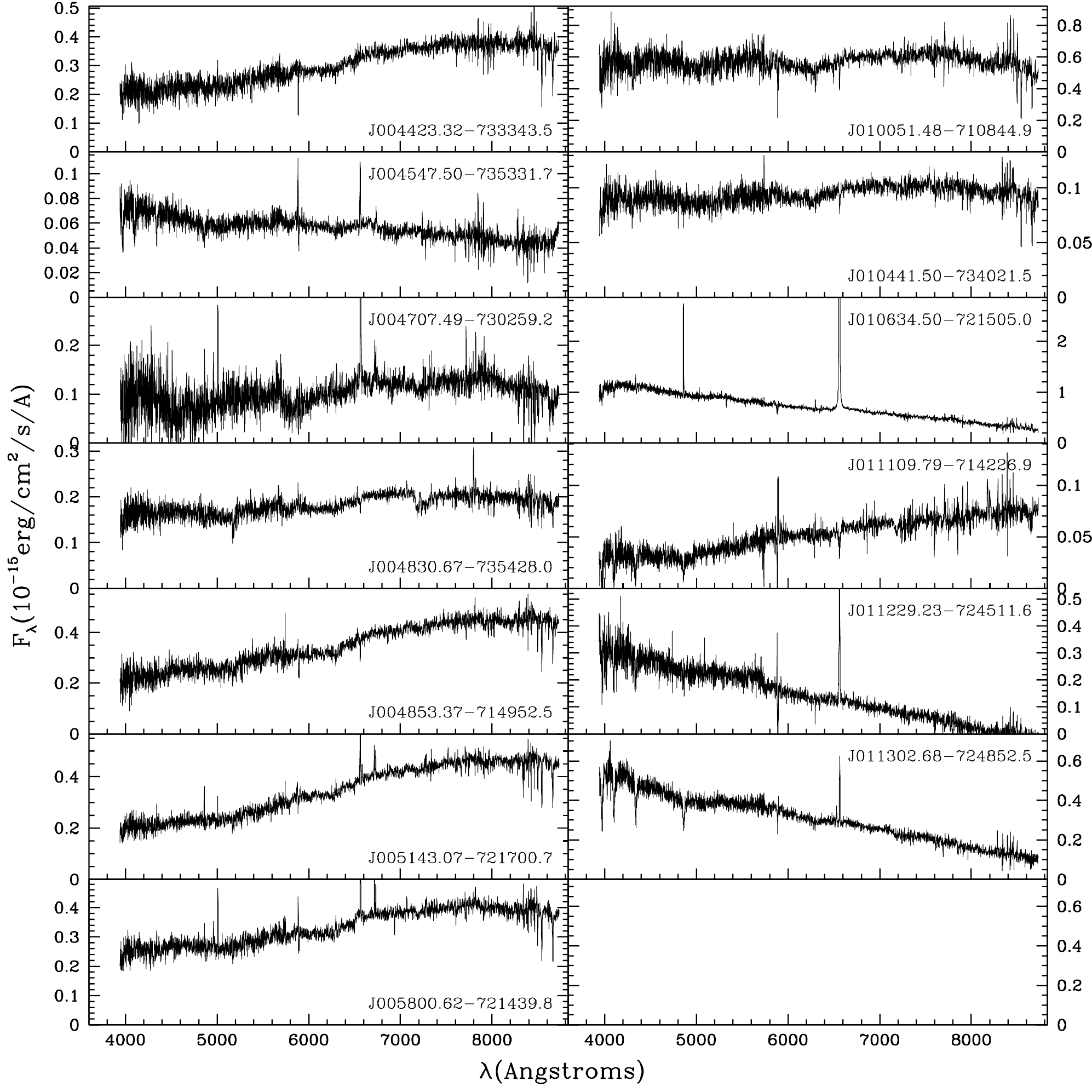}}
\caption{Same as Figure~\ref{pagb1spec} but for the 
high probability Quality 2 YSO candidates.}
\label{yso2spec}
\end{figure*}
\clearpage
\label{lastpage}
\end{document}